\documentclass[acmsmall,screen]{acmart} 

\setcopyright{rightsretained}
\acmJournal{PACMPL}
\acmYear{2023} \acmVolume{7} \acmNumber{PLDI} \acmArticle{127} \acmMonth{6} \acmPrice{}\acmDOI{10.1145/3591241}
\copyrightyear{2023}
\acmSubmissionID{pldi23main-p134-p}
\received{2022-11-10}
\received[accepted]{2023-03-31}

\usepackage[most]{tcolorbox}
\usepackage{agda/agda}
\usepackage{nameref}
\usepackage{cleveref}
\usepackage{afterpage}
\usepackage{fancyvrb}
\usepackage{apxproof}
\usepackage{ottalt,ottlayout}
\usepackage{subcaption}
\usepackage{bussproofs}
\usepackage{amsmath}
\usepackage{tikz}
\usepackage{tikz-cd}
\usepackage{listings}
\usetikzlibrary{calc}

\newenvironment{ottdefnblock}[3][]{ \framebox{\mbox{#2}} \quad #3 \\[0pt]}{}

\newcommand{\ottafterlastrule}{\\}

  \renewottcommands[ott]

\renewenvironment{ottdefnblock}[3][]{\defnblockSTY[#1]{#2}{#3}}{\enddefnblockSTY}

\ottstyledefaults{premiselayout=justify,premisenamelayout=right}
\DeclareMathAlphabet{\mathsf}{OT1}{\sfdefault}{m}{n}
\SetMathAlphabet{\mathsf}{normal}{OT1}{\sfdefault}{m}{n}
\DeclareUnicodeCharacter{03A0}{$\Pi$}
\DeclareUnicodeCharacter{2200}{$\forall$}
\DeclareUnicodeCharacter{03BB}{$\lambda$}
\DeclareUnicodeCharacter{2115}{$\mathbb{N}$}
\DeclareUnicodeCharacter{2080}{$_0$}
\DeclareUnicodeCharacter{2081}{$_1$}
\DeclareUnicodeCharacter{2082}{$_2$}
\DeclareUnicodeCharacter{2098}{$_m$}
\DeclareUnicodeCharacter{2099}{$_n$}
\DeclareUnicodeCharacter{1D48}{$_d$}
\DeclareUnicodeCharacter{1D62}{$_i$}
\DeclareUnicodeCharacter{219D}{$\leadsto$}

\crefformat{section}{\S#2#1#3}
\crefmultiformat{section}{\S\S#2#1#3}{ and~#2#1#3}{, #2#1#3}{, and~#2#1#3}
\crefformat{figure}{Fig.~#2#1#3}

\newenvironment{syntax}{\[\begin{array}{l@{\quad}lcl}}{\end{array}\]}

\newcommand{\ignore}[1]{}

\newtheoremrep{theorem}{Theorem}[section]
\newtheoremrep{lemma}[theorem]{Lemma}
\newtheoremrep{definition}[theorem]{Definition}
\newtheoremrep{coro}[theorem]{Corollary}

\newcommand{\flabel}{\mathfrak{L}}
\newcommand{\fdef}{\mathfrak{D}}

\newcommand{\pitype}[3]{\Pi #1{\goodcolon} #2 . #3}
\newcommand{\lam}[3]{\lambda#1\goodcolon#2.#3}

\newcommand{\sub}[2]{\text{[\ensuremath{#1}/\ensuremath{#2}]}}

\newcommand{\target}[1]{{\color{blue!95!white}{\mathsf{#1}}}}

\newcommand{\targetpi}[3]{\mathsf{\color{blue!95!white}\Pi\color{black} #1 {:} #2 . #3 }}
\newcommand{\targetapp}[2]{\target{#1\mathop{@}#2}}
\newcommand{\targetlab}[2]{\target{#1}\{\target{#2}\}}
\newcommand{\targetsub}[2]{\text{[\ensuremath{\target{#1}}/\ensuremath{\target{#2}}]}}
\newcommand{\targetcon}{\target{\fdef} \semicolon \target{\Gamma}}
\newcommand{\targettext}[1]{{\color{blue!95!white}{\textsf{#1}}}\color{black}}

\newcommand{\itemtype}[2]{\mathsf{\color{blue!95!white}#1 \color{black} {\goodcolon} \color{blue!95!white}#2}}

\newcommand{\sfcomma}{\textsf{, }}
\newcommand{\sfpl}{\textsf{(}}
\newcommand{\sfpr}{\textsf{)}}
\newcommand{\bbracket}[1]{{[\![ #1 ]\!]}}
\newcommand{\bluebr}[1]{\color{blue!95!white}{[\![ \color{black}#1\color{blue!95!white} ]\!]}\color{black}}

\newcommand{\ccs}{CC$_S$}
\newcommand{\helper}[1]{{\color{teal}{#1}}}
\newcommand{\hbracket}[1]{\helper{[\![}{#1}\helper{]\!]}}
\newcommand{\hbracketd}[1]{\helper{[\![}{#1}\helper{]\!]_d}}
\newcommand{\hsub}[2]{\helper{ \{#2 \mapsto #1\} }}

\newcommand{\semicolon}[0]{\hspace{0.5mm}\textsf{;}\hspace{0.5mm}}
\newcommand{\goodcolon}[0]{\hspace{0.5mm}{:}\hspace{0.5mm}}

\newcommand{\smalldisplayskips}{%
  \setlength{\abovedisplayskip}{3.5pt}%
  \setlength{\belowdisplayskip}{3.5pt}%
  \setlength{\abovedisplayshortskip}{3.5pt}%
  \setlength{\belowdisplayshortskip}{3.5pt}}
\appto{\normalsize}{\smalldisplayskips}
\appto{\small}{\smalldisplayskips}
\appto{\footnotesize}{\smalldisplayskips}

\begin{document}

\title{Defunctionalization with Dependent Types}         
\titlenote{We use different colours and fonts to distinguish different languages, and recommend reading this paper in colour.}             


\author{Yulong Huang}
\orcid{0009-0001-4288-5690}             
\affiliation{
  \institution{University of Cambridge}            
  \city{Cambridge}
  \country{UK}                    
}
\email{yh419@cam.ac.uk}          

\author{Jeremy Yallop}
\orcid{0009-0002-1650-6340}             
\affiliation{
  \institution{University of Cambridge}           
  \city{Cambridge}
  \country{UK}                   
}
\email{jeremy.yallop@cl.cam.ac.uk}         

\begin{abstract}
  The \emph{defunctionalization} translation that eliminates
  higher-order functions from programs forms a key part of many
  compilers.
  However, defunctionalization for dependently-typed languages has not
  been formally studied.

  We present the first formally-specified defunctionalization translation
  for a dependently-typed language and establish key metatheoretical properties 
  such as soundness and type preservation.
  The translation is suitable for incorporation into type-preserving
  compilers for dependently-typed languages.
\end{abstract}



\maketitle

\section{Introduction}
\label{section:introduction}

Types are increasingly used not merely for \emph{classification}
(i.e.~in identifying a subset of programs with desirable
properties), but for \emph{compilation}.
A \emph{type-preserving compiler}, organised as a series of
translations between two or more typed
languages~\cite[e.g.][]{DBLP:conf/pldi/TarditiMCSHL96,DBLP:journals/toplas/MorrisettWCG99,DBLP:conf/icfp/XiH01,DBLP:journals/pacmpl/BowmanCRA18,DBLP:conf/pldi/BowmanA18},
can support features such as type-driven elaboration of source
programs into more explicit core
calculi~\cite{DBLP:conf/popl/WadlerB89,DBLP:journals/pacmpl/Kovacs20},
translation of disparate source features into a simple uniform
core~\cite{DBLP:conf/tldi/SulzmannCJD07} and type-driven
optimizations~\cite{DBLP:conf/pldi/TarditiMCSHL96}.
More generally, type-preserving translations between intermediate
languages can increase confidence in correctness of the compilation
process~\cite{DBLP:journals/csur/PatrignaniAC19}.

As type systems increase in sophistication,
defining type-preserving presents new challenges.
Some of the most significant arise in the compilation of
\emph{dependently-typed} languages such as
Agda~\cite{DBLP:conf/tphol/BoveDN09},
Idris~\cite{DBLP:journals/jfp/Brady13},
and Coq~\cite{CoqDoc},
whose type systems are sufficiently expressive to support arbitrary computation.
It has proved difficult to adapt long-studied translations such as
\emph{continuation-passing style} conversion, \emph{closure
conversion}, and conversion into \emph{administrative normal form} to the dependently-typed
setting~\cite{DBLP:conf/pepm/BartheU02,DBLP:journals/lisp/BartheHS99,DBLP:journals/pacmpl/BowmanCRA18,DBLP:conf/pldi/BowmanA18,koronkevich_rakow_ahmed_bowman_2022}.

Another such translation, \emph{defunctionalization}, which eliminates
higher-order functions from programs, forms a key part of compilers
for several higher-order
languages~\cite[e.g.][]{mlton,generalized-stack-inspection,curry-implementation,grin}.
Type-preserving variants of defunctionalization are available for a
variety of type
systems~\cite{nielsen2000denotational,DBLP:conf/icfp/BellBH97,DBLP:conf/popl/PottierG04}.
Defunctionalization is also useful in the compilation of dependently-typed languages,
such as Idris~\footnote{\url{https://github.com/idris-lang/Idris-dev/blob/v1.3.4/src/IRTS/Defunctionalise.hs}}.
However, to date no type-preserving variant of the defunctionalization
translation for dependently-typed languages has been developed.

This work meets that need, introducing a typed defunctionalization
translation for a dependently-typed language, and establishing its
fundamental properties.
As with previous work that has adapted similar program translations to
support dependent types, we have encountered and resolved various
difficulties that do not arise in simply-typed settings.
In particular,
the need to preserve universe sizes
  (used by dependently-typed languages to avoid inconsistencies), and
to preserve reduction
  (used to establish type equality)
make a straightforward adaption of the standard defunctionalization unfeasible.

\subsection{Contributions}
\label{section:contributions}

The central contribution of this paper is the first type-preserving
defunctionalization translation for a dependently typed language.  In
more detail,

\begin{itemize}
\item
  \Cref{section:overview} shows that the type-preserving
defunctionalization translations used for simply typed
languages (\Cref{section:defunctionalization}) do not extend to a
dependently-typed setting (\Cref{section:size-problems}),
and presents an abstract translation suited to dependently-typed langauges
(\Cref{section:abstract-defunctionalization}).

\item
  \Cref{section:cc-defunctionalization} has the technical development
of our type-preserving defunctionalization translation.
\Cref{section:translation} formally defines the abstract translation
and \Cref{section:soundness} and \Cref{section:consistency} establish
key meta-theoretical properties such as soundness, type preservation,
and consistency.
\end{itemize}

Finally, \Cref{section:implementation} describes an implementation of
our translation (included as supplementary material), and 
\Cref{section:related} summarises related work on
type-preserving compilation and on defunctionalization.

\section{Overview}
\label{section:overview}

\subsection{Defunctionalization}
\label{section:defunctionalization}

The \emph{defunctionalization} translation
 turns higher-order programs
 into first-order programs,
 by replacing the function arrow \lstinline!- -> -!
 with a first-order data type \lstinline!- ~> -!.
Defunctionalization
  replaces each abstraction $\lambda x.e_i$
           in the source program
  with a constructor application $C_i\;\overline{y}$
  where $C_i$ is a constructor of \lstinline!- ~> -!
  and $\overline{y}$ are the free variables of the abstraction,
and replaces
  each application $f\;x$
  with $f\;\$\;x$,
  where the infix operator $\$$ maps $C_i$ back to $e_i$.


Here is an example.
The polymorphic \lstinline!compose! function
  contains
    three abstractions,
    here labeled
     {\color{red}\small F1},
     {\color{red}\small F2}, and
     {\color{red}\small F3}.

\begingroup
\setlength{\parindent}{0.8em}
\begin{tikzpicture}
\node[inner sep=0pt] (compose) {
\begin{lstlisting}
compose :: (b -> c) -> (a -> b) -> (a -> c)
compose = $\lambda$f$\;$->$\;\;\lambda$g$\;$->$\;\;\lambda$x$\;$->$\;\;$f (g x)
\end{lstlisting}
};
\node[anchor=south west,rectangle,draw,red,minimum height=3.5ex,minimum width=16ex] (f3) at ($(compose.south west) + (25.5ex,-1.1ex)$)  { };
\node[anchor=south west,rectangle,draw,red,minimum height=5.6ex,minimum width=24ex] (f2) at ($(compose.south west) + (18.5ex,-2.9ex)$)  { };
\node[anchor=south west,rectangle,draw,red,minimum height=7.7ex,minimum width=31.5ex] (f1) at ($(compose.south west) + (12ex,-4.7ex)$)  { };
\node[fill=white] at ($(f1.south) + (0,0.2ex)$) {\color{red}\scriptsize F1 };
\node[fill=white] at ($(f2.south) + (0,0.2ex)$) {\color{red}\scriptsize F2 };
\node[fill=white] at ($(f3.south) + (0,0.2ex)$) {\color{red}\scriptsize F3 };

\end{tikzpicture} 
\endgroup

Defunctionalizing \lstinline!compose! produces a data type \lstinline!~>! with
one constructor for each abstraction.
Here \lstinline!->! separates constructor arguments:
  \lstinline!F2! has one argument of type \lstinline!b ~> c!,
    corresponding to \lstinline!f! in {\color{red}\small F2} above, and
  \lstinline!F3! has two arguments,
    corresponding to \lstinline!f! and \lstinline!g! in {\color{red}\small F3}.

\begin{center}
\begin{tabular}{c}
\begin{lstlisting}
data (~>) a b where
  F1 :: (b ~> c) ~> (a ~> b) ~> (a ~> c)
  F2 :: (b ~> c) -> (a ~> b) ~> (a ~> c)
  F3 :: (b ~> c) -> (a ~> b) -> (a ~> c)
\end{lstlisting}
\end{tabular}
\end{center}

\noindent
Following~\citet{DBLP:conf/popl/PottierG04}, the data type \lstinline!~>! produced by defunctionalization is a \emph{generalized algebraic data type} (GADT),
in which
  the return type of each constructor
     can have a distinct instantiation of the type parameters, and
  constructor types can involve type variables
     (such as \lstinline!b! in the type of \lstinline!F3!)
     that do not appear in return types.

Defunctionalization also produces an operator \lstinline!$\$$!
that maps the constructors of \lstinline!~>! to the bodies
of the corresponding abstractions:

\begin{center}
\begin{tabular}{c}
\begin{lstlisting}
($\$$) :: (a ~> b) -> a -> b
F1     $\$$ f = F2 f
F2 f   $\$$ g = F3 f g
F3 f g $\$$ x = f $\$$ (g $\$$ x)
\end{lstlisting}
\end{tabular}
\end{center}

\noindent
Here \lstinline!$\$$! maps \lstinline!F1! and the argument $x$ to \lstinline!F2 x!,
since the body of the abstraction {\color{red}\small F1} is {\color{red}\small F2}, with \lstinline!x! free.
Similarly, it maps \lstinline!F3! to \lstinline!f $\$$ (g $\$$ x)!
because the body of {\color{red}\small F3} is \lstinline!f (g x)!.

Finally, defunctionalization
  replaces 
     \lstinline!->! with \lstinline!~>! and
     {\color{red}\small F1} with \lstinline!F1!
   in \lstinline!compose! itself:

\begin{center}
\begin{tabular}{c}
\begin{lstlisting}
compose_ :: (b ~> c) ~> (a ~> b) ~> (a ~> c)
compose_ = F1
\end{lstlisting}
\end{tabular}
\end{center}

Previous work on defunctionalization in typed settings has examined a variety of languages,
from
simply-typed~\cite{nielsen2000denotational} and
monomorphizable~\cite{DBLP:conf/icfp/BellBH97}
to fully polymorphic~\cite{DBLP:conf/popl/PottierG04}.
In each case,
defunctionalization represents functions
  using inductive data types,
   from simple algebraic datatypes for simply-typed functions,
   to the more powerful \emph{generalized} algebraic data types for polymorphism.

Might the same approach,
extended to yet more powerful data types,
support defunctionalization of dependently-typed programs?
Polymorphic functions abstract over types
 and defunctionalize to GADTs indexed by types.
By analogy, might dependent functions,
   which abstract over expressions,
 defunctionalize to inductive families indexed by expressions?

\subsection{Inductive families}
\label{section:inductive-families}

We briefly recall inductive families and their associated
restrictions.
Inductive families~\cite{DBLP:journals/fac/Dybjer94} generalize both
ordinary inductive data types and generalized algebraic data types by
permitting indexing by expressions.

The constructor of each inductive family may return an instantiation
of the family instantiated with arbitrary indexes.
For example, in the following Agda~\cite{DBLP:conf/afp/Norell08} definition the
constructors of the inductive family of finite sets, \AgdaInductiveConstructor{Fin},
are indexed by natural numbers \AgdaBound{n}:
\begin{code}[hide]%
\>[0]\AgdaKeyword{data}\AgdaSpace{}%
\AgdaDatatype{ℕ}\AgdaSpace{}%
\AgdaSymbol{:}\AgdaSpace{}%
\AgdaPrimitive{Set}\AgdaSpace{}%
\AgdaKeyword{where}\<%
\\
\>[0][@{}l@{\AgdaIndent{0}}]%
\>[2]\AgdaInductiveConstructor{Z}\AgdaSpace{}%
\AgdaSymbol{:}\AgdaSpace{}%
\AgdaDatatype{ℕ}\<%
\\
\>[2]\AgdaInductiveConstructor{suc}\AgdaSpace{}%
\AgdaSymbol{:}\AgdaSpace{}%
\AgdaDatatype{ℕ}\AgdaSpace{}%
\AgdaSymbol{→}\AgdaSpace{}%
\AgdaDatatype{ℕ}\<%
\end{code}
\begin{code}%
\>[0]\AgdaKeyword{data}\AgdaSpace{}%
\AgdaDatatype{Fin}\AgdaSpace{}%
\AgdaSymbol{:}\AgdaSpace{}%
\AgdaDatatype{ℕ}\AgdaSpace{}%
\AgdaSymbol{→}\AgdaSpace{}%
\AgdaPrimitive{Set₀}\AgdaSpace{}%
\AgdaKeyword{where}\<%
\\
\>[0][@{}l@{\AgdaIndent{0}}]%
\>[2]\AgdaInductiveConstructor{fzero}\AgdaSpace{}%
\AgdaSymbol{:}\AgdaSpace{}%
\AgdaSymbol{\{}\AgdaBound{n}\AgdaSpace{}%
\AgdaSymbol{:}\AgdaSpace{}%
\AgdaDatatype{ℕ}\AgdaSymbol{\}}\AgdaSpace{}%
\AgdaSymbol{→}\AgdaSpace{}%
\AgdaDatatype{Fin}\AgdaSpace{}%
\AgdaSymbol{(}\AgdaInductiveConstructor{suc}\AgdaSpace{}%
\AgdaBound{n}\AgdaSymbol{)}\<%
\\
\>[2]\AgdaInductiveConstructor{fsuc}%
\>[8]\AgdaSymbol{:}\AgdaSpace{}%
\AgdaSymbol{\{}\AgdaBound{n}\AgdaSpace{}%
\AgdaSymbol{:}\AgdaSpace{}%
\AgdaDatatype{ℕ}\AgdaSymbol{\}}\AgdaSpace{}%
\AgdaSymbol{→}\AgdaSpace{}%
\AgdaDatatype{Fin}\AgdaSpace{}%
\AgdaBound{n}\AgdaSpace{}%
\AgdaSymbol{→}\AgdaSpace{}%
\AgdaDatatype{Fin}\AgdaSpace{}%
\AgdaSymbol{(}\AgdaInductiveConstructor{suc}\AgdaSpace{}%
\AgdaBound{n}\AgdaSymbol{)}\<%
\end{code}

The constructor \AgdaInductiveConstructor{fzero} constructs an element of type
{\AgdaInductiveConstructor{Fin} (\AgdaInductiveConstructor{suc} \AgdaBound{n})} for any \AgdaBound{n}, while \AgdaInductiveConstructor{fsuc}
constructs an element of type {\AgdaInductiveConstructor{Fin} (\AgdaInductiveConstructor{suc} \AgdaBound{n})} from an
element of type {\AgdaInductiveConstructor{Fin}\;\AgdaBound{n}}.

In general, inductive families (without parameters) have the following form:
\begin{Verbatim}[commandchars=\\\{\}]
\AgdaKeyword{data} \AgdaDatatype{D} \AgdaSymbol{:} (\AgdaBound{y₁} \AgdaSymbol{:} \AgdaDatatype{T₁}) \AgdaSymbol{→} \AgdaSymbol{…} \AgdaSymbol{→} (\AgdaBound{yₙ} \AgdaSymbol{:} \AgdaDatatype{Tₙ}) \AgdaSymbol{→} \AgdaPrimitive{Setᵈ} \AgdaKeyword{where}
    \AgdaBound{c₁} \AgdaSymbol{:} \AgdaDatatype{A₁}
    \AgdaBound{c₁} \AgdaSymbol{:} \AgdaDatatype{A₂}
    \AgdaSymbol{…}
\end{Verbatim}
where \AgdaDatatype{D} is an inductive family in \AgdaPrimitive{Setᵈ} indexed by
expressions of type \AgdaDatatype{T₁}, $\ldots$, \AgdaDatatype{Tₙ}.
For each constructor \AgdaBound{cᵢ},
it takes a number of arguments $(\AgdaBound{zᵢ}\;\AgdaSymbol{:}\;\AgdaDatatype{Sᵢ})$
and constructs an element of type $\AgdaDatatype{D}\;\AgdaBound{t₁}\;\ldots\;\AgdaBound{tₙ}$,
where each \AgdaBound{tᵢ} is an expression of type \AgdaDatatype{Tᵢ}.
Concretely, each \AgdaDatatype{Aᵢ} takes the following form:

\begin{Verbatim}[commandchars=\\\{\}]
(\AgdaBound{z₁} \AgdaSymbol{:} \AgdaDatatype{S₁}) \AgdaSymbol{→} \AgdaSymbol{…} \AgdaSymbol{→} (\AgdaBound{zₘ} \AgdaSymbol{:} \AgdaDatatype{Sₘ}) \AgdaSymbol{→} \AgdaDatatype{D} \AgdaBound{t₁} \AgdaSymbol{…} \AgdaBound{tₙ}.
\end{Verbatim}

The arguments to \AgdaDatatype{D} and \AgdaBound{c} are dependently typed:
\AgdaDatatype{T$_{i+1}$} can mention \AgdaBound{y$_1$}$\;\ldots\;$\AgdaBound{y$_i$},
and \AgdaDatatype{S$_{i+1}$} can mention \AgdaBound{z$_1$}$\;\ldots\;$\AgdaBound{z$_i$}.

To ensure that inductive family definitions are consistent, Agda imposes additional restrictions.

First, \emph{universe checking} rejects inductive definitions with
impredicative constructors --- that is, definitions whose constructors
inhabit a larger universe than the data types themselves.
More concretely, for a data type such as \AgdaDatatype{D} above, the
universe of every argument type \AgdaDatatype{S$_i$} (i.e.~the type of
\AgdaDatatype{S$_i$}) should be smaller than \AgdaPrimitive{Set$_d$} to
pass the universe check.
Without this restriction, inductive families can be used to encode
Girard's paradox.

Second, \emph{positivity checking} rejects inductive families that
contain references to themselves in non-strictly-positive positions.
Without this restriction, inductive families such as the following
\AgdaDatatype{Fix} can be used to build recursive definitions
that violate consistency, such as \AgdaFunction{bad}:
\begin{code}[hide]%
\>[0]\AgdaSymbol{\{-\#}\AgdaSpace{}%
\AgdaKeyword{NO\AgdaUnderscore{}POSITIVITY\AgdaUnderscore{}CHECK}\AgdaSpace{}%
\AgdaSymbol{\#-\}}\<%
\end{code}
\begin{code}%
\>[0]\AgdaKeyword{data}\AgdaSpace{}%
\AgdaDatatype{Fix}\AgdaSpace{}%
\AgdaSymbol{:}\AgdaSpace{}%
\AgdaPrimitive{Set}\AgdaSpace{}%
\AgdaSymbol{→}\AgdaSpace{}%
\AgdaPrimitive{Set}\AgdaSpace{}%
\AgdaKeyword{where}\<%
\\
\>[0][@{}l@{\AgdaIndent{0}}]%
\>[2]\AgdaInductiveConstructor{fix}\AgdaSpace{}%
\AgdaSymbol{:}\AgdaSpace{}%
\AgdaSymbol{∀}\AgdaSpace{}%
\AgdaSymbol{\{}\AgdaBound{a}\AgdaSymbol{\}}\AgdaSpace{}%
\AgdaSymbol{→}\AgdaSpace{}%
\AgdaSymbol{(}\AgdaDatatype{Fix}\AgdaSpace{}%
\AgdaBound{a}\AgdaSpace{}%
\AgdaSymbol{→}\AgdaSpace{}%
\AgdaBound{a}\AgdaSymbol{)}\AgdaSpace{}%
\AgdaSymbol{→}\AgdaSpace{}%
\AgdaDatatype{Fix}\AgdaSpace{}%
\AgdaBound{a}\<%
\\
\\[\AgdaEmptyExtraSkip]%
\>[0]\AgdaFunction{bad}\AgdaSpace{}%
\AgdaSymbol{:}\AgdaSpace{}%
\AgdaSymbol{∀}\AgdaSpace{}%
\AgdaSymbol{\{}\AgdaBound{a}\AgdaSymbol{\}}\AgdaSpace{}%
\AgdaSymbol{→}\AgdaSpace{}%
\AgdaBound{a}\<%
\\
\>[0]\AgdaFunction{bad}\AgdaSpace{}%
\AgdaSymbol{=}\AgdaSpace{}%
\AgdaFunction{f}\AgdaSpace{}%
\AgdaSymbol{(}\AgdaInductiveConstructor{fix}\AgdaSpace{}%
\AgdaFunction{f}\AgdaSymbol{)}\AgdaSpace{}%
\AgdaKeyword{where}%
\>[29I]\AgdaFunction{f}\AgdaSpace{}%
\AgdaSymbol{:}\AgdaSpace{}%
\AgdaSymbol{∀}\AgdaSpace{}%
\AgdaSymbol{\{}\AgdaBound{a}\AgdaSymbol{\}}\AgdaSpace{}%
\AgdaSymbol{→}\AgdaSpace{}%
\AgdaDatatype{Fix}\AgdaSpace{}%
\AgdaBound{a}\AgdaSpace{}%
\AgdaSymbol{→}\AgdaSpace{}%
\AgdaBound{a}\<%
\\
\>[.][@{}l@{}]\<[29I]%
\>[22]\AgdaFunction{f}\AgdaSpace{}%
\AgdaSymbol{(}\AgdaInductiveConstructor{fix}\AgdaSpace{}%
\AgdaBound{g}\AgdaSymbol{)}\AgdaSpace{}%
\AgdaSymbol{=}\AgdaSpace{}%
\AgdaBound{g}\AgdaSpace{}%
\AgdaSymbol{(}\AgdaInductiveConstructor{fix}\AgdaSpace{}%
\AgdaBound{g}\AgdaSymbol{)}\<%
\end{code}

Strict positivity imposes two conditions on the constructor types
\AgdaDatatype{A$_i$} of an inductive family definition
\AgdaDatatype{D}.
First, where \AgdaDatatype{D} appears, it must
not be indexed by expressions involving \AgdaDatatype{D} itself.
Second, in the argument types \AgdaDatatype{S$_i$}, 
\AgdaDatatype{D} must not occur to the left of function arrows.

These requirements around strict positivity and universes are shared
by many dependently-typed languages that support inductive families,
like Coq's Gallina~\cite{CoqDoc},
Lean~\cite{DBLP:conf/cade/MouraKADR15},
and Timany and Sozeau's
pCuIC~\cite{DBLP:journals/corr/abs-1710-03912}.
   
\subsection{Problems extending defunctionalization to dependent types}
\label{section:size-problems}

At first glance, extending defunctionalization to support dependent
functions, targeting inductive families, appears straightforward.
As an example, we consider the defunctionalization of the following fully-dependent 
\AgdaFunction{compose} function, written in Agda, with all arguments explicit for clarity:
\begin{code}[hide]%
\>[0]\AgdaKeyword{module}\AgdaSpace{}%
\AgdaModule{compose}\AgdaSpace{}%
\AgdaKeyword{where}\<%
\end{code}

\begin{code}%
\>[0]\AgdaFunction{compose}\AgdaSpace{}%
\AgdaSymbol{:}%
\>[3I]\AgdaSymbol{(}\AgdaBound{A}\AgdaSpace{}%
\AgdaSymbol{:}\AgdaSpace{}%
\AgdaPrimitive{Set}\AgdaSymbol{)}\AgdaSpace{}%
\AgdaSymbol{→}%
\>[23]\AgdaSymbol{(}\AgdaBound{B}\AgdaSpace{}%
\AgdaSymbol{:}\AgdaSpace{}%
\AgdaBound{A}\AgdaSpace{}%
\AgdaSymbol{→}\AgdaSpace{}%
\AgdaPrimitive{Set}\AgdaSymbol{)}\AgdaSpace{}%
\AgdaSymbol{→}\AgdaSpace{}%
\AgdaSymbol{(}\AgdaBound{C}\AgdaSpace{}%
\AgdaSymbol{:}\AgdaSpace{}%
\AgdaSymbol{(}\AgdaBound{x}\AgdaSpace{}%
\AgdaSymbol{:}\AgdaSpace{}%
\AgdaBound{A}\AgdaSymbol{)}\AgdaSpace{}%
\AgdaSymbol{→}\AgdaSpace{}%
\AgdaBound{B}\AgdaSpace{}%
\AgdaBound{x}\AgdaSpace{}%
\AgdaSymbol{→}\AgdaSpace{}%
\AgdaPrimitive{Set}\AgdaSymbol{)}\AgdaSpace{}%
\AgdaSymbol{→}\<%
\\
\>[.][@{}l@{}]\<[3I]%
\>[10]\AgdaSymbol{(}\AgdaBound{f}\AgdaSpace{}%
\AgdaSymbol{:}\AgdaSpace{}%
\AgdaSymbol{(}\AgdaBound{y}\AgdaSpace{}%
\AgdaSymbol{:}\AgdaSpace{}%
\AgdaBound{A}\AgdaSymbol{)}\AgdaSpace{}%
\AgdaSymbol{→}\AgdaSpace{}%
\AgdaSymbol{(}\AgdaBound{z}\AgdaSpace{}%
\AgdaSymbol{:}\AgdaSpace{}%
\AgdaBound{B}\AgdaSpace{}%
\AgdaBound{y}\AgdaSymbol{)}\AgdaSpace{}%
\AgdaSymbol{→}\AgdaSpace{}%
\AgdaBound{C}\AgdaSpace{}%
\AgdaBound{y}\AgdaSpace{}%
\AgdaBound{z}\AgdaSymbol{)}\AgdaSpace{}%
\AgdaSymbol{→}\AgdaSpace{}%
\AgdaSymbol{(}\AgdaBound{g}\AgdaSpace{}%
\AgdaSymbol{:}\AgdaSpace{}%
\AgdaSymbol{(}\AgdaBound{x}\AgdaSpace{}%
\AgdaSymbol{:}\AgdaSpace{}%
\AgdaBound{A}\AgdaSymbol{)}\AgdaSpace{}%
\AgdaSymbol{→}\AgdaSpace{}%
\AgdaBound{B}\AgdaSpace{}%
\AgdaBound{x}\AgdaSymbol{)}\AgdaSpace{}%
\AgdaSymbol{→}\AgdaSpace{}%
\AgdaSymbol{(}\AgdaBound{x}\AgdaSpace{}%
\AgdaSymbol{:}\AgdaSpace{}%
\AgdaBound{A}\AgdaSymbol{)}\AgdaSpace{}%
\AgdaSymbol{→}\<%
\\
\>[10]\AgdaBound{C}\AgdaSpace{}%
\AgdaBound{x}\AgdaSpace{}%
\AgdaSymbol{(}\AgdaBound{g}\AgdaSpace{}%
\AgdaBound{x}\AgdaSymbol{)}\<%
\\
\>[0]\AgdaFunction{compose}\AgdaSpace{}%
\AgdaSymbol{=}\AgdaSpace{}%
\AgdaSymbol{λ}\AgdaSpace{}%
\AgdaBound{A}\AgdaSpace{}%
\AgdaSymbol{→}\AgdaSpace{}%
\AgdaSymbol{λ}\AgdaSpace{}%
\AgdaBound{B}\AgdaSpace{}%
\AgdaSymbol{→}\AgdaSpace{}%
\AgdaSymbol{λ}\AgdaSpace{}%
\AgdaBound{C}\AgdaSpace{}%
\AgdaSymbol{→}\AgdaSpace{}%
\AgdaSymbol{λ}\AgdaSpace{}%
\AgdaBound{f}\AgdaSpace{}%
\AgdaSymbol{→}\AgdaSpace{}%
\AgdaSymbol{λ}\AgdaSpace{}%
\AgdaBound{g}\AgdaSpace{}%
\AgdaSymbol{→}\AgdaSpace{}%
\AgdaSymbol{λ}\AgdaSpace{}%
\AgdaBound{x}\AgdaSpace{}%
\AgdaSymbol{→}\AgdaSpace{}%
\AgdaBound{f}\AgdaSpace{}%
\AgdaBound{x}\AgdaSpace{}%
\AgdaSymbol{(}\AgdaBound{g}\AgdaSpace{}%
\AgdaBound{x}\AgdaSymbol{)}\<%
\end{code}

\noindent
Adapting Pottier and Gauthier's recipe, we start  
by defining an inductive family $\Pi$ to represent dependent functions,
just as the GADT \lstinline!~>! represents non-dependent functions:
\begin{code}[hide]%
\>[0]\AgdaSymbol{\{-\#}\AgdaSpace{}%
\AgdaKeyword{OPTIONS}\AgdaSpace{}%
\AgdaPragma{--type-in-type}\AgdaSpace{}%
\AgdaSymbol{\#-\}}\<%
\end{code}

\begin{code}%
\>[0]\AgdaKeyword{data}\AgdaSpace{}%
\AgdaDatatype{Π}\AgdaSpace{}%
\AgdaSymbol{:}\AgdaSpace{}%
\AgdaSymbol{(}\AgdaBound{A}\AgdaSpace{}%
\AgdaSymbol{:}\AgdaSpace{}%
\AgdaPrimitive{Set}\AgdaSymbol{)}\AgdaSpace{}%
\AgdaSymbol{→}\AgdaSpace{}%
\AgdaSymbol{(}\AgdaBound{A}\AgdaSpace{}%
\AgdaSymbol{→}\AgdaSpace{}%
\AgdaPrimitive{Set}\AgdaSymbol{)}\AgdaSpace{}%
\AgdaSymbol{→}\AgdaSpace{}%
\AgdaPrimitive{Set}\AgdaSpace{}%
\AgdaKeyword{where}\<%
\end{code}

Each dependent function type $\Pi x{:}A.f x$ (written \lstinline!(x : A) $\to$ f x! in Agda) in the original program will be defunctionalized
to \lstinline!$\Pi$ A f!.

Next, we add a constructor to $\Pi$ for each lambda abstraction in the
original program.
For example, the \lstinline!F6! constructor corresponds to the
innermost abstraction, with free variables \lstinline!A!,
\lstinline!B!, \lstinline!C!, \lstinline!f! and \lstinline!g!:
\begin{code}[hide]%
\>[0]\AgdaSymbol{\{-\#}\AgdaSpace{}%
\AgdaKeyword{OPTIONS}\AgdaSpace{}%
\AgdaPragma{--type-in-type}\AgdaSpace{}%
\AgdaSymbol{\#-\}}\<%
\end{code}

\begin{code}[hide]%
\>[0]\AgdaKeyword{data}\AgdaSpace{}%
\AgdaDatatype{Π}\AgdaSpace{}%
\AgdaSymbol{:}\AgdaSpace{}%
\AgdaSymbol{(}\AgdaBound{A}\AgdaSpace{}%
\AgdaSymbol{:}\AgdaSpace{}%
\AgdaPrimitive{Set}\AgdaSymbol{)}\AgdaSpace{}%
\AgdaSymbol{→}\AgdaSpace{}%
\AgdaSymbol{(}\AgdaBound{A}\AgdaSpace{}%
\AgdaSymbol{→}\AgdaSpace{}%
\AgdaPrimitive{Set}\AgdaSymbol{)}\AgdaSpace{}%
\AgdaSymbol{→}\AgdaSpace{}%
\AgdaPrimitive{Set}\<%
\\
\>[0]\AgdaKeyword{infixl}\AgdaSpace{}%
\AgdaNumber{9}\AgdaSpace{}%
\AgdaOperator{\AgdaFunction{\AgdaUnderscore{}\$\AgdaUnderscore{}}}\<%
\\
\>[0]\AgdaOperator{\AgdaFunction{\AgdaUnderscore{}\$\AgdaUnderscore{}}}\AgdaSpace{}%
\AgdaSymbol{:}\AgdaSpace{}%
\AgdaSymbol{∀}\AgdaSpace{}%
\AgdaSymbol{\{}\AgdaBound{A}\AgdaSpace{}%
\AgdaSymbol{:}\AgdaSpace{}%
\AgdaPrimitive{Set}\AgdaSymbol{\}}\AgdaSpace{}%
\AgdaSymbol{\{}\AgdaBound{p}\AgdaSpace{}%
\AgdaSymbol{:}\AgdaSpace{}%
\AgdaBound{A}\AgdaSpace{}%
\AgdaSymbol{→}\AgdaSpace{}%
\AgdaPrimitive{Set}\AgdaSymbol{\}}\AgdaSpace{}%
\AgdaSymbol{→}\AgdaSpace{}%
\AgdaDatatype{Π}\AgdaSpace{}%
\AgdaBound{A}\AgdaSpace{}%
\AgdaBound{p}\AgdaSpace{}%
\AgdaSymbol{→}\AgdaSpace{}%
\AgdaSymbol{(}\AgdaBound{x}\AgdaSpace{}%
\AgdaSymbol{:}\AgdaSpace{}%
\AgdaBound{A}\AgdaSymbol{)}\AgdaSpace{}%
\AgdaSymbol{→}\AgdaSpace{}%
\AgdaBound{p}\AgdaSpace{}%
\AgdaBound{x}\<%
\\
\\[\AgdaEmptyExtraSkip]%
\>[0]\AgdaSymbol{\{-\#}\AgdaSpace{}%
\AgdaKeyword{NO\AgdaUnderscore{}POSITIVITY\AgdaUnderscore{}CHECK}\AgdaSpace{}%
\AgdaSymbol{\#-\}}\<%
\\
\>[0]\AgdaKeyword{data}\AgdaSpace{}%
\AgdaDatatype{Π}\AgdaSpace{}%
\AgdaKeyword{where}\<%
\end{code}
\begin{code}%
\>[0][@{}l@{\AgdaIndent{1}}]%
\>[3]\AgdaInductiveConstructor{F6}\AgdaSpace{}%
\AgdaSymbol{:}%
\>[42I]\AgdaSymbol{(}\AgdaBound{A}\AgdaSpace{}%
\AgdaSymbol{:}\AgdaSpace{}%
\AgdaPrimitive{Set}\AgdaSymbol{)}\AgdaSpace{}%
\AgdaSymbol{→}\<%
\\
\>[.][@{}l@{}]\<[42I]%
\>[8]\AgdaSymbol{(}\AgdaBound{B}\AgdaSpace{}%
\AgdaSymbol{:}\AgdaSpace{}%
\AgdaDatatype{Π}\AgdaSpace{}%
\AgdaBound{A}\AgdaSpace{}%
\AgdaSymbol{(λ}\AgdaSpace{}%
\AgdaBound{\AgdaUnderscore{}}\AgdaSpace{}%
\AgdaSymbol{→}\AgdaSpace{}%
\AgdaPrimitive{Set}\AgdaSymbol{))}\AgdaSpace{}%
\AgdaSymbol{→}\<%
\\
\>[8]\AgdaSymbol{(}\AgdaBound{C}\AgdaSpace{}%
\AgdaSymbol{:}\AgdaSpace{}%
\AgdaDatatype{Π}\AgdaSpace{}%
\AgdaBound{A}\AgdaSpace{}%
\AgdaSymbol{(λ}\AgdaSpace{}%
\AgdaBound{x}\AgdaSpace{}%
\AgdaSymbol{→}\AgdaSpace{}%
\AgdaDatatype{Π}\AgdaSpace{}%
\AgdaSymbol{(}\AgdaBound{B}\AgdaSpace{}%
\AgdaOperator{\AgdaFunction{\$}}\AgdaSpace{}%
\AgdaBound{x}\AgdaSymbol{)}\AgdaSpace{}%
\AgdaSymbol{(λ}\AgdaSpace{}%
\AgdaBound{\AgdaUnderscore{}}\AgdaSpace{}%
\AgdaSymbol{→}\AgdaSpace{}%
\AgdaPrimitive{Set}\AgdaSymbol{)))}\AgdaSpace{}%
\AgdaSymbol{→}\<%
\\
\>[8]\AgdaSymbol{(}\AgdaBound{f}\AgdaSpace{}%
\AgdaSymbol{:}\AgdaSpace{}%
\AgdaDatatype{Π}\AgdaSpace{}%
\AgdaBound{A}\AgdaSpace{}%
\AgdaSymbol{(λ}\AgdaSpace{}%
\AgdaBound{y}\AgdaSpace{}%
\AgdaSymbol{→}\AgdaSpace{}%
\AgdaDatatype{Π}\AgdaSpace{}%
\AgdaSymbol{(}\AgdaBound{B}\AgdaSpace{}%
\AgdaOperator{\AgdaFunction{\$}}\AgdaSpace{}%
\AgdaBound{y}\AgdaSymbol{)}\AgdaSpace{}%
\AgdaSymbol{(λ}\AgdaSpace{}%
\AgdaBound{z}\AgdaSpace{}%
\AgdaSymbol{→}\AgdaSpace{}%
\AgdaBound{C}\AgdaSpace{}%
\AgdaOperator{\AgdaFunction{\$}}\AgdaSpace{}%
\AgdaBound{y}\AgdaSpace{}%
\AgdaOperator{\AgdaFunction{\$}}\AgdaSpace{}%
\AgdaBound{z}\AgdaSymbol{)))}\AgdaSpace{}%
\AgdaSymbol{→}\<%
\\
\>[8]\AgdaSymbol{(}\AgdaBound{g}\AgdaSpace{}%
\AgdaSymbol{:}\AgdaSpace{}%
\AgdaDatatype{Π}\AgdaSpace{}%
\AgdaBound{A}\AgdaSpace{}%
\AgdaSymbol{(λ}\AgdaSpace{}%
\AgdaBound{x}\AgdaSpace{}%
\AgdaSymbol{→}\AgdaSpace{}%
\AgdaBound{B}\AgdaSpace{}%
\AgdaOperator{\AgdaFunction{\$}}\AgdaSpace{}%
\AgdaBound{x}\AgdaSymbol{))}\AgdaSpace{}%
\AgdaSymbol{→}\<%
\\
\>[8]\AgdaDatatype{Π}\AgdaSpace{}%
\AgdaBound{A}\AgdaSpace{}%
\AgdaSymbol{(λ}\AgdaSpace{}%
\AgdaBound{x}\AgdaSpace{}%
\AgdaSymbol{→}\<%
\\
\>[8]\AgdaBound{C}\AgdaSpace{}%
\AgdaOperator{\AgdaFunction{\$}}\AgdaSpace{}%
\AgdaBound{x}\AgdaSpace{}%
\AgdaOperator{\AgdaFunction{\$}}\AgdaSpace{}%
\AgdaSymbol{(}\AgdaBound{g}\AgdaSpace{}%
\AgdaOperator{\AgdaFunction{\$}}\AgdaSpace{}%
\AgdaBound{x}\AgdaSymbol{))}\<%
\end{code}
\begin{code}[hide]%
\>[0]\AgdaSymbol{\{-\#}\AgdaSpace{}%
\AgdaKeyword{TERMINATING}\AgdaSpace{}%
\AgdaSymbol{\#-\}}\<%
\\
\>[0]\AgdaInductiveConstructor{F6}\AgdaSpace{}%
\AgdaBound{A}\AgdaSpace{}%
\AgdaBound{B}\AgdaSpace{}%
\AgdaBound{C}\AgdaSpace{}%
\AgdaBound{f}\AgdaSpace{}%
\AgdaBound{g}\AgdaSpace{}%
\AgdaOperator{\AgdaFunction{\$}}\AgdaSpace{}%
\AgdaBound{x}\AgdaSpace{}%
\AgdaSymbol{=}\AgdaSpace{}%
\AgdaBound{f}\AgdaSpace{}%
\AgdaOperator{\AgdaFunction{\$}}\AgdaSpace{}%
\AgdaBound{x}\AgdaSpace{}%
\AgdaOperator{\AgdaFunction{\$}}\AgdaSpace{}%
\AgdaSymbol{(}\AgdaBound{g}\AgdaSpace{}%
\AgdaOperator{\AgdaFunction{\$}}\AgdaSpace{}%
\AgdaBound{x}\AgdaSymbol{)}\<%
\end{code}

\noindent
Finally, we add a case for each constructor to the definition of
\lstinline!$\$$!:
\begin{code}[hide]%
\>[0]\AgdaSymbol{\{-\#}\AgdaSpace{}%
\AgdaKeyword{OPTIONS}\AgdaSpace{}%
\AgdaPragma{--type-in-type}\AgdaSpace{}%
\AgdaSymbol{\#-\}}\<%
\\
\\[\AgdaEmptyExtraSkip]%
\>[0]\AgdaKeyword{data}\AgdaSpace{}%
\AgdaDatatype{Π}\AgdaSpace{}%
\AgdaSymbol{:}\AgdaSpace{}%
\AgdaSymbol{(}\AgdaBound{A}\AgdaSpace{}%
\AgdaSymbol{:}\AgdaSpace{}%
\AgdaPrimitive{Set}\AgdaSymbol{)}\AgdaSpace{}%
\AgdaSymbol{→}\AgdaSpace{}%
\AgdaSymbol{(}\AgdaBound{A}\AgdaSpace{}%
\AgdaSymbol{→}\AgdaSpace{}%
\AgdaPrimitive{Set}\AgdaSymbol{)}\AgdaSpace{}%
\AgdaSymbol{→}\AgdaSpace{}%
\AgdaPrimitive{Set}\<%
\\
\>[0]\AgdaKeyword{infixl}\AgdaSpace{}%
\AgdaNumber{9}\AgdaSpace{}%
\AgdaOperator{\AgdaFunction{\AgdaUnderscore{}\$\AgdaUnderscore{}}}\<%
\\
\>[0]\AgdaOperator{\AgdaFunction{\AgdaUnderscore{}\$\AgdaUnderscore{}}}\AgdaSpace{}%
\AgdaSymbol{:}\AgdaSpace{}%
\AgdaSymbol{∀}\AgdaSpace{}%
\AgdaSymbol{\{}\AgdaBound{A}\AgdaSpace{}%
\AgdaSymbol{:}\AgdaSpace{}%
\AgdaPrimitive{Set}\AgdaSymbol{\}}\AgdaSpace{}%
\AgdaSymbol{\{}\AgdaBound{p}\AgdaSpace{}%
\AgdaSymbol{:}\AgdaSpace{}%
\AgdaBound{A}\AgdaSpace{}%
\AgdaSymbol{→}\AgdaSpace{}%
\AgdaPrimitive{Set}\AgdaSymbol{\}}\AgdaSpace{}%
\AgdaSymbol{→}\AgdaSpace{}%
\AgdaDatatype{Π}\AgdaSpace{}%
\AgdaBound{A}\AgdaSpace{}%
\AgdaBound{p}\AgdaSpace{}%
\AgdaSymbol{→}\AgdaSpace{}%
\AgdaSymbol{(}\AgdaBound{x}\AgdaSpace{}%
\AgdaSymbol{:}\AgdaSpace{}%
\AgdaBound{A}\AgdaSymbol{)}\AgdaSpace{}%
\AgdaSymbol{→}\AgdaSpace{}%
\AgdaBound{p}\AgdaSpace{}%
\AgdaBound{x}\<%
\\
\\[\AgdaEmptyExtraSkip]%
\>[0]\AgdaSymbol{\{-\#}\AgdaSpace{}%
\AgdaKeyword{NO\AgdaUnderscore{}POSITIVITY\AgdaUnderscore{}CHECK}\AgdaSpace{}%
\AgdaSymbol{\#-\}}\<%
\\
\>[0]\AgdaKeyword{data}\AgdaSpace{}%
\AgdaDatatype{Π}\AgdaSpace{}%
\AgdaKeyword{where}\<%
\\
\>[0][@{}l@{\AgdaIndent{0}}]%
\>[3]\AgdaInductiveConstructor{F6}\AgdaSpace{}%
\AgdaSymbol{:}%
\>[42I]\AgdaSymbol{(}\AgdaBound{A}\AgdaSpace{}%
\AgdaSymbol{:}\AgdaSpace{}%
\AgdaPrimitive{Set}\AgdaSymbol{)}\AgdaSpace{}%
\AgdaSymbol{→}\<%
\\
\>[.][@{}l@{}]\<[42I]%
\>[8]\AgdaSymbol{(}\AgdaBound{B}\AgdaSpace{}%
\AgdaSymbol{:}\AgdaSpace{}%
\AgdaDatatype{Π}\AgdaSpace{}%
\AgdaBound{A}\AgdaSpace{}%
\AgdaSymbol{(λ}\AgdaSpace{}%
\AgdaBound{\AgdaUnderscore{}}\AgdaSpace{}%
\AgdaSymbol{→}\AgdaSpace{}%
\AgdaPrimitive{Set}\AgdaSymbol{))}\AgdaSpace{}%
\AgdaSymbol{→}\<%
\\
\>[8]\AgdaSymbol{(}\AgdaBound{C}\AgdaSpace{}%
\AgdaSymbol{:}\AgdaSpace{}%
\AgdaDatatype{Π}\AgdaSpace{}%
\AgdaBound{A}\AgdaSpace{}%
\AgdaSymbol{(λ}\AgdaSpace{}%
\AgdaBound{x}\AgdaSpace{}%
\AgdaSymbol{→}\AgdaSpace{}%
\AgdaDatatype{Π}\AgdaSpace{}%
\AgdaSymbol{(}\AgdaBound{B}\AgdaSpace{}%
\AgdaOperator{\AgdaFunction{\$}}\AgdaSpace{}%
\AgdaBound{x}\AgdaSymbol{)}\AgdaSpace{}%
\AgdaSymbol{(λ}\AgdaSpace{}%
\AgdaBound{\AgdaUnderscore{}}\AgdaSpace{}%
\AgdaSymbol{→}\AgdaSpace{}%
\AgdaPrimitive{Set}\AgdaSymbol{)))}\AgdaSpace{}%
\AgdaSymbol{→}\<%
\\
\>[8]\AgdaSymbol{(}\AgdaBound{f}\AgdaSpace{}%
\AgdaSymbol{:}\AgdaSpace{}%
\AgdaDatatype{Π}\AgdaSpace{}%
\AgdaBound{A}\AgdaSpace{}%
\AgdaSymbol{(λ}\AgdaSpace{}%
\AgdaBound{y}\AgdaSpace{}%
\AgdaSymbol{→}\AgdaSpace{}%
\AgdaDatatype{Π}\AgdaSpace{}%
\AgdaSymbol{(}\AgdaBound{B}\AgdaSpace{}%
\AgdaOperator{\AgdaFunction{\$}}\AgdaSpace{}%
\AgdaBound{y}\AgdaSymbol{)}\AgdaSpace{}%
\AgdaSymbol{(λ}\AgdaSpace{}%
\AgdaBound{z}\AgdaSpace{}%
\AgdaSymbol{→}\AgdaSpace{}%
\AgdaBound{C}\AgdaSpace{}%
\AgdaOperator{\AgdaFunction{\$}}\AgdaSpace{}%
\AgdaBound{y}\AgdaSpace{}%
\AgdaOperator{\AgdaFunction{\$}}\AgdaSpace{}%
\AgdaBound{z}\AgdaSymbol{)))}\AgdaSpace{}%
\AgdaSymbol{→}\<%
\\
\>[8]\AgdaSymbol{(}\AgdaBound{g}\AgdaSpace{}%
\AgdaSymbol{:}\AgdaSpace{}%
\AgdaDatatype{Π}\AgdaSpace{}%
\AgdaBound{A}\AgdaSpace{}%
\AgdaSymbol{(λ}\AgdaSpace{}%
\AgdaBound{x}\AgdaSpace{}%
\AgdaSymbol{→}\AgdaSpace{}%
\AgdaBound{B}\AgdaSpace{}%
\AgdaOperator{\AgdaFunction{\$}}\AgdaSpace{}%
\AgdaBound{x}\AgdaSymbol{))}\AgdaSpace{}%
\AgdaSymbol{→}\<%
\\
\>[8]\AgdaDatatype{Π}\AgdaSpace{}%
\AgdaBound{A}\AgdaSpace{}%
\AgdaSymbol{(λ}\AgdaSpace{}%
\AgdaBound{x}\AgdaSpace{}%
\AgdaSymbol{→}\<%
\\
\>[8]\AgdaBound{C}\AgdaSpace{}%
\AgdaOperator{\AgdaFunction{\$}}\AgdaSpace{}%
\AgdaBound{x}\AgdaSpace{}%
\AgdaOperator{\AgdaFunction{\$}}\AgdaSpace{}%
\AgdaSymbol{(}\AgdaBound{g}\AgdaSpace{}%
\AgdaOperator{\AgdaFunction{\$}}\AgdaSpace{}%
\AgdaBound{x}\AgdaSymbol{))}\<%
\\
\\[\AgdaEmptyExtraSkip]%
\>[0]\AgdaSymbol{\{-\#}\AgdaSpace{}%
\AgdaKeyword{TERMINATING}\AgdaSpace{}%
\AgdaSymbol{\#-\}}\<%
\end{code}
\begin{code}%
\>[0]\AgdaInductiveConstructor{F6}\AgdaSpace{}%
\AgdaBound{A}\AgdaSpace{}%
\AgdaBound{B}\AgdaSpace{}%
\AgdaBound{C}\AgdaSpace{}%
\AgdaBound{f}\AgdaSpace{}%
\AgdaBound{g}\AgdaSpace{}%
\AgdaOperator{\AgdaFunction{\$}}\AgdaSpace{}%
\AgdaBound{x}\AgdaSpace{}%
\AgdaSymbol{=}\AgdaSpace{}%
\AgdaBound{f}\AgdaSpace{}%
\AgdaOperator{\AgdaFunction{\$}}\AgdaSpace{}%
\AgdaBound{x}\AgdaSpace{}%
\AgdaOperator{\AgdaFunction{\$}}\AgdaSpace{}%
\AgdaSymbol{(}\AgdaBound{g}\AgdaSpace{}%
\AgdaOperator{\AgdaFunction{\$}}\AgdaSpace{}%
\AgdaBound{x}\AgdaSymbol{)}\<%
\end{code}

\noindent
\Cref{appendix:agda-defun} gives the full definitions.
Unfortunately, although these definitions are type-correct,
they do not satisfy Agda's additional checks.

First, \emph{universe checking} rejects the definition of \lstinline!F6! because 
in the type of \AgdaBound{B} the second argument of \AgdaDatatype{$\Pi$} (i.e.~$\lambda \_ \to\;$\AgdaPrimitive{Set})
inhabits the universe \AgdaPrimitive{Set$_1$}, which is larger than \AgdaPrimitive{Set}.

Second, \emph{positivity checking} rejects the definition of
\lstinline!F6! because in the type of \AgdaBound{C}, \AgdaDatatype{$\Pi$} is indexed by an expression involving \AgdaDatatype{$\Pi$}.

Finally, Agda's \emph{termination checking} rejects the definition of
\lstinline!$\$$! because the case for \AgdaInductiveConstructor{F6} is
not structurally terminating.

\subsubsection{A simpler example}

The example above suggests that although defunctionalization
apparently extends naturally to dependent types, the extension suffers
from consistency problems.
In fact, the situation is more grave: even if we do not make use of
dependency, the same problems with universes and positivity arise.

For example, here is a simply-typed
\AgdaFunction{compose} function, based on fixed types
\AgdaGeneralizable{A}, \AgdaGeneralizable{B}, and
\AgdaGeneralizable{C}:
\begin{code}[hide]%
\>[0]\AgdaKeyword{variable}\AgdaSpace{}%
\AgdaGeneralizable{A}\AgdaSpace{}%
\AgdaSymbol{:}\AgdaSpace{}%
\AgdaPrimitive{Set}\<%
\\
\>[0]\AgdaKeyword{variable}\AgdaSpace{}%
\AgdaGeneralizable{B}\AgdaSpace{}%
\AgdaSymbol{:}\AgdaSpace{}%
\AgdaPrimitive{Set}\<%
\\
\>[0]\AgdaKeyword{variable}\AgdaSpace{}%
\AgdaGeneralizable{C}\AgdaSpace{}%
\AgdaSymbol{:}\AgdaSpace{}%
\AgdaPrimitive{Set}\<%
\end{code}

\begin{code}%
\>[0]\AgdaFunction{compose}\AgdaSpace{}%
\AgdaSymbol{:}\AgdaSpace{}%
\AgdaSymbol{(}\AgdaGeneralizable{B}\AgdaSpace{}%
\AgdaSymbol{→}\AgdaSpace{}%
\AgdaGeneralizable{C}\AgdaSymbol{)}\AgdaSpace{}%
\AgdaSymbol{→}\AgdaSpace{}%
\AgdaSymbol{(}\AgdaGeneralizable{A}\AgdaSpace{}%
\AgdaSymbol{→}\AgdaSpace{}%
\AgdaGeneralizable{B}\AgdaSymbol{)}\AgdaSpace{}%
\AgdaSymbol{→}\AgdaSpace{}%
\AgdaSymbol{(}\AgdaGeneralizable{A}\AgdaSpace{}%
\AgdaSymbol{→}\AgdaSpace{}%
\AgdaGeneralizable{C}\AgdaSymbol{)}\<%
\\
\>[0]\AgdaFunction{compose}\AgdaSpace{}%
\AgdaSymbol{=}\AgdaSpace{}%
\AgdaSymbol{λ}\AgdaSpace{}%
\AgdaBound{f}\AgdaSpace{}%
\AgdaSymbol{→}\AgdaSpace{}%
\AgdaSymbol{λ}\AgdaSpace{}%
\AgdaBound{g}\AgdaSpace{}%
\AgdaSymbol{→}\AgdaSpace{}%
\AgdaSymbol{λ}\AgdaSpace{}%
\AgdaBound{x}\AgdaSpace{}%
\AgdaSymbol{→}\AgdaSpace{}%
\AgdaBound{f}\AgdaSpace{}%
\AgdaSymbol{(}\AgdaBound{g}\AgdaSpace{}%
\AgdaBound{x}\AgdaSymbol{)}\<%
\end{code}

\noindent
Defunctionalizing \AgdaFunction{compose} produces an inductive family
\AgdaOperator{\AgdaDatatype{↝}} and corresponding \emph{apply}
function \AgdaOperator{\AgdaFunction{\$}}:
\begin{code}[hide]%
\>[0]\AgdaSymbol{\{-\#}\AgdaSpace{}%
\AgdaKeyword{OPTIONS}\AgdaSpace{}%
\AgdaPragma{--type-in-type}\AgdaSpace{}%
\AgdaSymbol{\#-\}}\<%
\\
\>[0]\AgdaKeyword{variable}\AgdaSpace{}%
\AgdaGeneralizable{A}\AgdaSpace{}%
\AgdaSymbol{:}\AgdaSpace{}%
\AgdaPrimitive{Set}\<%
\\
\>[0]\AgdaKeyword{variable}\AgdaSpace{}%
\AgdaGeneralizable{B}\AgdaSpace{}%
\AgdaSymbol{:}\AgdaSpace{}%
\AgdaPrimitive{Set}\<%
\\
\>[0]\AgdaKeyword{variable}\AgdaSpace{}%
\AgdaGeneralizable{C}\AgdaSpace{}%
\AgdaSymbol{:}\AgdaSpace{}%
\AgdaPrimitive{Set}\<%
\\
\>[0]\AgdaKeyword{infixr}\AgdaSpace{}%
\AgdaNumber{0}\AgdaSpace{}%
\AgdaOperator{\AgdaDatatype{\AgdaUnderscore{}↝\AgdaUnderscore{}}}\<%
\\
\>[0]\AgdaSymbol{\{-\#}\AgdaSpace{}%
\AgdaKeyword{NO\AgdaUnderscore{}POSITIVITY\AgdaUnderscore{}CHECK}\AgdaSpace{}%
\AgdaSymbol{\#-\}}\<%
\end{code}

\begin{code}%
\>[0]\AgdaKeyword{data}\AgdaSpace{}%
\AgdaOperator{\AgdaDatatype{\AgdaUnderscore{}↝\AgdaUnderscore{}}}\AgdaSpace{}%
\AgdaSymbol{:}\AgdaSpace{}%
\AgdaPrimitive{Set}\AgdaSpace{}%
\AgdaSymbol{→}\AgdaSpace{}%
\AgdaPrimitive{Set}\AgdaSpace{}%
\AgdaSymbol{→}\AgdaSpace{}%
\AgdaPrimitive{Set}\AgdaSpace{}%
\AgdaKeyword{where}\<%
\\
\>[0][@{}l@{\AgdaIndent{0}}]%
\>[3]\AgdaInductiveConstructor{F1}\AgdaSpace{}%
\AgdaSymbol{:}\AgdaSpace{}%
\AgdaSymbol{(}\AgdaGeneralizable{B}\AgdaSpace{}%
\AgdaOperator{\AgdaDatatype{↝}}\AgdaSpace{}%
\AgdaGeneralizable{C}\AgdaSymbol{)}\AgdaSpace{}%
\AgdaOperator{\AgdaDatatype{↝}}\AgdaSpace{}%
\AgdaSymbol{(}\AgdaGeneralizable{A}\AgdaSpace{}%
\AgdaOperator{\AgdaDatatype{↝}}\AgdaSpace{}%
\AgdaGeneralizable{B}\AgdaSymbol{)}\AgdaSpace{}%
\AgdaOperator{\AgdaDatatype{↝}}\AgdaSpace{}%
\AgdaSymbol{(}\AgdaGeneralizable{A}\AgdaSpace{}%
\AgdaOperator{\AgdaDatatype{↝}}\AgdaSpace{}%
\AgdaGeneralizable{C}\AgdaSymbol{)}\<%
\\
\>[3]\AgdaInductiveConstructor{F2}\AgdaSpace{}%
\AgdaSymbol{:}\AgdaSpace{}%
\AgdaSymbol{(}\AgdaGeneralizable{B}\AgdaSpace{}%
\AgdaOperator{\AgdaDatatype{↝}}\AgdaSpace{}%
\AgdaGeneralizable{C}\AgdaSymbol{)}\AgdaSpace{}%
\AgdaSymbol{→}\AgdaSpace{}%
\AgdaSymbol{(}\AgdaGeneralizable{A}\AgdaSpace{}%
\AgdaOperator{\AgdaDatatype{↝}}\AgdaSpace{}%
\AgdaGeneralizable{B}\AgdaSymbol{)}\AgdaSpace{}%
\AgdaOperator{\AgdaDatatype{↝}}\AgdaSpace{}%
\AgdaSymbol{(}\AgdaGeneralizable{A}\AgdaSpace{}%
\AgdaOperator{\AgdaDatatype{↝}}\AgdaSpace{}%
\AgdaGeneralizable{C}\AgdaSymbol{)}\<%
\\
\>[3]\AgdaInductiveConstructor{F3}\AgdaSpace{}%
\AgdaSymbol{:}\AgdaSpace{}%
\AgdaSymbol{(}\AgdaGeneralizable{B}\AgdaSpace{}%
\AgdaOperator{\AgdaDatatype{↝}}\AgdaSpace{}%
\AgdaGeneralizable{C}\AgdaSymbol{)}\AgdaSpace{}%
\AgdaSymbol{→}\AgdaSpace{}%
\AgdaSymbol{(}\AgdaGeneralizable{A}\AgdaSpace{}%
\AgdaOperator{\AgdaDatatype{↝}}\AgdaSpace{}%
\AgdaGeneralizable{B}\AgdaSymbol{)}\AgdaSpace{}%
\AgdaSymbol{→}\AgdaSpace{}%
\AgdaSymbol{(}\AgdaGeneralizable{A}\AgdaSpace{}%
\AgdaOperator{\AgdaDatatype{↝}}\AgdaSpace{}%
\AgdaGeneralizable{C}\AgdaSymbol{)}\<%
\end{code}
\begin{code}[hide]%
\>[0]\AgdaSymbol{\{-\#}\AgdaSpace{}%
\AgdaKeyword{TERMINATING}\AgdaSpace{}%
\AgdaSymbol{\#-\}}\<%
\end{code}
\begin{code}%
\>[0]\AgdaOperator{\AgdaFunction{\AgdaUnderscore{}\$\AgdaUnderscore{}}}\AgdaSpace{}%
\AgdaSymbol{:}\AgdaSpace{}%
\AgdaSymbol{∀}\AgdaSpace{}%
\AgdaSymbol{\{}\AgdaBound{A}\AgdaSpace{}%
\AgdaBound{B}\AgdaSymbol{\}}\AgdaSpace{}%
\AgdaSymbol{→}\AgdaSpace{}%
\AgdaSymbol{(}\AgdaBound{A}\AgdaSpace{}%
\AgdaOperator{\AgdaDatatype{↝}}\AgdaSpace{}%
\AgdaBound{B}\AgdaSymbol{)}\AgdaSpace{}%
\AgdaSymbol{→}\AgdaSpace{}%
\AgdaBound{A}\AgdaSpace{}%
\AgdaSymbol{→}\AgdaSpace{}%
\AgdaBound{B}\<%
\\
\>[0]\AgdaInductiveConstructor{F1}\AgdaSpace{}%
\AgdaOperator{\AgdaFunction{\$}}\AgdaSpace{}%
\AgdaBound{f}\AgdaSpace{}%
\AgdaSymbol{=}\AgdaSpace{}%
\AgdaInductiveConstructor{F2}\AgdaSpace{}%
\AgdaBound{f}\<%
\\
\>[0]\AgdaInductiveConstructor{F2}\AgdaSpace{}%
\AgdaBound{f}\AgdaSpace{}%
\AgdaOperator{\AgdaFunction{\$}}\AgdaSpace{}%
\AgdaBound{g}\AgdaSpace{}%
\AgdaSymbol{=}\AgdaSpace{}%
\AgdaInductiveConstructor{F3}\AgdaSpace{}%
\AgdaBound{f}\AgdaSpace{}%
\AgdaBound{g}\<%
\\
\>[0]\AgdaInductiveConstructor{F3}\AgdaSpace{}%
\AgdaBound{f}\AgdaSpace{}%
\AgdaBound{g}\AgdaSpace{}%
\AgdaOperator{\AgdaFunction{\$}}\AgdaSpace{}%
\AgdaBound{x}\AgdaSpace{}%
\AgdaSymbol{=}\AgdaSpace{}%
\AgdaBound{f}\AgdaSpace{}%
\AgdaOperator{\AgdaFunction{\$}}\AgdaSpace{}%
\AgdaSymbol{(}\AgdaBound{g}\AgdaSpace{}%
\AgdaOperator{\AgdaFunction{\$}}\AgdaSpace{}%
\AgdaBound{x}\AgdaSymbol{)}\<%
\end{code}

Unfortunately, the simple definition \AgdaOperator{\AgdaDatatype{↝}}
suffers from the same problems as the more dependent \AgdaDatatype{$\Pi$}.
First, universe checking rejects the constructor \AgdaInductiveConstructor{F1}, because
the type \AgdaGeneralizable{B}\AgdaSpace{}\AgdaOperator{\AgdaDatatype{↝}}\AgdaSpace{}\AgdaGeneralizable{C}
inhabits the universe \AgdaPrimitive{Set$_1$}, which is larger than \AgdaPrimitive{Set}.
Second, in the type of \AgdaInductiveConstructor{F1},
\AgdaOperator{\AgdaDatatype{↝}} is indexed by
\AgdaOperator{\AgdaDatatype{↝}} itself, so the definition fails positivity checking.
Finally, the \AgdaInductiveConstructor{F3} case of
\AgdaOperator{\AgdaFunction{\$}} fails termination checking because
the arguments to the recursive call are not structurally smaller than
the parameters.

\subsubsection{An expressivity mismatch}

We might note
 that the Agda's restrictions are only fairly crude syntactic approximations
   of semantic properties,
 that programs that breach them are not necessarily ``incorrect''.
A similar approach has been taken
by~\citet{DBLP:journals/lmcs/AhrensLV18} (for universe checking), and
by \citet{DBLP:conf/ssgip/WeirichC10} (for all three checks), among
others.

However, we do not favour taking off the safety guards in this way for
the code generated by defunctionalization.
In our view, the fact that Agda rejects the inductive families
generated by defunctionalization suggests that inductive families are
ill suited to the task.
For example, the universe restriction that rejects the constructors of
\AgdaDatatype{$\Pi$} does not apply to the closures that correspond to those
constructors in the source program: there is nothing requiring a free
variable in an abstraction body to inhabit a smaller universe than the
function itself.
The additional restriction arises from an expressivity mismatch: the
universe restriction is only needed when inductive families are not
used in a closure-like fashion --- e.g.~when constructor
arguments are extracted.

\subsection{Abstract defunctionalization}
\label{section:abstract-defunctionalization}



  The examples above suggest that the extension of defunctionalization
  to dependent types is \emph{type-preserving}.
  It is also possible to show that it is \emph{meaning-preserving}.
  As \citet{DBLP:journals/lisp/PottierG06} observe, when
  defunctionalization produces a single polymorphic apply function, it
  coincides with the untyped defunctionalization translation.
  Pottier and Gauthier use this coincidence to prove that the typed
  translation is meaning-preserving by lifting a proof about the untyped
  translation.
  We might similarly lift the proof to the dependently-typed setting to
  establish the correctness of the extended translation.

Since the extended defunctionalization translation appears to preserve
types and meanings, it is disappointing that it falls foul of Agda's
various restrictions.
How might we build a translation that does not violate these checks?

We choose to follow the direction taken by
\citet{DBLP:conf/popl/MinamideMH96} and
\citet{DBLP:conf/pldi/BowmanA18} for \emph{abstract closure conversion},
which studies closure conversion for a specialized target language
with new constructs for representing closures and closure types.
Closure conversion into these constructs captures the essence of the
translation, while avoiding the unnecessary restrictions imposed by more
concrete settings.
Similarly, we will define a target language, the
\emph{Defunctionalized Calculus of Constructions} (DCC), in the style
of lambda calculus, but with a new construct for defunctionalized
\emph{labels} (representing indexes into a \emph{label context}) in
place of lambda abstractions.

\Cref{fig:compose-example} shows the result of defunctionalizing
the simply-typed \lstinline!compose! function to DCC~\footnote{We assume that $A$, $B$, and
$C$ are base types here.}, which looks and behaves like the
conventional defunctionalization presented in
\Cref{section:defunctionalization}.
In our translation into DCC, each abstraction $\lambda x.e_i$ is
replaced with a \emph{label expression} $[[LL_[I] {dy^}]]$ where $\target{\flabel_i}$ is
the label's identifier and $\target{\overline{y}}$ are the
abstraction's free variables.
The function body $e_i$ is stored in a separate \emph{label context}
$\target{\fdef}$ indexed by the label identifier, along with its
typing information.

In \Cref{fig:compose-example}, the label context $\target{\fdef}$ has
three entries, one for each abstraction in the original \lstinline!compose! function.
Each entry corresponds to one case of the $\$$ function in the
conventional defunctionalization.  For example, $[[ LL_[3] ]]$ arises from the translation of $\lambda x \mathrel{:} A .\ f\ (g\ x)$,
and corresponds to the \lstinline!F3! case in the definition of $\$$:
it has two free variables $[[dxf]] \mathrel{:} [[dxB]] \rightarrow [[dxC]] $ and 
$[[dxg]] \mathrel{:} [[dxA]] \rightarrow [[dxB]]$,
a bound variable $\target{x}$, and a body $[[dxf @ (! dxg @ dx !)]]$. 
As we shall see, a label application $[[LL_[3] {dxf , dxg} @ dN]]$
reduces to $[[dxf @ (! dxg @ dN !)]]$, just as the application
\lstinline!(F3 f g) $\$$ x! reduces to the corresponding right hand
side \lstinline!f $\$$ (g $\$$ x)!.

\begin{figure}[t]
$\begin{array}{rclcl}
  [[D]] &::=& 
\target{\flabel_3}(\{\target{f}:[[dxB]] \rightarrow [[dxC]], \target{g}:[[dxA]] \rightarrow [[dxB]]\}, \target{x}:\target{A} &\mapsto& [[dxf @ (! dxg @ dx !)]]:\target{C}),\\
&&\target{\flabel_2}(\{\target{f}:[[dxB]] \rightarrow [[dxC]]\},
\target{g}:([[dxA]] \rightarrow [[dxB]]) &\mapsto& \target{\flabel_3}\{\target{f}, \target{g}\}:[[dxA]] \rightarrow [[dxC]]),\\
&&\target{\flabel_1}(\{\}, \target{f}:([[dxB]] \rightarrow [[dxC]]) &\mapsto& [[LL_[2] {dxf}]]:
\rightarrow ([[dxA]] \rightarrow [[dxB]]) \rightarrow [[dxA]] \rightarrow [[dxC]]) \\
  \targettext{compose} &::=& [[LL_[1] {dnone}]] &&\\
\end{array}$
\caption{Defunctionalized simply-typed composition}
\label{fig:compose-example}
\end{figure}

It is straightforward to add dependent types to this scheme,
but some care is needed to define the transformation and
show that it has the desired meta-theoretical properties.
In particular, as we shall see, the transformation needs to consider 
the entire derivation tree rather than just the source language expression (\Cref{section:extracting-function-definitions}),
and we need to use a version of the source language with explicit substitutions (\Cref{section:soundness})
to make the type-preservation proof go through.
These challenges arise only in defunctionalization in a dependently typed setting, 
which has not been previously studied.

\section{Defunctionalizing with dependent types}
\label{section:cc-defunctionalization}

Having informally introduced the key concepts and motivated our abstract
defunctionalization translation, we now turn to the technical details.
The next few sections introduce
 our source language, the calculus of constructions (\Cref{section:cc}),
 our target language, the \emph{defunctionalized} calculus of constructions (\Cref{section:dcc}),
 and the defunctionalizion translation that links them (\Cref{section:translation}).
We then establish the soundness of the translation (\Cref{section:soundness})
 and prove the consistency of the target language (\Cref{section:consistency}).

\subsection{Calculus of Constructions}
\label{section:cc}

Our source language is a variant of the Calculus of Constructions
(CC)~\cite{DBLP:journals/iandc/CoquandH88}, an expressive
dependently-typed lambda calculus that serves as a basis for several
programming languages and proof assistants.
Our main departure from the original presentation of CC is in following the
approach taken by~\citet{DBLP:phd/ethos/Luo90}, and by many
dependently-typed languages such as Agda, Lean, Coq, and F*, by extending
CC with a Martin-Löf style hierarchy of universes.

Here is an example CC definition, $\text{compose}$, which represents the fully dependent composition
function for functions $f$ and $g$:
%
\[
\begin{array}{rcl}
  \text{compose} &::=& \lam{A}{U_0}
  {\enspace \lam{B}{(\pitype{x}{A}{U_0})}
  {\enspace \lam{C}{(\pitype{x}{A}{\pitype{y}{B\ x}{U_0}})}{} }}\\
    && \quad \lam{f}{(\pitype{y}{A}{(\pitype{z}{B\ y}{C\ y\ z})})}
    {\enspace \lam{g}{(\pitype{x}{A}{B\ x})}{} }\\
      && \qquad \lam{x}{A}{\enspace f\ x\ (g\ x)}
\end{array}
\]
\noindent
The expression component $\lambda A. \lambda B. \lambda C. \lambda f. \lambda g. \lambda x. f\;x\;(g\;x)$ of this definition  is unremarkable;
all the interest is in the dependencies of types on arguments.
In particular, the result type $C\ y\ z$ of $f$ depends on $f$'s arguments $y$ and $z$
and
the result type $B\ x$ of $g$ depends on $g$'s argument $x$.
(In a practical programming language, both $y$ and the type arguments
$A$, $B$ and $C$ would be passed implicitly,
but our minimal calculus does not support implicit arguments.)

\begin{figure*}[ht]
\begin{subfigure}{0.9\textwidth}
  \begin{syntax}
    \textrm{Expressions} & A,B,L,M,N  & ::= & {x}  \mid  {U}  \mid  [[Pi x : A . B]]  \mid [[L M]] \mid  [[\ x : A . M]]\\
    \textrm{Universes}   & U       & ::= & {U_i} \\
    \textrm{Contexts} & {\Gamma} & ::= & \cdot \mid  [[env , x : A]] \\
  \end{syntax}
  \caption{Syntax}
  \label{fig:cc syntax}

  \medskip
  \centering

  \begin{drulepar}[red]{$M [[|>]] N$}{Reduction}
    \drule{Beta}
  \end{drulepar}

  \caption{Reduction}
  \label{fig:cc reduction}

  \begin{drulepar}[ty]{$[[env |- M : A]]$}{Typing}
    \drule{Var}

    \drule{Universe}

    \drule{Pi}

    \drule{Apply}

    \drule{Lambda}

    \drule{Equiv}
  \end{drulepar}

  \caption{Typing}
  \label{fig:cc typing}

  \begin{drulepar}[wf]{$[[|- env]]$}{Well-formedness}
    \drule{Empty}

    \drule{Cons}
  \end{drulepar}

  \caption{Context formation}
  \label{fig:cc well-formedness}

  \begin{drulepar}[eq]{$[[|- M == N]]$}{Equivalence}
    \drule{Reduce}

    \drule{EtaOne} 

    \drule{EtaTwo}
  \end{drulepar}

  \caption{Equivalence}
  \label{fig:cc equivalence}
\end{subfigure}
  \caption{The Calculus of Constructions (CC)}
  \label{fig:cc}
\end{figure*}

\Cref{fig:cc syntax} shows the syntax of CC.
The expressions of CC are variables $[[x]]$ (drawn from an infinite set of names), universes $U$, dependent function types $[[Pi x : A . B]]$, applications $[[L M]]$, and abstractions $[[\ x : A . M]]$. A CC context $[[env]]$ is a telescope of variable-expression pairs.

CC has four judgements:
\begin{enumerate}
\item reduction (\Cref{fig:cc reduction})
\[
  M [[|>]] N
\]

\item type membership (\Cref{fig:cc typing})
\[
  [[env |- M : A]]
\]

\item context formation (\Cref{fig:cc well-formedness})
\[
  [[|- env]]
\]

\item equivalence (\Cref{fig:cc equivalence})
\[
  [[|- A == B]]
\]
\end{enumerate}

There is a single reduction rule (\Cref{fig:cc reduction}), for  $\beta$-reduction, $[[ (\ x : A . N) M]] \mathrel{\triangleright} [[N[M/x] ]]$.
We write $[[L |>*M]]$ to mean that $[[L]]$ reduces to $[[M]]$ in a sequence with zero or more steps.
%

CC's rules for typing (\Cref{fig:cc typing}) and context formation (\Cref{fig:cc well-formedness}) are defined by mutual induction.

The type of a variable $x$ is $A$ if $x:A$ is present in the well-formed context $[[env]]$ (\rref*{ty-Var}).
The type of a universe $U_i$ is $U_{i+1}$ (\rref*{ty-Universe}),
and the type of $[[Pi x : A . B]]$ is the higher universe among universes of $A$ and $B$ (\rref*{ty-Pi}).
If $M$ has type $B$ in some context $[[env]]$ extended with $x{:}A$,
then $[[ \ x : A . M]]$ has the dependent function type $[[Pi x : A . B]]$ (\rref*{ty-Lambda}).
Applications have types $B[N/x]$, since the output of a function type may depend on the argument $N$ (\rref*{ty-Apply}).
Finally, if an expression $M$ has type $A$ and $A$ is equivalent to $B$, then $M$ also has type $B$ (\rref*{ty-Equiv}).

A context $[[env]]$ is \emph{well-formed} (written $[[|- env]]$) if
every variable in it is associated with a valid type --- that is, the
associated expression’s type is a universe in the context $[[env]]$.

We make use of two shorthands, writing $[[env |- A : U]]$ to mean that $[[env |- A : U_i]]$ for \emph{some} $i$ (which means that $A$ is a type),
and $[[A]] \to [[B]]$ to stand for the $\Pi$-type $[[Pi x : A . B]]$ where $B$ does not depend on $x$.
For simplicity, we omit base types such as the unit type $1$ and the
natural numbers $\Bbb{N}$ from the formal definition, but we will use
them freely in examples.

In CC, two expressions are \emph{equivalent} (\Cref{fig:cc equivalence}) if they reduce to the same expression (\rref*{eq-Reduce}) or are $\eta$-equivalent
as defined by two symmetric rules \rref*{eq-Eta1} and \rref*{eq-Eta2}.
Under \rref*{eq-Eta1}, $L$ and $M$ are equivalent if $L$ reduces to an abstraction $[[\ x : A . L']]$, $M$ reduces to some $M'$, and $[[L']] \equiv [[M' x]]$, and \rref*{eq-Eta2} corresponds symmetrically~\cite{DBLP:journals/pacmpl/BowmanCRA18,DBLP:conf/pldi/BowmanA18}.

One useful property of CC is as follows: if $[[env |- M : A]]$, then $[[env |- A : U]]$.
Furthemore, CC is type safe and consistent, and type-checking in CC is decidable~\cite{DBLP:journals/iandc/CoquandH88,DBLP:phd/ethos/Luo90}.


\subsection{Defunctionalized Calculus of Constructions}
\label{section:dcc}

\begin{figure}[!h]
\begin{subfigure}{0.95\textwidth}
  \begin{syntax}
    \textrm{Universes}   & \target{U}        & ::= & \target{U_i} \\
    \textrm{Expressions} & \target{A},\target{B},\target{L},\target{M},\target{N}  & ::= & 
                         \target{x}
                         \mid \target{U}
                         \mid [[Pi dx : dA . dB]]
                         \mid \targetapp{L}{M}
                         \mid \targetlab{\flabel}{\overline{M}}\\
    \textrm{Type contexts} & \target{\Gamma} & ::= &
                             \target{\cdot}  
                             \mid \target{\Gamma}\sfcomma \itemtype{x}{A} \\
    \textrm{Label contexts} & \target{\fdef} & ::= &
                              \target{\cdot} 
                              \mid [[D , LL ( {dx^ : dA^} , dx : dA --> dM : dB )]]\\
    \textrm{DCC contexts} &\targetcon
    \end{syntax}%
  \caption{Syntax}
  \label{fig:dcc syntax}  

  \begin{drulepar}[d-red]{\small$[[D |- dM |> dN ]]$\vspace{-3ex}}{Reduction}
    \drule{Beta}
  \end{drulepar}

  \caption{Reduction}
  \label{fig:dcc reduction}

  \begin{drulepar}[d-ty]{\small$[[D ; denv |- dM : dA]]$}{Typing}
    \drule{Var}
    \drule{Universe}

    \drule{Pi}
    \drule{Apply}

    \drule{Label}
    \drule{Equiv}
  \end{drulepar}

  \caption{Typing}
  \label{fig:dcc typing}

  \begin{drulepar}[d-wf]{\small$[[|- D ; denv]]$}{Well-formedness}
    \drule{Empty}
    \drule{Label}
    \drule{Type}
  \end{drulepar}

  \caption{Context and label context formation}
  \label{fig:dcc context}

  \begin{drulepar}[d-eq]{\small$[[D |- dM == dN ]]$}{Equivalence}
    \drule{Reduce}

    \drule{EtaOne}
    \drule{EtaTwo}
  \end{drulepar}

  \caption{Equivalence}
  \label{fig:dcc equivalence}
\end{subfigure}

  \caption{The Defunctionalized Calculus of Constructions (DCC)}
  \label{fig:dcc}

\end{figure}
\afterpage{\clearpage}

\Cref{fig:dcc syntax} shows the syntax of our target language, the
Defunctionalized Calculus of Constructions (DCC).
As in CC, DCC expressions include
variables $\target{x}$,
universes $\target{U}$,
dependent function types $[[Pi dx : dA . dB]]$,
and
applications $\targetapp{L}{M}$.
Unlike CC, DCC contains first-class function labels
$[[LL{dM^}]]$ instead of lambda abstractions.

A label expression $[[LL{dM^}]]$ is a label name
$\target{\flabel}$ supplied with a list of zero or more expressions
$\target{\overline{M}}$ (standing for $\target{M_1}, \cdots,
\target{M_n}$) assigned to its free variables.
Label names $\target{\flabel_1}, \target{\flabel_2}, \cdots$ are
disjoint from variable names, as we emphasize using a different font.

There are two varieties of context in DCC.
As in CC, type contexts $\target{\Gamma}$ associate variables
$\target{x}$ with types $\target{A}$.
Label definition contexts $\target{\fdef}$ pair label names with their associated data: 
$[[LL ( {dx^ : dA^} , dx : dA --> dM : dB )]]$.
Here $[[dx^ : dA^]]$ records the type of the (possibly empty)
telescope of free variables that the label takes, $([[dx]]
\mathrel{:} [[dA]]) \to [[dB]]$ specifies the label type, and
$[[dM]]$ is the expression to which the label reduces when applied to
an argument.
Note that types in a type context $\target{\Gamma}$
may refer to labels $\target{\flabel_1}, \target{\flabel_2}, \cdots$
in the label context $\target{\fdef}$, but not vice versa.

DCC has four judgements:

\begin{enumerate}
\item reduction (\Cref{fig:dcc reduction})
\[
  [[ D |- dM |> dN ]]
\]

\item type membership (\Cref{fig:dcc typing})
\[
  [[D ; denv |- dM : dA]]
\]

\item context formation (\Cref{fig:dcc context})
\[
  [[|- D ; denv]]
\]

\item equivalence (\Cref{fig:dcc equivalence})
\[
  [[D |- dA == dB]]
\]
\end{enumerate}

\subsubsection{Reduction}

There is a single reduction rule (\Cref{fig:dcc reduction}), for label
application: the application of the label
$[[ LL { dM^ }]]$ to the argument $[[ dN ]]$ reduces to
$[[ dL [ dM^ / dx^ , dN / dx ] ]]$,
where $[[ dL ]]$ is the body of the entry for $[[ LL ]]$ in the label context
and $[[ dM^ ]]$ is the closure of $[[ LL ]]$.
A reduction sequence is noted
as $[[D |- dM |>* dN]]$, which means
$\target{M}$ reduces to $\target{N}$ in zero or more steps.
%

Substitutions for variables, universes, $\Pi$-types and applications
in DCC follow the conventional definition. Substitutions for labels
are
\begin{equation*}
[[ LL{dM^} [ dN / dx ] ]] \triangleq [[ LL{ dM^ [dN / dx] } ]],
\end{equation*}
where $[[ dM^ [dN / dx] ]]$ is syntactic sugar for
$\target{M_1}\targetsub{N}{x}, \cdots, \target{M_n}\targetsub{N}{x}$.

\subsubsection{Type judgements}

DCC's type judgements are of the form $[[D ; denv |- dM : dA]]$, 
and typing rules are given in \Cref{fig:dcc typing}. 
Rules for variables, universes, $\Pi$-types, applications,
and conversion are identical to their counterpart rules in CC, so we
focus on the rule for labels. A label term
$[[LL{dM^}]]$  is well-typed in $\targetcon$ if the
following conditions are satisfied.
\begin{enumerate}
	\item The context $\targetcon$ is \emph{well-formed}.
	\item $[[LL ( {dx^ : dA^} , dx : dA --> dM : dB )]]$ is present in $\target{\fdef}$.
	\item The length of the two lists $\target{\overline{M}}$ and
          $\itemtype{\overline{x}}{\overline{A}}$ are equal.
	\item All expressions in $\target{\overline{M}}$ are well-typed,
          and their types match the specified types of free variables
          $\target{\overline{A}}$.
\end{enumerate}
Specifically, condition (4) means:
\begin{align*}
  &[[ D ; denv |- dM_[1] : dA_[1] ]], \\[-0.1ex]
  &[[ D ; denv |- dM_[2] : dA_[2] [ dM_[1] / dx_[1] ] ]], \\[-0.1ex]
  &\cdots,\\[-0.1ex] \
  &\targetcon \vdash 
    \itemtype{M_n}{A_n}[\target{M_1} \slash \target{x_1},
          \cdots, \target{M_{n-1}} \slash \target{M_{n-1}}].
\end{align*}
Each $\target{A_{i+1}}$ depends on $\target{x_1}, \cdots,
\target{x_i}$, so $\target{M_1}, \cdots, \target{M_i}$ need to be
substituted in $\target{A_{i+1}}$ in the type judgement for
$\target{M_{i+1}}$. 
The type of $[[LL{dM^}]]$ is $[[ Pi dx : (dA [dM^ / dx^]) . (dB [dM^ / dx^]) ]]$.

Note that values of free variables $\target{\overline{M}}$ are substituted
in $[[Pi dx : dA . dB]]$, the specified type of the label. We use
$\targetsub{\overline{M}}{\overline{x}}$ as a syntactic sugar of
$[\target{M_1}\slash\target{x_1}\sfcomma \cdots\sfcomma
  \target{M_n}\slash\target{x_n}]$, and conditions (3) and (4) are
abbreviated to $[[D ; denv |- dM^ : dA^]]$ as a
convention.

The DCC judgement for well-formed contexts is $[[|- D ; denv]]$ 
and its rules are given in \Cref{fig:dcc context}.
A context is \emph{well-formed} if every variable in the type context 
is associated with a valid type (in the previous context $\targetcon$),
and every label is associated with a well-typed data.
In other words, if we have $[[LL ( {dx^ : dA^} , dx : dA --> dM : dB )]]$,
$[[dM]]$ should have the type $[[dB]]$ as specified
in the context formed by the previous label context and the free variables in $[[dM]]$
(namely $[[D]] ; [[dx^ : dA^ , dx : dA]]$).

Two terms $\target{L}$ and $\target{M}$ are equivalent (\Cref{fig:dcc equivalence}) if they
both reduce to the same term $\target{N}$ in a reduction sequence or
they are $\eta$-equivalent. DCC's $\eta$-equivalence rules are similar
to that of CC.  Rule (\rref*{d-eq-Eta1}) defines that
$\target{L}$ and $\target{M}$ are equivalent if 
$\target{L}$ reduces to a label $[[LL{dN^}]]$,
$\target{M}$ reduces to $\target{M'}$,
$[[LL ( {dx^ : dA^} , dx : dA --> dN : dB )]]$ is found in the label context $\target{\fdef}$,
and $[[dM' @ dx]]$ is equivalent to $[[dN [ dN^ / dx^ ] ]]$. 
Rules (\rref*{d-eq-Eta1}) and (\rref*{d-eq-Eta2}) are
symmetrical.

Both the type context and the label context have the weakening property: a well-typed expression is still well-typed in an extended type or label context (by induction on the type derivation rules). 

\begin{lemma}[Type weakening]\label{lemma:type-weakening}
If $[[D ; denv |- dM : dA]]$, $[[D ; denv |- dB : dU[I] ]]$, and $[[dx]]$ is fresh,
then ${[[D ; denv, dx : dB |- dM : dA]]}$.
\end{lemma}

\begin{lemma}[Label weakening]\label{lemma:label-weakening}
If  $[[D ; denv |- dM : dC]]$, $[[D ; dx^ : dA^ , dx : dA |- dN : dB]]$, and $[[ LL_[I] ]]$ is fresh,
then ${[[D , LL_[I] ( {dx^ : dA^} , dx : dA --> dN : dB ); denv |- dM : dC]]}$.
\end{lemma}

DCC is type-safe and consistent
(we establish these properties in \Cref{section:consistency}).
In addition, it is sufficiently expressive to support the $\text{compose}$ function,
but we must write it in defunctionalized style, since the calculus
does not support lambda abstraction.
There is one entry in the label context $[[D]]$ for each $\lambda$ in the CC definition of $\text{compose}$:

\smallskip
$\begin{array}{rcl}
  [[D]]                 &::=& 
  \target{\flabel_5}(\{\target{A},\target{B},\target{C},\target{f},\target{g}\}\,, [[dx]] : [[dxA]] \, \mapsto 
  [[((!dxf @ dx!) @ dxg @ dx)]] : [[((!dxC @ dx!) @ dxg @ dx)]]),\\
  && \cdots,\\
  && \target{\flabel_1}(\{[[dxA]] : \target{U_0}\}\,, [[dB]] : (\targetpi{\target{x}}{\target{A}}{\target{U_0}}) \, \mapsto 
  [[LL_[2] {dxA , dxB} ]] : \target{\Pi C}.\target{\Pi f}.\target{\Pi g}.\target{\Pi x}.[[((!dxC @ dx!) @ dxg @ dx)]]),\\
  && \target{\flabel_0}(\{\}\,, [[dxA]] : \target{U_0} \, \mapsto 
  [[LL_[1] {dxA} ]] : \target{\Pi B}.\target{\Pi C}.\target{\Pi f}.\target{\Pi g}.\target{\Pi x}.[[((!dxC @ dx!) @ dxg @ dx)]])\\
\end{array}$
\smallskip

\noindent
and the definition of $\text{compose}$ itself is simply a projection of a closed label from the context:

\smallskip
$\begin{array}{rcl}
  \targettext{compose}  &::=& [[LL_[0] {dnone} ]]\\
\end{array}$
\smallskip

\noindent
The full definition appears in \Cref{appendix:compose-defunctionalization}.

\subsection{The Defunctionalization Translation}
\label{section:translation}

\begin{figure*}[ht]
\begin{subfigure}{\textwidth}
  
  \begin{drulepar}[t]{$[[env |- M : A ~> dM]]$}{Expression transformation}
    \drule{Var}

    \drule{Universe}

    \drule{Pi}

    \drule{Apply}

    \drule{Lambda}

    \drule{Equiv}
  \end{drulepar}

  \caption{Defunctionalization of expressions}
  \label{fig:dcc transformation}

  \begin{drulepar}[d]{$[[env |- M : A ~>d D]]$}{Function extraction}
    \drule{Var}

    \drule{Universe}

    \drule{Pi}

    \drule{Lambda}

    \drule{Apply}

    \drule{Equiv}
  \end{drulepar}
  
  \caption{Extraction of function definitions}
  \label{fig:dcc def}

\end{subfigure}
  \caption{The Defunctionalization Translation}
  \label{fig:defunctionalization}

\end{figure*}

\Cref{fig:defunctionalization} shows the translation. It consists of two parts: a transformation $\bbracket{-}$ for expressions and a meta-function $\bbracket{-}_d$ that extracts function definitions from the source program. The expression transformation produces the target program and the meta-function $\bbracket{-}_d$ gives a label context.

\subsubsection{Expression transformation}

\begin{definition} The expression transformation $\bbracket{-}$ 
takes a well-typed term in CC and an implicit argument of that term's type derivation.
We define $\bbracket{M} \triangleq [[dM]]$, where $[[dM]]$ is given by a new judgement 
of the form $[[env |- M : A ~> dM]]$ (\Cref{fig:dcc transformation}).
\end{definition}

The transformation simply transcribes the variables, universes, $\Pi$-types, applications, and base types and values in CC to their counterparts in DCC compositionally. Functions in the source language are translated into labels in the target language.

Defunctionalization requires a unique correspondence between each label and each source-program function.
We use a convention that every lambda in the transformation's input $M$ is tagged with a unique identifier $i$ ($i \in \mathbb{N}$), and its corresponding label's name is $\target{\flabel_i}$.

The transformation turns a function $[[\^I x:A.M]]$ into a label $[[LL_[I] {dx^}]]$, where $\target{\overline{x}}$ come from the function's free variables $\overline{x}$ (\rref*{t-Lambda}). The meta-function FV (see \Cref{def:FV}) computes all free variables and their types involved in a well-typed CC-expression. Note that FV is different from \emph{fv}, the conventional free variable function that computes all the \emph{unbound variables} in an expression. In dependently typed languages, the type of a free variable may contain other free variables, and their types may still contain other free variables, and so on! Therefore, FV$(M)$ must recursively work out all the variables needed for $M$ to be well-typed. 

\begin{definition}FV$(M)$ takes $[[env |- M : A]]$, the type judgement of $M$, as an implicit argument. It firstly computes all the unbound variables $x_1, \cdots, x_n$ in $M$ and in $A$, then calls itself recursively on types of these variables, and finally returns the union of all free variables and their types it found. 
\begin{equation*}
\renewcommand{\arraystretch}{1.1}
\begin{array}{l l l}
	\text{FV}(M) & = & \text{FV}(A_1) \cup \cdots \cup \text{FV}(A_n) \cup \Gamma_{fv}\\
	& \text{where} & \emph{fv }(M) \cup \emph{fv }(A) = x_1, \cdots, x_n\\
	& & \Gamma \vdash x_1 \goodcolon A_1, \cdots, \Gamma \vdash x_n \goodcolon A_n\\
	& & \Gamma_{fv} \triangleq x_1 \goodcolon A_1, \cdots, x_n \goodcolon A_n.\\
\end{array}
\end{equation*}
\label{def:FV}
\end{definition}

Here, the union of two type contexts ${\Gamma_1} \cup {\Gamma_2}$ is ${\Gamma_1}$ appended with all the variable-expression pairs ${x} \goodcolon {A}$ that only appear in ${\Gamma_2}$, preserving their order. 
%
%
Intuitively, $[[FV M]]$ computes all the variables needed to correctly type $M$. 
Therefore, $M$ is still well-typed in its free-variable context $[[FV M]]$.

\begin{lemma}If $[[env |- M : A]]$, then $[[FV M |- M : A]]$.
\label{lem:fv}
\end{lemma}

\subsubsection{Extracting function definitions}
\label{section:extracting-function-definitions}

\begin{definition} 
$\bbracket{-}_d$ takes a well-typed CC term and implicitly its type derivation.
We define $\bbracket{M}_d \triangleq [[D]]$, where $[[D]]$ is given by a new judgement 
of the form $[[env |- M : A ~>d D]]$ (\Cref{fig:dcc def}).
\end{definition}

In a simply typed system, the only thing $\bbracket{-}_d$ has to do is
finding every function $[[\^I x:A.M]]$ in the source program
and placing them in the label context $\target{\fdef}$ in the following form
\begin{equation*}
  [[LL ( { dx^ : dA^ } , dx : dA --> dM : dB ) ]]
\end{equation*}
where
$\{\itemtype{\overline{x}}{\overline{A}}\}$, $\target{x} :
\target{A}$, $[[dM]]$, and $[[dB]]$ respectively correspond to the free variables
$(\overline{x} \mathrel{:} \overline{A})$ in the function, the bound
variable $x : A$, the function body $M$, and the return type $B$.

Alas, types may index over functions in our dependent type theory, and 
functions may appear in the type of an expression, 
even if the expression itself does not contain that function! 
For example, consider the following triple $(\Gamma,M,N)$ in CC
(with built-in natural numbers and addition).
\begin{align*}
  \Gamma\ & \triangleq\ \cdot,\ A \goodcolon (Nat \rightarrow Nat) \rightarrow U_0,\ 
  a \goodcolon\ \pitype{f}{(Nat \rightarrow Nat)}{A\ (\lam{n}{Nat}{1 + (f\ n)})} \\
  M\ & \triangleq\ a\ (\lam{x}{Nat}{1+x}) \\
  N\ & \triangleq A\ (\lam{n}{Nat}{2+n})
\end{align*}
$A$ is a family of types indexed by $Nat \rightarrow Nat$ functions and 
$a\ f$ constructs an element of type $A\ (\lam{n}{Nat}{1 + (f\ n)})$. 
According to the rule (\rref*{ty-Apply}), the inferred type of $M$ is 
\begin{align*}
     & (A\ (\lam{n}{Nat}{1 + (f\ n)}))\sub{(\lam{x}{Nat}{1 + x})}{f}\\
  =\ & A\ (\lam{n}{Nat}{(1 + (\lam{x}{Nat}{1 + x})\ n)}),
\end{align*}
which reduces to $A\ (\lam{n}{Nat}{2+n})$.
We have $[[env |- M : N]]$, yet $N$ contains a function that is not
in $\Gamma$ or $M$!
We should include this new function in $[[D]]$,
as it guarantees that we will never be in a situation
where we need a non-existent label in $\bbracket{M}_d$ 
to type $\bbracket{M}$.
In other words, the transformation
defunctionalizes not just the source-language expression, but its
entire type derivation tree.

Hence, we arrive at the rules in \Cref{fig:dcc def}.
Type derivations of universes do not involve functions at all (\rref*{d-Universe}).
Function definitions in a variable $x$ are just the definitions in its type $A$ (\rref*{d-Var}).  Definitions in a dependent function type $\pitype{x}{A}{B}$ are the \emph{union} of definitions in $A$ and $B$ (\rref*{d-Pi}). The union here is defined in the same way as the union of contexts (see \Cref{def:FV}), and there is no ambiguity since different functions correspond to different label names.

Definitions in an application $[[M N]]$ are the union of definitions in $M$, $N$, and $B\sub{N}{x}$, 
since the substitution $B\sub{N}{x}$ may create new function definitions (\rref*{d-Apply}). 
For a lambda abstraction $[[\^I x:A.M]]$, the definitions it contains are the union of definitions in $M$ and in $A$ appended with $\target{\flabel_i}$, the definition of itself (\rref*{d-Lambda}).
If $M$ has type $B$ by the conversion rule, then the definitions involved in the derivation of $[[env |- M : B]]$ are the union of definitions in the derivation of $[[env |- M : A]]$ and definitions in $B$ (\rref*{d-Equiv}).

We define the \emph{subset relation} of label contexts to help state further definitions and theorems.

\begin{definition} For two well-formed label contexts $[[D1]]$ and $[[D2]]$, $[[D1]] \subseteq [[D2]]$
if for all
$[[ LL_[I] ( {dx^ : dA^} , dx : dA --> dN : dB ) ]]$
in
$[[D1]]$,
$[[ LL_[I] ( {dx^ : dA^} , dx : dA --> dN : dB ) ]]$
is also in
$[[D2]]$.
\end{definition}

The notion of subsets gives a stronger weakening property to DCC: a well-typed expression is still well-typed in a larger label context.

\begin{lemma}[Label context weakening (subsets)] \label{lemma:context-weakening}
If $[[D_[1] ; denv |- dM : dA]]$, $[[|- D_[2] ]]$, and $[[ D_[1] subseteq D_[2] ]]$,
then $[[D_[2] ; denv |- dM : dA]]$.
\end{lemma}

Since the transformation defunctionalizes the entire type derivation tree of an expression, 
if $[[env |- M : A]]$, then all elements in $[[ [|A|]d ]]$ are also in $[[ [|M|]d ]]$.
We can prove this property by induction on the type derivation rules.

\begin{lemmarep}\label{lemma:type-label-subset}
For any well-typed expression $[[env |- M : A]]$ in CC, $[[ [|A|]d subseteq [|M|]d ]]$.
\end{lemmarep}
\begin{appendixproof}
By induction on rules defined in \Cref{fig:dcc def}. 

\begin{description}
  \item[Case (\rref*{d-Lambda}).] The goal is to show that $[[ [|Pi x : A . B|]d subseteq [|\^I x : A . M |]d ]]$.
By assumption,
\begin{equation*}
[[env |- \^I x : A . M : Pi x : A . B ~>d D_[A] cup ( D_[M] , LL_[I] ( {dx^ : dA^} , dx : dA --> dM : dB ) )]].
\end{equation*}
In other words, $[[ [|\^I x : A . M |]d ]] = [[ [|A|]d cup [|M|]d ]]$ with definition of function $\lambda^i$ appended to it. 
By definition, $[[ [|Pi x : A . B|]d ]] = [[ [|A|]d cup [|B|]d ]]$. We have $[[ [|A|]d subseteq [|\^I x : A . M |]d ]]$ and $[[ [|B|]d subseteq [|\^I x : A . M |]d ]]$ by the induction hypothesis (since $[[env, x : A |- M : B]]$, $[[ [|B|]d subseteq [|M|]d ]]$).
\end{description}

Other cases are either trivial or follows simply from definitions.

\end{appendixproof}

The expression transformation and the process of extracting function definitions ($\bbracket{-}$ and $\bbracket{-}_d$) act pointwise on CC contexts. In other words,
\begin{equation*}
\begin{array}{l @{\qquad} l}
	\bbracket{\cdot}\ \triangleq\ \target{\cdot},\  & \bbracket{\Gamma, x \goodcolon A}\ \triangleq \bbracket{\Gamma}\sfcomma \target{x} \goodcolon \bbracket{A},\\
	\bbracket{\cdot}_d \triangleq\ \target{\cdot},\  & \bbracket{\Gamma, x \goodcolon A}_d \triangleq \bbracket{\Gamma}_d\cup\bbracket{A}_d.\\
\end{array}
\end{equation*}

Now, we can see that the (tagged) composition function
$\lambda^0 A. \lambda^1 B. \lambda^2 C. \lambda^3 f. \lambda^4 g. \lambda^5 x. f\;x\;(g\;x)$
transforms to $[[LL_[0] { dnone }]]$, a label with no free variables supplied, 
since the function is closed.
The label context $[[D]]$ for composition can be derived from the function extraction judgements
with the sketch derivation tree shown below.  

\begin{prooftree}
\AxiomC{$A, B, C, f, g \vdash \lambda^5x.f\;x\;(g\;x)
\leadsto_d \target{\flabel_5}(\{\target{A},\target{B},\target{C},\target{f},\target{g}\}, \target{x}:\_ \mapsto [[(!dxf @ dx!) @ dxg @ dx]] : \_ )
$}
\RightLabel{\rref*{d-Lambda}}
\UnaryInfC{$A, B, C, f \vdash \lambda^4g.\lambda^5x.f\;x\;(g\;x)
\leadsto_d \cdots, \target{\flabel_4}(\{\target{A},\target{B},\target{C},\target{f}\}, \target{g}:\_ \mapsto [[LL_[5] {dxA,dxB,dxC,dxf,dxg}]] : \_ )
$}
\RightLabel{\rref*{d-Lambda}}
\UnaryInfC{$A, B, C \vdash \lambda^3f.\lambda^4g.\lambda^5x.f\;x\;(g\;x)
\leadsto_d \cdots, \target{\flabel_3}(\{\target{A},\target{B},\target{C}\}, \target{f}:\_ \mapsto [[LL_[4] {dxA,dxB,dxC,dxf}]] : \_ )
$}
\RightLabel{\rref*{d-Lambda}}
\UnaryInfC{$A, B\vdash \lambda^2C.\lambda^3f.\lambda^4g.\lambda^5x.f\;x\;(g\;x)
\leadsto_d \cdots, \target{\flabel_2}(\{\target{A},\target{B}\}, \target{C}:\_ \mapsto [[LL_[3] {dxA,dxB,dxC}]] : \_ )
$}
\RightLabel{\rref*{d-Lambda}}
\UnaryInfC{$A \vdash \lambda^1B.\lambda^2C.\lambda^3f.\lambda^4g.\lambda^5x.f\;x\;(g\;x)
\leadsto_d \cdots, \target{\flabel_1}(\{\target{A}\}, \target{B}:\_ \mapsto [[LL_[2] {dxA,dxB}]] : \_ )
$}
\RightLabel{\rref*{d-Lambda}}
\UnaryInfC{$\cdot \vdash \lambda^0A.\lambda^1B.\lambda^2C.\lambda^3f.\lambda^4g.\lambda^5x.f\;x\;(g\;x)
\leadsto_d \cdots, \target{\flabel_0}(\{\}, \target{A}:\_ \mapsto [[LL_[1] {dxA}]] : \_ )
$}
\end{prooftree}

\subsection{Soundness}
\label{section:soundness}

\begin{figure}[ht]
  \begin{equation*}
    Expressions \quad ::= \quad \cdots\ |\ [[sM { sx |-> sN }]] 
  \end{equation*}

  \begin{drulepar}[s-ty]{$[[senv |- sM : sA ]]$}{Typing}
    \drule{Apply}

    \drule{Subst}
  \end{drulepar}

  \begin{drulepar}[s-red]{$[[sM |> sN]]$}{Reduction}
    \drule{VarOne}
    \drule{VarTwo}
    \drule{Universe}

    \drule{Beta}
    \drule{Apply}

    \drule{Closure}
  \end{drulepar}

  \begin{drulepar}[s-eq]{$[[|- sM == sN ]]$}{Equivalence}
    \drule{ClosureOne}
    \drule{ClosureTwo}
  \end{drulepar}

  \caption{\label{fig:ccs}New syntax and rules in {\ccs}}
\end{figure}

We consider dependently typed defunctionalization \emph{correct} if for all base types $A$, values $v$,
and programs $M$ of type $A$,
\begin{equation*}
  [[empty |- M : A]] \land \ M\ \triangleright^* v\ \Longrightarrow\ 
	\target{\fdef_{\Gamma}} \cup \target{\fdef_{M}} \vdash \target{M}\ \triangleright^* \target{v'} \text{ where } \target{v'} \equiv \target{v}.
\end{equation*}
In other words, if a closed program $M$ evaluates to a base-type value $v$, then $\target{M}$ evaluates to a base-type value $\target{v'}$ that is equivalent to $\target{v}$. This property follows as a corollary of the \emph{preservation of reduction sequences}, which states that
\begin{align*}
	M\ \triangleright^* N\ \Longrightarrow\ 
	& \target{\fdef_{\Gamma}} \cup \target{\fdef_{M}} \cup \target{\fdef_{N}} \vdash
	\target{M}\ \triangleright^* \target{M'},\\ 
	&\text{ where } \target{\fdef_{\Gamma}} \cup \target{\fdef_{M}} \cup \target{\fdef_{N}} \vdash \target{M'} \equiv \target{N}.
\end{align*}

Ideally, we could show this property by showing that the
transformation preserves all small-step reductions $M\mathrel{\triangleright}N$, followed by an induction on the number of
reduction steps in a sequence. However, CC's meta-language
substitution $(\lam{x}{A}{M})\sub{N}{y}$ creates a new function
definition when it substitutes an expression into a free variable of a
function.  So, for a reduction sequence $M_1\ \triangleright \cdots \ \triangleright\ M_n$ in CC, some $M_i$ may contain function
definitions that do not exist in $M_1$ or $M_n$. Consequently, not
all $M_i$ translate into well-typed DCC expressions $\target{M_i}$ in $\sfpl \target{\fdef_{\Gamma}} \cup \target{\fdef_{M_1}} \cup \target{\fdef_{M_n}} \sfpr \semicolon \target{\Gamma}$, which makes
the standard approach infeasible. Moreover, preservation of reduction
sequences is a key lemma for showing type preservation, since CC's
typing rules involve equivalence and the equivalence rule (\rref*{eq-Reduce})
is defined with reductions.

Fortunately, meta-theoretic substitution is the only means of creating
new function definitions in CC's reduction sequences. There would be
no problem if the source language did not evaluate substitutions into
functions but kept them as primitive expressions. To apply
this observation we define a helper language {\ccs}, which is an
extension of CC with \emph{explicit
substitutions}~\cite{DBLP:journals/jfp/AbadiCCL91}. In addition,
{\ccs} does not reduce substitutions of expressions into functions.

Since {\ccs} extends CC, every CC expression is trivially a {\ccs} expression. We denote this trivial transformation from CC to {\ccs} as $\sigma$. Then, we define the defunctionalization transformation from {\ccs} to DCC in a similar way as that from CC to DCC -- an expression transformation $\hbracket{-}$ and a meta-function $\hbracketd{-}$ for extracting definitions. Next, we show that $\sigma$ and defunctionalization for {\ccs} preserve reduction sequences and they commute with the transformation from CC into DCC. As a corollary, defunctionalization from CC to DCC preserves reduction sequences. In other words, we show that the following diagram commutes for all CC-expressions $M$ and $N$ (contexts omitted).

\[
\begin{tikzcd}
M \arrow[d, "\sigma"] \arrow[dd, "\bbracket{-}"', bend right, shift right=2] \arrow[rr, "\triangleright^*"] &                                               & N \arrow[d, "\sigma"'] \arrow[dd, "\bbracket{-}", bend left, shift left=2] \\
\helper{M} \arrow[r, "\triangleright^*"] \arrow[d, "\hbracket{-}"]                                           & \helper{M'} \arrow[r, "\equiv"] \arrow[d, "\hbracket{-}"] & \helper{N} \arrow[d, "\hbracket{-}"']                                       \\
\target{M} \arrow[r, "\triangleright^*"]                                                                    & \target{M'} \arrow[r, "\equiv"]                & \target{N}                                                                
\end{tikzcd}
\]

{\ccs} is an extension of CC with new syntax, type derivation rules,
reduction rules, and equivalence rules (\Cref{fig:ccs}). We write
{\ccs} expressions in a $\helper{teal, mathematical}$ $\helper{font}$
to avoid ambiguity. {\ccs} extends the CC syntax with
\emph{syntactic substitutions} of the form
$[[sM { sx |-> sN }]]$.

Type rules for variables, universes, $\Pi$-types, functions, and
equivalence in {\ccs} are the same as the standard rules in CC, except
that the type of an application $[[sM sN]]$ is
$[[sB { sx |-> sN }]]$ with the syntactic substitution.  The type of
a substitution $[[sM { sx |-> sN }]]$ is the type of
$\helper{M}$ with $\helper{x}$ substituted by $\helper{N}$
(\rref*{s-ty-Subst}).

{\ccs} has five reduction rules for 
substitutions, which are the standard meta-theoretic substitution
rules for variables, universes, $\Pi$-types, and applications being
internalised into the language. Note that the meta-theoretic
substitution in the CC's original beta-reduction rule
$[[(\sx : sA . sM) sN |> sM { sx |-> sN } ]]$
is also replaced by the syntactic one. {\ccs} does not reduce
substitutions into functions, but it $\beta$-reduces them when they
are applied to arguments (\rref*{s-red-Closure}). 
We write
$\helper{M}\helper{\{x_1 \mapsto N_1, x_2 \mapsto N_2\}}$ for a
substitution followed by another substitution
$(\helper{M}\hsub{N_1}{x_1})\hsub{N_2}{x_2}$, and
$\helper{M}\hsub{\overline{N}}{\overline{y}}$ for a sequence of substitutions
$(((\helper{M}\hsub{N_1}{y_1})\hsub{N_2}{x_2}) \cdots
)\hsub{N_n}{y_n}$.

Like in CC, two terms in {\ccs} are equivalent if they $\beta$-reduce to
the same expression or are $\eta$-equivalent.
In addition, {\ccs} has two symmetric rules
(\rref*{s-eq-Closure1}) and (\rref*{s-eq-Closure2}) for determining when a
sequence of substitutions into a function 
$[[(\sx : sA. sM) { sy^ |-> sN^}]]$
is equivalent to another expression. This is essentially a variant of
the $\eta$-equivalence rules that is compatible with substitutions --
$[[(\sx : sA. sM) { sy^ |-> sN^}]]$ is equivalent to
$\helper{N}$ if applying $\helper{N}$ to $\helper{x}$ is
equivalent to the function body $\helper{M}$ with $\helper{\overline{y}}$
being substituted for $\helper{\overline{N}}$.

Now, we define the defunctionalization transformation from {\ccs} to
DCC, which is the transformation from CC to DCC extended with the
following two rules. We use $\hbracket{-}$ and $\hbracketd{-}$ to stand
for the expression transformation and the metafunction for extracting
function definitions, and we apply the convention of tagging lambdas
with unique identifiers $i\ (i \in \mathbb{N})$ as usual.

  \begin{drulepar}[s-t]{$[[senv |- sM : sA ~>s dM ]]$}{Expression transformation}
    \drule{Subst}
  \end{drulepar}

  \begin{drulepar}[s-d]{$[[senv |- sM : sA ~>sd D ]]$}{Function extraction}
    \drule{Subst}
  \end{drulepar}

The transformation turns a syntactic substitution in {\ccs} into a
meta-theoretic substitution in DCC (\rref*{s-t-Subst}); the function
definitions in a substitution $[[sM {sx |-> sN}]]$ are the
union of the definitions in $\helper{M}$ and $\helper{N}$
(\rref*{s-d-Subst}).  Since substitutions into functions do not reduce
in {\ccs}, the transformation from it into DCC have the following
strong properties by definition, which are not true for the
transformation from CC into DCC.
\begin{align}
	\label{reduction def}
  [[sM |>* sN]] \Longrightarrow\ \helper{\bbracket{N}_d} \subseteq\ \helper{\bbracket{M}_d}\\
	\label{ccs substitution}
	\hbracket{[[sM {sx |-> sN}]]} = \hbracket{\helper{M}}\sub{\hbracket{\helper{N}}}{\target{x}}.
\end{align}

Next, we show that the transformation preserves small step reductions
in {\ccs} -- if a {\ccs} program $\helper{M}$ reduces to
$\helper{N}$ in one step, then the translated program $\target{M}$
evaluates to $\target{N}$ in a sequence.

\begin{lemmarep}[Preservation of small step reductions]\label{lemma:ss-reduction-preservation}
If $[[ senv |- sM : sA ]]$ and $[[ sM |> sN ]]$, then 
$\helper{\bbracket{\Gamma}_d} \cup \helper{\bbracket{M}_d} \vdash \target{M}\ \triangleright^* \target{N}$.
\end{lemmarep}
\begin{appendixproof}
By induction on the reduction rules of {\ccs}. 

\begin{description}
  \item[Case (\rref*{s-red-Beta}).] Assuming
  $(\helper{\lambda^i x} \goodcolon \helper{A} . \helper{M})\ \helper{N}\ \triangleright\ \helper{M\hsub{N}{x}}$, the goal is to show that 
  \[
  [[ D |- LL_[ I ] { dx^ } @ dN |>* dM [ dN / dx ] ]],
  \]
  where $\target{\fdef} = \helper{\bbracket{\Gamma}_d} \cup \helper{\bbracket{M}_d}$ 
  and $\target{\overline{x}}$ corresponds to the free variables in $\helper{\lambda^i}$. 
  By definition of $\hbracketd{-}$, we have $[[ LL_[ I ] ( {dx^ : d _} , dx : d _ --> dM : d _ )  in D]]$
  (ignoring type information).
  So, $[[ D |- LL_[ I ] { dx^ } @ dN |> dM [ dx^ / dx^ , dN / dx ] ]] = [[ dM [ dN / dx] ]]$.
  \item[Case (\rref*{s-red-Beta-Eta}).] Similar.
\end{description}

Other cases are trivial. 

\end{appendixproof}

The transformation preserves sequences of reductions, and the proof follows from a trivial induction on the number of small steps in the sequence.

\begin{lemma}[Preservation of reduction sequences ({\ccs})]\label{lem:ccs prev sequence}
If $[[senv |- sM : sA]]$ and $[[sM |>* sN]]$, 
then $\helper{\bbracket{\Gamma}_d} \cup \helper{\bbracket{M}_d} \vdash \target{M}\ \triangleright^* \target{N}$.
\end{lemma}

The transformation is also coherent, i.e.~it preserves the equivalence
relation in {\ccs}.

\begin{lemmarep}[Coherence ({\ccs})]
\label{lem:ccs coherence}
If $ [[senv |- sM : sA]] $, $ [[senv |- sN : sA]] $, and $[[ |- sM == sN ]]$,
then $[[D |- dM == dN]]$, where 
$\target{\fdef} = \helper{\bbracket{\Gamma}_d} \cup \helper{\bbracket{M}_d} \cup \helper{\bbracket{N}_d}$.
\end{lemmarep}
\begin{appendixproof}
By induction on the equivalence rules of {\ccs}. 

\begin{description}
  \item[Case (\rref*{s-eq-Reduce}).] By \Cref{lem:ccs prev sequence}.
  \item[Case (\rref*{s-eq-Eta1}).] Assume that we have $[[senv |- sM : Pi sx : sA . sB]]$, $[[senv |- sN : Pi sx : sA . sB]]$, and $[[|- sM == sN ]]$ by \rref{s-eq-Eta1},
  then we must have $[[sM |>* \^ I sx : sA . sM' ]]$, $[[sN |>* sN']]$, and $[[|- sM' == sN' sx]]$.
  
  By \Cref{lem:ccs prev sequence}, we have 
  $[[D_[1] |- dM |>* LL_[I] { dx^ } ]]$,
  $[[D_[2] |- dN |>* dN']]$,
  where $\target{\fdef_1} = \helper{\bbracket{\Gamma}_d} \cup \helper{\bbracket{M}_d}$,
  $\target{\fdef_2} = \helper{\bbracket{\Gamma}_d} \cup \helper{\bbracket{N}_d}$,
  and $[[ LL_[ I ] ( {dx^ : d _} , dx : d _ --> dM : d _ )  in D_[1] ]]$.
  By the induction hypothesis, $[[D_[3] |- dM' |>* dN' @ dx ]]$, 
  where $\target{\fdef_3} = \hbracketd{\helper{\Gamma}} \cup \helper{\bbracket{M'}_d} \cup \helper{\bbracket{N'}_d}$.  
  
  The goal is to show that $[[D |- dM == dN]]$ using \rref{d-eq-Eta1} (shown below).
  \begin{center}
  \drule{d-eq-EtaOne}
  \end{center}

  We already have all the premises like $\target{M}\ \triangleright^* \targetlab{\flabel_i}{\overline{x}}$, etc., but they are not judged in the desired label context $\target{\fdef} = \helper{\bbracket{\Gamma}_d} \cup \helper{\bbracket{M}_d} \cup \helper{\bbracket{N}_d}$. We only need to show that $\target{\fdef_1}$, $\target{\fdef_2}$, and $\target{\fdef_3}$ are subsets of $\target{\fdef}$ and to apply the weakening theorem. $\target{\fdef_1}$ and $\target{\fdef_2}$ are clearly subsets of $\target{\fdef}$. By \eqref{reduction def}, we have
  $\hbracketd{\helper{M'}} \subseteq \hbracketd{\helper{\Gamma}} \cup \hbracketd{\helper{M}}$ and
  $\hbracketd{\helper{N'}} \subseteq \hbracketd{\helper{\Gamma}} \cup \hbracketd{\helper{N}}$.
  Therefore, $\target{\fdef_3}$ is also a subset of $\target{\fdef}$.
  \item[Case (\rref*{s-eq-Clo1}).] Similar, but more tedious.
  \item[Case (\rref*{s-eq-Eta2, s-eq-Clo2}).] By symmetry.
\end{description}

(Note that rules \rref*{s-eq-Reduce}, \rref*{s-eq-Eta1}, and \rref*{s-eq-Eta2} are not listed in \Cref{fig:ccs}, but they are exactly the same as \rref*{eq-Reduce}, \rref*{eq-Eta1}, and \rref*{eq-Eta2} in \Cref{fig:cc}).

\end{appendixproof}

Recall that $\sigma$ denotes the trivial transformation from CC to {\ccs}. This trivial transformation commutes with the two term transformations by definition.
\begin{equation}
\hbracket{\sigma(M)} = \bbracket{M}
\label{eq:commute}
\end{equation}
In addition, function definitions in $\hbracketd{\sigma(M)}$ is a subset of the definitions in $\bbracket{M}_d$, because new function definitions appear in CC's type derivation trees as results of substitutions, but this does not happen in {\ccs}. 
\begin{equation}
\hbracketd{\sigma(M)} \subseteq \bbracket{M}_d
\label{eq:commute_d}
\end{equation}
We show that $\sigma$ also preserves sequences of reductions. As a convention, we write $\helper{M}$ for $\sigma(M)$ when there is no ambiguity.

\begin{lemmarep}[Preservation of reduction sequences ($\sigma$)]
\label{lem:sigma prev sequence}
If $\Gamma \vdash M\ \triangleright^* N$, then
$\helper{\Gamma} \vdash \helper{M}\ \triangleright^* \helper{M'} \text{ where } \helper{\Gamma} \vdash \helper{M'} \equiv \helper{N}$.
\end{lemmarep}
\begin{appendixproof}

We firstly proof that $[[env , x : A |- M : B]]$ and $[[env |- N : A]]$ implies that
$\vdash \sigma(M\sub{N}{x}) \equiv \helper{M\hsub{N}{x}}$. We prove this by induction on the type derivations of $M$.

\begin{description}
  \item[Case (\rref*{ty-lambda}).] By definition, $[[(\ y : A . M) [ N / x] ]] = [[\ y : A [ N / x] . M [ N / x] ]]$.
  So, $\sigma([[(\ y : A . M) [ N / x] ]]) = \helper{\lambda y :}\sigma([[A [N / x] ]])\helper{.}\sigma([[M [N / x] ]])$.
  We have
  \[
  [[((\ sy : sB . sM) { sx |-> sN}) sy]] \triangleright [[(sM { sx |-> sN}) {sy |-> sy}]]
  \equiv [[sM { sx |-> sN}]].
  \]
  By the induction hypothesis, $[[sM { sx |-> sN}]] \equiv \sigma([[M [ N / x] ]])$. Hence, we have
  $\sigma([[(\ y : B . M) [ N / x] ]]) \equiv [[((\ sy : sB . sM) { sx |-> sN})]] $ by \rref{s-eq-Eta1}.
\end{description}

Other cases are trivial. Using this result, we have 
$[[(\ sx : sA . sM ) sN |> sM { sx |-> sN } ]] \equiv \sigma([[M [ N / x] ]])$, 
so, $\sigma$ preserves small step reductions. Again, the preservation of reduction sequences follows immediately from this.
\end{appendixproof}

We can finally prove the preservation of reduction sequences for
dependently typed defunctionalization (from CC to DCC) using the
lemmas above.

\begin{lemmarep}[Preservation of reduction sequences]
\label{lem:prev sequence}
For all $M$ and $N$, if $[[env |- M : A]]$ and $[[M |>* N]]$, then we have
\begin{align}
& [[ D_[env] cup D_[M] cup D_[N]  |- dM |>* dM' ]],\\
& [[ D_[env] cup D_[M] cup D_[N]  |- dM' |>* dN ]]
\end{align}
for some $[[dM']]$ where
$([[ D_[env] cup D_[M] cup D_[N] ]])$ = 
$([[ [|env|]d ]], [[ [|M|]d ]], [[ [|N|]d ]])$ and 
$([[dM]], [[dN]])$ = $([[ [|M|] ]], [[ [|N|] ]])$
.
\end{lemmarep}
\begin{appendixproof}
Suppose that $[[env |- M : A]]$ and $[[M |>* N]]$. Then, we have 
$\vdash \helper{M}\ \triangleright^* \helper{M'} \text{ and } \vdash \helper{M'} \equiv \helper{N}$ for some $\helper{M'}$, since $\sigma$ preserves reduction sequences. By \Cref{lem:ccs prev sequence} and \Cref{lem:ccs coherence}, 
\begin{align*}
\hbracketd{\helper{\Gamma}} \cup \hbracketd{\helper{M}} &\vdash
\target{M} \triangleright^* \target{M'}\\
\hbracketd{\helper{\Gamma}} \cup \hbracketd{\helper{M}} \cup \hbracketd{\helper{N}} &\vdash \target{M'} \equiv \target{N}.
\end{align*}
Finally, since $(\hbracketd{\helper{\Gamma}} \cup \hbracketd{\helper{M}})$ and $(\hbracketd{\helper{\Gamma}} \cup \hbracketd{\helper{M}} \cup \hbracketd{\helper{N}})$ are both subsets of $([[ D_[env] cup D_[M] cup D_[N] ]])$, we have 
$[[ D_[env] cup D_[M] cup D_[N]  |- dM |>* dM' ]]$ and 
$[[ D_[env] cup D_[M] cup D_[N]  |- dM' |>* dN ]]$ 
by weakening.
\end{appendixproof}

Since ground types and values do not contain functions,
$\bbracket{v}_d = \target{\cdot}$, and the correctness of the
transformation is just a special case of \Cref{lem:prev
  sequence}.

\begin{cororep}(Correctness)\label{coro:correctness}
For all ground types $A$ and values $v$ of type $A$,
\begin{equation*}
  [[empty |- M : A]] \land [[M |>* v]]
  \Longrightarrow
  [[ D_[env] cup D_[M] |- dM |>* dv' ]]
  \text{ where } [[dv']] \equiv [[dv]].
\end{equation*}
\end{cororep}

The proof of type-preservation requires three lemmas:
\emph{substitution}, \emph{preservation of reduction sequences}, and
\emph{coherence}. \Cref{lem:prev sequence} established that
dependently-typed defunctionalization preserves reduction sequences
with the help of {\ccs}, and now we prove the remaining two lemmas in
a similar way. The substitution lemma states that defunctionalization
is compatible with substitutions.

\begin{lemmarep}[Substitution] 
\label{lem:sub}
If $[[env , x : A |- M : B]]$ and $[[env |- N : A]]$, then $[[D |- [| M [ N / x] |] == dM [ dN / dx ] ]]$, 
where $[[D]] = [[ D_[env] cup D_[M] cup D_[N] cup D_[ M [ N / x] ] ]]$.
\end{lemmarep}
\begin{appendixproof}
From the proof of \Cref{lem:sigma prev sequence}, we know that
$[[env , x : A |- M : B]]$ and $[[env |- N : A]]$ implies that 
$\sigma(M\sub{N}{x}) \equiv \helper{M\hsub{N}{x}}$. 
This further implies that $\target{\fdef'} \vdash \hbracket{\sigma(M\sub{N}{x})} \equiv \hbracket{\helper{M\hsub{N}{x}}}$, where 
$\target{\fdef'} = \hbracketd{\helper{\Gamma}} \cup \hbracketd{\helper{M}} \cup \hbracketd{\helper{N}} \cup \hbracketd{\sigma(M\sub{N}{x})}$.

By (\Cref{eq:commute_d}), $\target{\fdef'} \subseteq \target{\fdef}$. Also, $\hbracket{\sigma(M\sub{N}{x})} = \bbracket{M\sub{N}{x}}$, $\hbracket{\helper{\Gamma}} = \target{\Gamma}$, and $\hbracket{\helper{M}}\sub{\hbracket{\helper{N}}}{\target{x}} = \target{M}\targetsub{N}{x}$ by \eqref{ccs substitution} and \eqref{eq:commute}. 
Therefore, we have $[[D |- [| M [ N / x] |] == dM [ dN / dx ] ]]$ by weakening.
\end{appendixproof}

The coherence lemma states that defunctionalization is compatible with CC's coherence judgements.

\begin{lemmarep}[Coherence] 
\label{lem:coherence}
If $[[env |- M : A]]$, $[[env |- N : A]]$, and $[[|- M == N]]$, then $[[D |- dM == dN]]$, 
where $[[D]] = [[ D_[env] cup D_[M] cup D_[N] ]]$.
\end{lemmarep}
\begin{appendixproof}
Assume that $[[env |- M : A]]$, $[[env |- N : A]]$, and $[[|- M == N]]$. 
We have $\vdash \helper{M} \equiv \helper{N}$ by definition, 
since the source language does not contain explicit substitution. 
By \Cref{lem:ccs coherence}, we have $[[D' |- dM == dN]]$, where
$\target{\fdef'} = \helper{\bbracket{\Gamma}_d} \cup \helper{\bbracket{M}_d} \cup \helper{\bbracket{N}_d}$.
Since $\target{\fdef'} \subseteq \target{\fdef}$, we have $[[D |- dM == dN]]$ by weakening.
\end{appendixproof}

Finally, we show type preservation with an induction on CC's type derivation rules. 

\begin{theoremrep}[Type preservation] 
\label{theorem:type preservation}
For all well-typed programs $[[M]]$,
\begin{equation*}
	[[env |- M : A]] \Longrightarrow [[D_[env] cup D_[M] ; denv |- dM : dA ]],
\end{equation*}
where $([[D_[env] ]], [[D_[M] ]]) = ([[ [|env|]d ]], [[ [|M|]d ]])$ and 
$([[ denv ]], [[ dM ]], [[ dA ]]) = (\bbracket{\Gamma}, [[ [|M|] ]], [[ [|A|] ]])$.
\end{theoremrep}
\begin{appendixproof}
By proving the following two statements together with a simultaneous induction on mutually-defined judgements 
$[[|- env]]$ and $[[env |- M : A]]$.
\begin{enumerate}
	\item $[[|- env]] \Longrightarrow\ [[|- D_[env] ; denv ]]$.
	\item $[[env |- M : A]] \Longrightarrow [[D_[env] cup D_[M] ; denv |- dM : dA ]]$.
\end{enumerate}
Statement $1$ follows trivially from the inductive hypothesis. We look at the cases for statement $2$.
\begin{description}
  \item[Case (\rref*{ty-Var, ty-Universe, ty-Pi}).] Trivial by the induction hypothesis.
  
  \item[Case (\rref*{ty-Apply})] Suppose $[[env |- M N : B[ N / x] ]]$. The goal is to show that 
  $\targetcon \vdash \targetapp{M}{N} \goodcolon \bbracket{B\sub{N}{x}}$, where 
  $\target{\fdef} = \sfpl \target{\fdef_{\Gamma}} \cup \bbracket{M\ N}_d \sfpr$. 
  We have
  $[[D_[env] cup D_[M] ; denv |- dM : Pi dx : dA . dB]]$ and
  $[[D_[env] cup D_[N] ; denv |- dN : dA]]$ by the induction hypothesis.
  By \rref{d-ty-Apply} and weakening,
  $[[D_[env] ; denv |- dM @ dN : dB [dN / dx] ]]$.
  By the substitution lemma, 
  $[[D' |- dB [dN / dx] == [|B [N / x]|] ]]$, 
  where $[[D']] = [[D_[env] cup D_[B] cup D_[N] cup D_[ B [N / x] ] ]]$.
  We have the goal if $[[D' subseteq D]]$. 
  Indeed,
  \begin{enumerate}
  \item $[[D_[env] subseteq D]]$ by def.
  \item $[[D_[N] subseteq D]]$ since $[[D_[N] subseteq [|M N|]d ]]$ by def.
  \item $[[D_[ B [N / x] ] subseteq D]]$ since $[[D_[ B [N / x] ] subseteq [|M N|]d ]]$ by def.
  \item $[[D_[B] subseteq D]]$, because $[[D_[N] subseteq [|Pi x : A . B|]d ]]$ by def., 
    $[[ [|Pi x : A . B|]d subseteq D_[M] ]]$ by $[[env |- M : Pi x : A . B]]$ and \Cref{lemma:type-label-subset}, 
    and $[[D_[M] subseteq D]]$ by def.
  \end{enumerate}

  \item[Case (\rref*{ty-Lambda})] Suppose $[[env |- \^I x : A . M : Pi x : A . B]]$.
  The goal is $[[D ; denv |- LL_[I] { dx^ } : Pi dx : dA . dB ]]$, where 
  $[[D]] = [[ (D_[env] cup D_[M]) , (LL_[I] ( {dx^ : dA^} , dx : dA --> dM : dB )) ]]$.
  We have $[[env |- x^ : A^ ]]$ since $\overline{x}$ are well-typed free variables, therefore, 
  we have $[[D ; denv |- dx^ : dA^]]$ by the inductive hypothesis and weakening.
  The definition of $[[LL_[I] ]]$ is in $[[D]]$.

  If $[[ |- D ; denv]]$, then we have our goal by \rref{d-ty-Label}.
  We have $[[|- D_[env] cup D_[M] ; denv]]$ by the induction hypothesis and our assumption that
  $[[env , x : A |- M : B]]$.
  Letting $[[denv fv]] = [[ empty , dx^ : dA^ ]]$, if the following statements are true, 
  then we have $[[ |- D ; denv]]$ by \rref{d-wf-Label} and weakening.

  \begin{enumerate}
    \item[(a)] $[[D_[env] cup D_[M] ; denv fv |- Pi dx : dA . dB : dU ]] $.
    \item[(b)] $[[D_[env] cup D_[M] ; denv fv , dx : dA |- dM : dB ]] $.
  \end{enumerate}
  
  By \Cref{lem:fv}, $[[FV \^I x : A . M |- \^I x : A . M : Pi x : A . B]]$, 
  so, $[[FV \^I x : A . M |- Pi x : A . B : U]]$, and
  $[[FV \^I x : A . M , x : A |- M : B]]$. 
  Therefore, conditions (a) and (b) are true by the induction hypothesis and weakening.

  \item[Case (\rref*{ty-Equiv}).] By the coherence lemma.
\end{description}

\end{appendixproof}

\subsection{Consistency and Type Safety of DCC}
\label{section:consistency}

\begin{figure}[ht]

  \begin{drulepar}[b]{$[[D ; denv |- dM : dA ~>b M]]$}{Backward transformation}
    \drule{Var}

    \drule{Universe}

    \drule{Pi}

    \drule{Label}

    \drule{Apply}

    \drule{Equiv}
  \end{drulepar}

  \caption{Backward transformation}
  \label{fig:backward transformation}
\end{figure}

As a dependent type theory, DCC should be type-safe when it acts as a
programming language and consistent when interpreted as a logic.
Following \citet{DBLP:conf/cpp/BoulierPT17} and \citet{DBLP:conf/pldi/BowmanA18},
we prove these properties in this section by defining a
\emph{backward transformation} from DCC to CC and showing that it
preserves reduction sequences, so that reducing an expression in DCC is
equivalent to reducing an expression in CC.
The transformation is type-preserving and turns the logical
interpretation of \emph{false} in DCC into that of CC, so that valid
proofs (i.e.~well-typed programs) in DCC correspond to valid proofs in CC.
This reduces the problem of proving the type safety and consistency of
DCC to proving that of CC, which is a standard result~\cite{DBLP:journals/iandc/CoquandH88}.
In other words, we show that DCC can be modelled by CC in a consistent
and meaning-preserving way.
Type preservation for the backward transformation also requires the
\emph{substitution}, \emph{preservation of reduction sequences},
and \emph{coherence} lemmas, similar to the proof of
\Cref{theorem:type preservation}, whose proofs are
straightforward.

We define the backward transformation $\bluebr{-}$ with a new
judgement (\Cref{fig:backward transformation}) of the form
$[[D ; denv |- dM : dA ~>b M]]$ and
$\bluebr{\target{M}} \triangleq M$.
The translation maps variables, universes, $\Pi$-types, and
applications back to their corresponding forms in CC,
and maps label expressions $[[LL { dM^} ]]$ where
$[[ LL ( {dx^ : dA^} , dx : dA --> dM : dB ) in D ]]$
into
$[[ \ x : A [ M^ / x^ ] . M [ M^ / x^ ] ]]$ --- a
function with all of its free-variable values substituted in, where
$A$, $M$, and $\overline{M}$ stand for $\bluebr{\target{A}}$,
$\bluebr{\target{M}}$, and $\bluebr{\target{\overline{M}}}$ respectively
(\rref*{b-Label}).
Intuitively, $\bluebr{-}$ decompiles a label back to the function it
represents.
The backward transformation also acts pointwise on type contexts.

In CC, the interpretation of the logical \emph{false} is $[[Pi x : U_[0] . x]]$. There is no closed expression with the \emph{false} type. In DCC, the interpretation of \emph{false} is $\targetpi{\target{x}}{\target{U_0}}{\target{x}}$, so the backward transformation preserves falseness by definition. 

Next, we show that the backward transformation is compatible with
substitutions.
As a convention in this section, we write $M$ for
$\bluebr{\target{M}}$ when there is no ambiguity.

\begin{lemmarep}[Backward transformation compatible with substitutions]\label{lemma:back-trans-compat}
 If $[[D ; denv , dx : dA |- dM : dB]]$ and $[[D ; denv |- dN : dA]]$, 
 then $\bluebr{[[dM [ dN / dx] ]]} = [[M [ N / x] ]]$.
\end{lemmarep}
\begin{appendixproof}
By induction on the type derivation of $[[dM]]$. 
\begin{description}
  \item[Case (\rref*{d-ty-Label}).] Assume that $[[D ; denv , dy : dC |- LL { dM^ } : Pi dx : dA . dB]]$, 
  $[[ LL ( {denv} , dx : dA --> dM : dB ) in D ]]$, and $[[D ; denv |- dN : dC]]$.
  The goal is to show that $\bluebr{[[(LL { dM^ }) [dN / dy] ]]} = [[(\ x : A [ M^ / x^ ] . M [ M^ / x^ ]) [N / y] ]]$.
  Indeed, we have
  \[
    \bluebr{[[(LL { dM^ }) [dN / dy] ]]} = \bluebr{[[LL { dM^ [dN / dy] } ]]}
    = [[\ x : A [ ( M^ [N / y] ) / x^ ] . M [ (M^ [N / y]) / x^ ] ]]
  \]
  by the induction hypothesis, and
  \begin{align*}
    [[(\ x : A [ M^ / x^ ] . M [ M^ / x^ ]) [N / y] ]] &= 
    [[\ x : ((A [ M^ / x^ ]) [N / y]) . ((M [ M^ / x^ ]) [N / y]) ]]\\
    &= [[\ x : A [ ( M^ [N / y] ) / x^ ] . M [ (M^ [N / y]) / x^ ] ]]
  \end{align*}
  since $y$ is not free in $A$ and in $M$ ($\overline{x}$ are all the free variables in them).
\end{description}

Ohter cases are either trivial or follows directly from the induction hypothesis.

\end{appendixproof}

Similar to proofs in Section \Cref{section:soundness}, we show preservation of reduction sequences by showing that the transformation preserves small-step reductions. 
Using that, we show the coherence lemma for the backward transformation,
and then the type preservation.
\begin{lemmarep} If $[[D ; denv |- dM : dA]]$ and $[[D |- dM |>* dN]]$, then $[[M |>* N]]$.
\label{lem:back-prev-sequence}
\end{lemmarep}
\begin{appendixproof}
We firstly show by induction that the backward transformation preserves small-step reductions. The only case is \rref{d-red-Beta}.

\begin{description}
  \item[Case (\rref*{d-red-Beta}).] Assume that $[[D ; denv |- LL { dM^ } @ dM' : dB ]]$, 
  $[[LL ( {denv} , dx : dA --> dN : dB ) in D]]$, and $[[D |- LL { dM^ } @ dN |>* dN [ dM^ / dx^ , dM' / dx ] ]]$.
  We have $\bluebr{[[LL { dM^ } @ dM']]} = [[\ x : A [ M^ / x^ ] . M [ M^ / x^ ] ]]$ by definition. Indeed, we have 
  \begin{align*}
    [[(\ x : A [ M^ / x^ ] . M [ M^ / x^ ]) M' ]] &\triangleright\
    [[(N [M^ / x^]) [M' / x] ]]\\
    &= [[N [M^ / x^ , M' / x] ]] \text{ since $x$ is not free in $\overline{M}$}\\
    &= \bluebr{[[dN [dM^ / dx^ , dM' / dx] ]]} \text{ by \Cref{lemma:back-trans-compat}.} 
  \end{align*}
\end{description}

Therefore, the backward transformation preserves reduction sequences by a trivial induction on the number of small steps in a reduction sequence.

\end{appendixproof}

\begin{lemmarep}If $[[D ; denv |- dM : dA]]$, $[[D ; denv |- dN : dA]]$, and $[[D |- dM == dN]]$, 
then $[[|- M == N]]$.
\label{lemma:back-trans-coherence}
\end{lemmarep}
\begin{appendixproof}
By induction on DCC's equivalence rules. 
\begin{description}
  \item[Case (\rref*{d-eq-Reduce}).] By \Cref{lem:back-prev-sequence}.
  \item[Case (\rref*{d-eq-Eta1}).] Assume that we have 
  \[
    \drule{d-eq-EtaOne}. 
  \]
  Our goal is to show that $[[|- L == M]]$. 
  By \Cref{lem:back-prev-sequence}, 
  $[[L |>* (\ x : A [N^ / x^] . N [N^ / x^]) ]]$ and $[[M |>* M']]$.
  By the induction hypothesis and \Cref{lemma:back-trans-compat},
  $[[|- N' [N^ / x^] == M' x ]]$.
  Therefore, $[[|- L == M]]$ by CC's \rref{eq-Eta1}.
  \item[Case (\rref*{d-eq-Eta2}).] By symmetry.
\end{description}

\end{appendixproof}

\begin{lemmarep} If $[[D ; denv |- dM : dA]]$, then $[[env |- M : A]]$.
\label{lemma:back-type-preservation}
\end{lemmarep}
\begin{appendixproof}
By proving the following two statements together with a simultaneous induction on mutually-defined judgements 
$[[|- D ; denv]]$ and $[[D ; denv |- dM : dA]]$.
\begin{enumerate}
	\item $[[|- D ; denv]] \Longrightarrow\ [[|- env]]$.
	\item $[[D ; denv |- dM : dA]] \Longrightarrow\ [[env |- M : A]]$.
\end{enumerate}
Statement $1$ follows trivially from the inductive hypothesis. We look at the cases for statement $2$.
\begin{description}
  \item [Case (\rref*{d-ty-Var, d-ty-Universe, d-ty-Pi}).] Trivial by induction.
  \item [Case (\rref*{d-ty-Apply}).] Assume that $[[D ; denv |- dM @ dN : dB [dN / dx] ]]$, the goal is to show that
  $\Gamma \vdash M N : \bluebr{[[dB [dN / dx] ]]}$. By the induction hypothesis, we have $[[env |- M : Pi x : A . B]]$
  and $[[env |- N : A]]$, which implies that $[[env |- M N : B [N / x] ]]$. By \Cref{lemma:back-trans-compat},
  $\bluebr{[[dB [dN / dx] ]]} \equiv [[B [N / x] ]]$. Therefore, we have $\Gamma \vdash M N : \bluebr{[[dB [dN / dx] ]]}$
  by \rref{Equiv}.
  \item [Case (\rref*{d-ty-Label}).] Similar.
  \item [Case (\rref*{d-ty-Equiv}).] By \Cref{lemma:back-trans-coherence}
\end{description}

\end{appendixproof}

As a corollary of \Cref{lem:back-prev-sequence} and \Cref{lemma:back-type-preservation}, 
DCC is type-safe and consistent since CC is.

\begin{theoremrep}[Type safety] If $[[D ; empty |- dM : dA]]$, then $[[ D |- dM |>* dv ]]$ 
for some irreducible value $[[dv]]$.
\label{theorem:dcc-type-safe}
\end{theoremrep}

That is, type safety guarantees that every well-typed closed DCC term reduces
to a value in a finite number of steps.

\begin{theoremrep}[Consistency] There is no pair of a label context $[[D]]$ and DCC term $[[dM]]$ such that $[[D ; empty |- dM : (Pi dxA : dU . dA)]]$.
\label{theorem:dcc-consistent}
\end{theoremrep}

Interpreting DCC as a logic, the term $[[Pi dxA : dU . dA]]$ (which
produces a term of any type $[[dA]]$) corresponds to \emph{false}; it
is backward-transformed to $\Pi A {:} U . A$, the representation of
\emph{false} in CC.  Consistency of DCC means that there is no closed
term of type $[[Pi dxA : dU . dA]]$; if there were, then translation
would yield a corresponding term in CC, and CC would also be
inconsistent.

\section{Implementation}
\label{section:implementation}

We provide a portable standalone implementation of the
defunctionalization translation of
\Cref{section:cc-defunctionalization}, written in OCaml and compiled
to run in a web browser using
\verb!js_of_ocaml!~\cite{DBLP:journals/spe/VouillonB14}.
The implementation performs
type checking of CC (\Cref{section:cc})
and DCC (\Cref{section:dcc}) terms,
abstract defunctionalization (\Cref{section:translation}) and
backwards translation from DCC to CC (\Cref{section:consistency}),
allowing the interested reader to experiment with the effects of the translation on real examples.
We include several ready-made examples, including dependent
composition, dependent pairs and finite sets.

\section{Related Work}
\label{section:related}

\paragraph{Type-preserving compilation}

Type-preserving compilation was initially developed for optimizing
compilation and verifying the compiled code; it has been used
extensively in compilers of simply-typed and polymorphic languages,
and occasionally for dependently-typed languages.
For example, \citet{DBLP:conf/pldi/TarditiMCSHL96} present TIL (typed
intermediate language), an ML compiler featuring type-directed code
optimization of loops, garbage collections, and polymorphic function
calls, and \citet{DBLP:journals/toplas/MorrisettWCG99} study a
type-preserving translation from System F to the typed assembly
language TAL.
\citet{DBLP:conf/icfp/XiH01} later extended TAL to DTAL, an assembly
language with a limited form of dependent types that serves as a
compilation target for Dependent ML.

\citet{DBLP:conf/icfp/GuillemetteM08} also present a type-preserving
compiler from System F, but take a different approach, building typed
intermediate representations using generalized algebraic data types
and using the type system of the host language (GHC Haskell) to verify
that each compiler phase preserves types.
Embedding typed transformations in this way is a popular technique in
the functional programming community, exemplified in work by
\citet{DBLP:journals/jfp/CaretteKS09}, which presents type-preserving
CPS transformations of an embedded language along with type-preserving optimizations
based on partial evaluation.

\citet{DBLP:conf/popl/Necula97}'s proof-carrying code is another early
method for generating reliable executables. It relies on an external
logical framework to check the correctness of proofs attached with the
code.

Bowman and collaborators have developed several type-preserving
translations for dependently-typed languages, including
CPS
transformation~\cite{DBLP:journals/pacmpl/BowmanCRA18},
closure
conversion~\cite{DBLP:conf/pldi/BowmanA18}
(building on typed closure conversion for System F
by~\citet{DBLP:conf/popl/MinamideMH96}), and 
translation to ANF~\cite{koronkevich_rakow_ahmed_bowman_2022}.

\paragraph{Defunctionalization}

Defunctionalization was first presented by
\citet{DBLP:conf/acm/Reynolds72} as a programming technique to
translate a higher-order interpreter into a first-order
one~\cite{DBLP:conf/acm/Reynolds72}.
It has been used in a variety of applications,
from ML
compilers~\cite{DBLP:journals/lisp/ChinD96,DBLP:conf/esop/CejtinJW00},
to type-safe garbage collectors~\cite{DBLP:conf/popl/WangA01}, and
encodings of higher-kinded
polymorphism~\cite{DBLP:conf/flops/YallopW14}.

Defunctionalization was originally presented as an untyped
translation.
Using a family
of monomorphic \emph{apply} functions to make simply-typed defunctionalization type-preserving is
a standard workaround  in the
literature~\cite{DBLP:conf/icfp/BellBH97,DBLP:journals/jfp/TolmachO98,DBLP:conf/esop/CejtinJW00,nielsen2000denotational}.

\citet{DBLP:conf/ppdp/DanvyN01} survey more examples of
defunctionalization in practice.

Formalization of defunctionalization has up to this point focused on
proving type preservation and correctness of the transformation.
\citet{DBLP:conf/icfp/BellBH97}~have shown that the translation for
simply typed programs is type preserving.
\citet{nielsen2000denotational} has proved its partial correctness with
denotational semantics, and \citet{DBLP:conf/tacs/BanerjeeHR01} have
established total correctness using operational
semantics. \citet{DBLP:conf/popl/PottierG04} have formalized
type-preserving polymorphic defunctionalization in System F extended
with GADTs.

\paragraph{Closure conversion}

Like defunctionalization, \emph{closure conversion} transformations also
involve representing a closure as a first-order value that pairs a
kind of code identifier with a collection of free variables.
The formulations of closure conversion in the work referenced
above~\cite{DBLP:conf/popl/MinamideMH96,DBLP:conf/pldi/BowmanA18}
differ markedly from defunctionalization: while defunctionalization
involves a globally-defined map indexed by code identifiers (such as
an \emph{apply} function or our label environment), these closure
conversions instead locally transform functions into
code-and-environment pairs that can then be applied using a standard
elimination rule.
%
However, other formulations of closure
conversion~\cite[e.g.][]{DBLP:books/cu/Appel1992,essence-closure-conversion}
additionally lift functions to top-level, making the transformation
more similar to defunctionalization.

Closure conversion plays a key role in compilers for many functional
languages, including Scheme~\cite{rabbit},
CAML~\cite{DBLP:conf/lfp/MaunyS86},
Standard ML~\cite{DBLP:conf/esop/CejtinJW00},
Haskell~\cite{DBLP:conf/iccS/LeshchinskiyCK06} and others.
Recent work has focused on establishing sophisticated semantic properties, such
as correctness of closure conversion in the presence of mutable state
and control effects (even when linked with foreign-language
code)~\cite{DBLP:conf/ppdp/MatesPA19}, and preservation of time and
space properties~\cite{DBLP:journals/pacmpl/Paraskevopoulou19}.

\paragraph{Refunctionalization}

Our backward translation (See \Cref{section:consistency}) 
is related to \emph{refunctionalization} \cite{DBLP:journals/scp/DanvyM09}, 
the left-inverse of defunctionalization. 
As in refunctionalization,
we replace target applications $[[dM @ dN]]$
with source applications $[[M N]]$,
and labels $[[LL{dM^}]]$ with abstractions $[[(\ x : A . M) [ M^ / x^ ] ]]$
based on their implementations $[[LL ( { dx^ : dA^ } , dx : dA --> dM : dB )]]$ in the label context.

\begin{acks}
We thank David Sheets, Andr{\'a}s Kov{\'a}cs, and Marcelo Fiore for helpful comments.
\end{acks}

\bibliography{defun}

\begin{thebibliography}{}

\end{thebibliography}



\begin{thebibliography}{54}


\ifx \showCODEN    \undefined \def \showCODEN     #1{\unskip}     \fi
\ifx \showDOI      \undefined \def \showDOI       #1{#1}\fi
\ifx \showISBNx    \undefined \def \showISBNx     #1{\unskip}     \fi
\ifx \showISBNxiii \undefined \def \showISBNxiii  #1{\unskip}     \fi
\ifx \showISSN     \undefined \def \showISSN      #1{\unskip}     \fi
\ifx \showLCCN     \undefined \def \showLCCN      #1{\unskip}     \fi
\ifx \shownote     \undefined \def \shownote      #1{#1}          \fi
\ifx \showarticletitle \undefined \def \showarticletitle #1{#1}   \fi
\ifx \showURL      \undefined \def \showURL       {\relax}        \fi
\providecommand\bibfield[2]{#2}
\providecommand\bibinfo[2]{#2}
\providecommand\natexlab[1]{#1}
\providecommand\showeprint[2][]{arXiv:#2}

\bibitem[Abadi et~al\mbox{.}(1991)]%
        {DBLP:journals/jfp/AbadiCCL91}
\bibfield{author}{\bibinfo{person}{Mart{\'{\i}}n Abadi}, \bibinfo{person}{Luca
  Cardelli}, \bibinfo{person}{Pierre{-}Louis Curien}, {and}
  \bibinfo{person}{Jean{-}Jacques L{\'{e}}vy}.}
  \bibinfo{year}{1991}\natexlab{}.
\newblock \showarticletitle{Explicit Substitutions}.
\newblock \bibinfo{journal}{\emph{J. Funct. Program.}} \bibinfo{volume}{1},
  \bibinfo{number}{4} (\bibinfo{year}{1991}), \bibinfo{pages}{375--416}.
\newblock
\urldef\tempurl%
\url{https://doi.org/10.1017/S0956796800000186}
\showDOI{\tempurl}


\bibitem[Ahrens et~al\mbox{.}(2018)]%
        {DBLP:journals/lmcs/AhrensLV18}
\bibfield{author}{\bibinfo{person}{Benedikt Ahrens},
  \bibinfo{person}{Peter~LeFanu Lumsdaine}, {and} \bibinfo{person}{Vladimir
  Voevodsky}.} \bibinfo{year}{2018}\natexlab{}.
\newblock \showarticletitle{Categorical structures for type theory in univalent
  foundations}.
\newblock \bibinfo{journal}{\emph{Log. Methods Comput. Sci.}}
  \bibinfo{volume}{14}, \bibinfo{number}{3} (\bibinfo{year}{2018}).
\newblock
\urldef\tempurl%
\url{https://doi.org/10.23638/LMCS-14(3:18)2018}
\showDOI{\tempurl}


\bibitem[Appel(1992)]%
        {DBLP:books/cu/Appel1992}
\bibfield{author}{\bibinfo{person}{Andrew~W. Appel}.}
  \bibinfo{year}{1992}\natexlab{}.
\newblock \bibinfo{booktitle}{\emph{Compiling with Continuations}}.
\newblock \bibinfo{publisher}{Cambridge University Press}.
\newblock
\showISBNx{0-521-41695-7}


\bibitem[Banerjee et~al\mbox{.}(2001)]%
        {DBLP:conf/tacs/BanerjeeHR01}
\bibfield{author}{\bibinfo{person}{Anindya Banerjee}, \bibinfo{person}{Nevin
  Heintze}, {and} \bibinfo{person}{Jon~G. Riecke}.}
  \bibinfo{year}{2001}\natexlab{}.
\newblock \showarticletitle{Design and Correctness of Program Transformations
  Based on Control-Flow Analysis}. In \bibinfo{booktitle}{\emph{Theoretical
  Aspects of Computer Software, 4th International Symposium, {TACS} 2001,
  Sendai, Japan, October 29-31, 2001, Proceedings}}
  \emph{(\bibinfo{series}{Lecture Notes in Computer Science},
  Vol.~\bibinfo{volume}{2215})}, \bibfield{editor}{\bibinfo{person}{Naoki
  Kobayashi} {and} \bibinfo{person}{Benjamin~C. Pierce}} (Eds.).
  \bibinfo{publisher}{Springer}, \bibinfo{pages}{420--447}.
\newblock
\urldef\tempurl%
\url{https://doi.org/10.1007/3-540-45500-0\_21}
\showDOI{\tempurl}


\bibitem[Barthe et~al\mbox{.}(1999)]%
        {DBLP:journals/lisp/BartheHS99}
\bibfield{author}{\bibinfo{person}{Gilles Barthe}, \bibinfo{person}{John
  Hatcliff}, {and} \bibinfo{person}{Morten~Heine S{\o}rensen}.}
  \bibinfo{year}{1999}\natexlab{}.
\newblock \showarticletitle{{CPS} Translations and Applications: The Cube and
  Beyond}.
\newblock \bibinfo{journal}{\emph{High. Order Symb. Comput.}}
  \bibinfo{volume}{12}, \bibinfo{number}{2} (\bibinfo{year}{1999}),
  \bibinfo{pages}{125--170}.
\newblock
\urldef\tempurl%
\url{https://doi.org/10.1023/A:1010000206149}
\showDOI{\tempurl}


\bibitem[Barthe and Uustalu(2002)]%
        {DBLP:conf/pepm/BartheU02}
\bibfield{author}{\bibinfo{person}{Gilles Barthe} {and} \bibinfo{person}{Tarmo
  Uustalu}.} \bibinfo{year}{2002}\natexlab{}.
\newblock \showarticletitle{{CPS} translating inductive and coinductive types}.
  In \bibinfo{booktitle}{\emph{Proceedings of the 2002 {ACM} {SIGPLAN} Workshop
  on Partial Evaluation and Semantics-Based Program Manipulation {(PEPM} '02),
  Portland, Oregon, USA, January 14-15, 2002}},
  \bibfield{editor}{\bibinfo{person}{Peter Thiemann}} (Ed.).
  \bibinfo{publisher}{{ACM}}, \bibinfo{pages}{131--142}.
\newblock
\urldef\tempurl%
\url{https://doi.org/10.1145/503032.503043}
\showDOI{\tempurl}


\bibitem[Bell et~al\mbox{.}(1997)]%
        {DBLP:conf/icfp/BellBH97}
\bibfield{author}{\bibinfo{person}{Jeffrey~M. Bell},
  \bibinfo{person}{Fran{\c{c}}oise Bellegarde}, {and} \bibinfo{person}{James
  Hook}.} \bibinfo{year}{1997}\natexlab{}.
\newblock \showarticletitle{Type-Driven Defunctionalization}. In
  \bibinfo{booktitle}{\emph{Proceedings of the 1997 {ACM} {SIGPLAN}
  International Conference on Functional Programming {(ICFP} '97), Amsterdam,
  The Netherlands, June 9-11, 1997}}, \bibfield{editor}{\bibinfo{person}{Simon
  L.~Peyton Jones}, \bibinfo{person}{Mads Tofte}, {and}
  \bibinfo{person}{A.~Michael Berman}} (Eds.). \bibinfo{publisher}{{ACM}},
  \bibinfo{pages}{25--37}.
\newblock
\urldef\tempurl%
\url{https://doi.org/10.1145/258948.258953}
\showDOI{\tempurl}


\bibitem[Boulier et~al\mbox{.}(2017)]%
        {DBLP:conf/cpp/BoulierPT17}
\bibfield{author}{\bibinfo{person}{Simon Boulier},
  \bibinfo{person}{Pierre{-}Marie P{\'{e}}drot}, {and} \bibinfo{person}{Nicolas
  Tabareau}.} \bibinfo{year}{2017}\natexlab{}.
\newblock \showarticletitle{The next 700 syntactical models of type theory}. In
  \bibinfo{booktitle}{\emph{Proceedings of the 6th {ACM} {SIGPLAN} Conference
  on Certified Programs and Proofs, {CPP} 2017, Paris, France, January 16-17,
  2017}}, \bibfield{editor}{\bibinfo{person}{Yves Bertot} {and}
  \bibinfo{person}{Viktor Vafeiadis}} (Eds.). \bibinfo{publisher}{{ACM}},
  \bibinfo{pages}{182--194}.
\newblock
\urldef\tempurl%
\url{https://doi.org/10.1145/3018610.3018620}
\showDOI{\tempurl}


\bibitem[Bove et~al\mbox{.}(2009)]%
        {DBLP:conf/tphol/BoveDN09}
\bibfield{author}{\bibinfo{person}{Ana Bove}, \bibinfo{person}{Peter Dybjer},
  {and} \bibinfo{person}{Ulf Norell}.} \bibinfo{year}{2009}\natexlab{}.
\newblock \showarticletitle{A Brief Overview of Agda - {A} Functional Language
  with Dependent Types}. In \bibinfo{booktitle}{\emph{Theorem Proving in Higher
  Order Logics, 22nd International Conference, TPHOLs 2009, Munich, Germany,
  August 17-20, 2009. Proceedings}} \emph{(\bibinfo{series}{Lecture Notes in
  Computer Science}, Vol.~\bibinfo{volume}{5674})},
  \bibfield{editor}{\bibinfo{person}{Stefan Berghofer}, \bibinfo{person}{Tobias
  Nipkow}, \bibinfo{person}{Christian Urban}, {and} \bibinfo{person}{Makarius
  Wenzel}} (Eds.). \bibinfo{publisher}{Springer}, \bibinfo{pages}{73--78}.
\newblock
\urldef\tempurl%
\url{https://doi.org/10.1007/978-3-642-03359-9\_6}
\showDOI{\tempurl}


\bibitem[Bowman and Ahmed(2018)]%
        {DBLP:conf/pldi/BowmanA18}
\bibfield{author}{\bibinfo{person}{William~J. Bowman} {and}
  \bibinfo{person}{Amal Ahmed}.} \bibinfo{year}{2018}\natexlab{}.
\newblock \showarticletitle{Typed closure conversion for the calculus of
  constructions}. In \bibinfo{booktitle}{\emph{Proceedings of the 39th {ACM}
  {SIGPLAN} Conference on Programming Language Design and Implementation,
  {PLDI} 2018, Philadelphia, PA, USA, June 18-22, 2018}},
  \bibfield{editor}{\bibinfo{person}{Jeffrey~S. Foster} {and}
  \bibinfo{person}{Dan Grossman}} (Eds.). \bibinfo{publisher}{{ACM}},
  \bibinfo{pages}{797--811}.
\newblock
\urldef\tempurl%
\url{https://doi.org/10.1145/3192366.3192372}
\showDOI{\tempurl}


\bibitem[Bowman et~al\mbox{.}(2018)]%
        {DBLP:journals/pacmpl/BowmanCRA18}
\bibfield{author}{\bibinfo{person}{William~J. Bowman}, \bibinfo{person}{Youyou
  Cong}, \bibinfo{person}{Nick Rioux}, {and} \bibinfo{person}{Amal Ahmed}.}
  \bibinfo{year}{2018}\natexlab{}.
\newblock \showarticletitle{Type-preserving {CPS} translation of {\(\Sigma\)}
  and {\(\Pi\)} types is not not possible}.
\newblock \bibinfo{journal}{\emph{Proc. {ACM} Program. Lang.}}
  \bibinfo{volume}{2}, \bibinfo{number}{{POPL}} (\bibinfo{year}{2018}),
  \bibinfo{pages}{22:1--22:33}.
\newblock
\urldef\tempurl%
\url{https://doi.org/10.1145/3158110}
\showDOI{\tempurl}


\bibitem[Brady(2013)]%
        {DBLP:journals/jfp/Brady13}
\bibfield{author}{\bibinfo{person}{Edwin~C. Brady}.}
  \bibinfo{year}{2013}\natexlab{}.
\newblock \showarticletitle{Idris, a general-purpose dependently typed
  programming language: Design and implementation}.
\newblock \bibinfo{journal}{\emph{J. Funct. Program.}} \bibinfo{volume}{23},
  \bibinfo{number}{5} (\bibinfo{year}{2013}), \bibinfo{pages}{552--593}.
\newblock
\urldef\tempurl%
\url{https://doi.org/10.1017/S095679681300018X}
\showDOI{\tempurl}


\bibitem[Bra{\ss}el(2011)]%
        {curry-implementation}
\bibfield{author}{\bibinfo{person}{Bernd Bra{\ss}el}.}
  \bibinfo{year}{2011}\natexlab{}.
\newblock \emph{\bibinfo{title}{Implementing Functional Logic Programs by
  Translation into Purely Functional Programs}}.
\newblock \bibinfo{thesistype}{Ph.\,D. Dissertation}.
\newblock
\urldef\tempurl%
\url{https://macau.uni-kiel.de/receive/diss_mods_00007056}
\showURL{%
\tempurl}


\bibitem[Carette et~al\mbox{.}(2009)]%
        {DBLP:journals/jfp/CaretteKS09}
\bibfield{author}{\bibinfo{person}{Jacques Carette}, \bibinfo{person}{Oleg
  Kiselyov}, {and} \bibinfo{person}{Chung{-}chieh Shan}.}
  \bibinfo{year}{2009}\natexlab{}.
\newblock \showarticletitle{Finally tagless, partially evaluated: Tagless
  staged interpreters for simpler typed languages}.
\newblock \bibinfo{journal}{\emph{J. Funct. Program.}} \bibinfo{volume}{19},
  \bibinfo{number}{5} (\bibinfo{year}{2009}), \bibinfo{pages}{509--543}.
\newblock
\urldef\tempurl%
\url{https://doi.org/10.1017/S0956796809007205}
\showDOI{\tempurl}


\bibitem[Cejtin et~al\mbox{.}(2000)]%
        {DBLP:conf/esop/CejtinJW00}
\bibfield{author}{\bibinfo{person}{Henry Cejtin}, \bibinfo{person}{Suresh
  Jagannathan}, {and} \bibinfo{person}{Stephen Weeks}.}
  \bibinfo{year}{2000}\natexlab{}.
\newblock \showarticletitle{Flow-Directed Closure Conversion for Typed
  Languages}. In \bibinfo{booktitle}{\emph{Programming Languages and Systems,
  9th European Symposium on Programming, {ESOP} 2000, Held as Part of the
  European Joint Conferences on the Theory and Practice of Software, {ETAPS}
  2000, Berlin, Germany, March 25 - April 2, 2000, Proceedings}}
  \emph{(\bibinfo{series}{Lecture Notes in Computer Science},
  Vol.~\bibinfo{volume}{1782})}, \bibfield{editor}{\bibinfo{person}{Gert
  Smolka}} (Ed.). \bibinfo{publisher}{Springer}, \bibinfo{pages}{56--71}.
\newblock
\urldef\tempurl%
\url{https://doi.org/10.1007/3-540-46425-5\_4}
\showDOI{\tempurl}


\bibitem[Chin and Darlington(1996)]%
        {DBLP:journals/lisp/ChinD96}
\bibfield{author}{\bibinfo{person}{Wei{-}Ngan Chin} {and} \bibinfo{person}{John
  Darlington}.} \bibinfo{year}{1996}\natexlab{}.
\newblock \showarticletitle{A Higher-Order Removal Method}.
\newblock \bibinfo{journal}{\emph{{LISP} Symb. Comput.}} \bibinfo{volume}{9},
  \bibinfo{number}{4} (\bibinfo{year}{1996}), \bibinfo{pages}{287--322}.
\newblock


\bibitem[{Coq} Development~Team(2022)]%
        {CoqDoc}
\bibfield{author}{\bibinfo{person}{The {Coq} Development~Team}.}
  \bibinfo{year}{2022}\natexlab{}.
\newblock \bibinfo{title}{The {Coq} Reference Manual}.
\newblock
  \bibinfo{howpublished}{\url{https://coq.inria.fr/distrib/current/refman/}}.
\newblock


\bibitem[Coquand and Huet(1988)]%
        {DBLP:journals/iandc/CoquandH88}
\bibfield{author}{\bibinfo{person}{Thierry Coquand} {and}
  \bibinfo{person}{G{\'{e}}rard~P. Huet}.} \bibinfo{year}{1988}\natexlab{}.
\newblock \showarticletitle{The Calculus of Constructions}.
\newblock \bibinfo{journal}{\emph{Inf. Comput.}} \bibinfo{volume}{76},
  \bibinfo{number}{2/3} (\bibinfo{year}{1988}), \bibinfo{pages}{95--120}.
\newblock
\urldef\tempurl%
\url{https://doi.org/10.1016/0890-5401(88)90005-3}
\showDOI{\tempurl}


\bibitem[Danvy and Millikin(2009)]%
        {DBLP:journals/scp/DanvyM09}
\bibfield{author}{\bibinfo{person}{Olivier Danvy} {and} \bibinfo{person}{Kevin
  Millikin}.} \bibinfo{year}{2009}\natexlab{}.
\newblock \showarticletitle{Refunctionalization at work}.
\newblock \bibinfo{journal}{\emph{Sci. Comput. Program.}} \bibinfo{volume}{74},
  \bibinfo{number}{8} (\bibinfo{year}{2009}), \bibinfo{pages}{534--549}.
\newblock
\urldef\tempurl%
\url{https://doi.org/10.1016/j.scico.2007.10.007}
\showDOI{\tempurl}


\bibitem[Danvy and Nielsen(2001)]%
        {DBLP:conf/ppdp/DanvyN01}
\bibfield{author}{\bibinfo{person}{Olivier Danvy} {and}
  \bibinfo{person}{Lasse~R. Nielsen}.} \bibinfo{year}{2001}\natexlab{}.
\newblock \showarticletitle{Defunctionalization at Work}. In
  \bibinfo{booktitle}{\emph{Proceedings of the 3rd international {ACM}
  {SIGPLAN} conference on Principles and practice of declarative programming,
  September 5-7, 2001, Florence, Italy}}. \bibinfo{publisher}{{ACM}},
  \bibinfo{pages}{162--174}.
\newblock
\urldef\tempurl%
\url{https://doi.org/10.1145/773184.773202}
\showDOI{\tempurl}


\bibitem[de~Moura et~al\mbox{.}(2015)]%
        {DBLP:conf/cade/MouraKADR15}
\bibfield{author}{\bibinfo{person}{Leonardo~Mendon{\c{c}}a de Moura},
  \bibinfo{person}{Soonho Kong}, \bibinfo{person}{Jeremy Avigad},
  \bibinfo{person}{Floris van Doorn}, {and} \bibinfo{person}{Jakob von
  Raumer}.} \bibinfo{year}{2015}\natexlab{}.
\newblock \showarticletitle{The Lean Theorem Prover (System Description)}. In
  \bibinfo{booktitle}{\emph{Automated Deduction - {CADE-25} - 25th
  International Conference on Automated Deduction, Berlin, Germany, August 1-7,
  2015, Proceedings}} \emph{(\bibinfo{series}{Lecture Notes in Computer
  Science}, Vol.~\bibinfo{volume}{9195})},
  \bibfield{editor}{\bibinfo{person}{Amy~P. Felty} {and} \bibinfo{person}{Aart
  Middeldorp}} (Eds.). \bibinfo{publisher}{Springer},
  \bibinfo{pages}{378--388}.
\newblock
\urldef\tempurl%
\url{https://doi.org/10.1007/978-3-319-21401-6\_26}
\showDOI{\tempurl}


\bibitem[Dybjer(1994)]%
        {DBLP:journals/fac/Dybjer94}
\bibfield{author}{\bibinfo{person}{Peter Dybjer}.}
  \bibinfo{year}{1994}\natexlab{}.
\newblock \showarticletitle{Inductive Families}.
\newblock \bibinfo{journal}{\emph{Formal Aspects Comput.}} \bibinfo{volume}{6},
  \bibinfo{number}{4} (\bibinfo{year}{1994}), \bibinfo{pages}{440--465}.
\newblock
\urldef\tempurl%
\url{https://doi.org/10.1007/BF01211308}
\showDOI{\tempurl}


\bibitem[Guillemette and Monnier(2008)]%
        {DBLP:conf/icfp/GuillemetteM08}
\bibfield{author}{\bibinfo{person}{Louis{-}Julien Guillemette} {and}
  \bibinfo{person}{Stefan Monnier}.} \bibinfo{year}{2008}\natexlab{}.
\newblock \showarticletitle{A type-preserving compiler in Haskell}. In
  \bibinfo{booktitle}{\emph{Proceeding of the 13th {ACM} {SIGPLAN}
  international conference on Functional programming, {ICFP} 2008, Victoria,
  BC, Canada, September 20-28, 2008}}, \bibfield{editor}{\bibinfo{person}{James
  Hook} {and} \bibinfo{person}{Peter Thiemann}} (Eds.).
  \bibinfo{publisher}{{ACM}}, \bibinfo{pages}{75--86}.
\newblock
\urldef\tempurl%
\url{https://doi.org/10.1145/1411204.1411218}
\showDOI{\tempurl}


\bibitem[Koronkevich et~al\mbox{.}(2022)]%
        {koronkevich_rakow_ahmed_bowman_2022}
\bibfield{author}{\bibinfo{person}{Paulette Koronkevich},
  \bibinfo{person}{Ramon Rakow}, \bibinfo{person}{Amal Ahmed}, {and}
  \bibinfo{person}{William~J. Bowman}.} \bibinfo{year}{2022}\natexlab{}.
\newblock \showarticletitle{ANF preserves dependent types up to extensional
  equality}.
\newblock \bibinfo{journal}{\emph{Journal of Functional Programming}}
  \bibinfo{volume}{32} (\bibinfo{year}{2022}), \bibinfo{pages}{e12}.
\newblock
\urldef\tempurl%
\url{https://doi.org/10.1017/S0956796822000090}
\showDOI{\tempurl}


\bibitem[Kov{\'{a}}cs(2020)]%
        {DBLP:journals/pacmpl/Kovacs20}
\bibfield{author}{\bibinfo{person}{Andr{\'{a}}s Kov{\'{a}}cs}.}
  \bibinfo{year}{2020}\natexlab{}.
\newblock \showarticletitle{Elaboration with first-class implicit function
  types}.
\newblock \bibinfo{journal}{\emph{Proc. {ACM} Program. Lang.}}
  \bibinfo{volume}{4}, \bibinfo{number}{{ICFP}} (\bibinfo{year}{2020}),
  \bibinfo{pages}{101:1--101:29}.
\newblock
\urldef\tempurl%
\url{https://doi.org/10.1145/3408983}
\showDOI{\tempurl}


\bibitem[Leshchinskiy et~al\mbox{.}(2006)]%
        {DBLP:conf/iccS/LeshchinskiyCK06}
\bibfield{author}{\bibinfo{person}{Roman Leshchinskiy}, \bibinfo{person}{Manuel
  M.~T. Chakravarty}, {and} \bibinfo{person}{Gabriele Keller}.}
  \bibinfo{year}{2006}\natexlab{}.
\newblock \showarticletitle{Higher Order Flattening}. In
  \bibinfo{booktitle}{\emph{Computational Science - {ICCS} 2006, 6th
  International Conference, Reading, UK, May 28-31, 2006, Proceedings, Part
  {II}}} \emph{(\bibinfo{series}{Lecture Notes in Computer Science},
  Vol.~\bibinfo{volume}{3992})}, \bibfield{editor}{\bibinfo{person}{Vassil~N.
  Alexandrov}, \bibinfo{person}{G.~Dick van Albada}, \bibinfo{person}{Peter
  M.~A. Sloot}, {and} \bibinfo{person}{Jack~J. Dongarra}} (Eds.).
  \bibinfo{publisher}{Springer}, \bibinfo{pages}{920--928}.
\newblock
\urldef\tempurl%
\url{https://doi.org/10.1007/11758525\_122}
\showDOI{\tempurl}


\bibitem[Luo(1990)]%
        {DBLP:phd/ethos/Luo90}
\bibfield{author}{\bibinfo{person}{Zhaohui Luo}.}
  \bibinfo{year}{1990}\natexlab{}.
\newblock \emph{\bibinfo{title}{An extended calculus of constructions}}.
\newblock \bibinfo{thesistype}{Ph.\,D. Dissertation}.
  \bibinfo{school}{University of Edinburgh, {UK}}.
\newblock
\urldef\tempurl%
\url{https://hdl.handle.net/1842/12487}
\showURL{%
\tempurl}


\bibitem[Mates et~al\mbox{.}(2019)]%
        {DBLP:conf/ppdp/MatesPA19}
\bibfield{author}{\bibinfo{person}{Phillip Mates}, \bibinfo{person}{Jamie
  Perconti}, {and} \bibinfo{person}{Amal Ahmed}.}
  \bibinfo{year}{2019}\natexlab{}.
\newblock \showarticletitle{Under Control: Compositionally Correct Closure
  Conversion with Mutable State}. In \bibinfo{booktitle}{\emph{Proceedings of
  the 21st International Symposium on Principles and Practice of Programming
  Languages, {PPDP} 2019, Porto, Portugal, October 7-9, 2019}},
  \bibfield{editor}{\bibinfo{person}{Ekaterina Komendantskaya}} (Ed.).
  \bibinfo{publisher}{{ACM}}, \bibinfo{pages}{16:1--16:15}.
\newblock
\urldef\tempurl%
\url{https://doi.org/10.1145/3354166.3354181}
\showDOI{\tempurl}


\bibitem[Mauny and Su{\'{a}}rez(1986)]%
        {DBLP:conf/lfp/MaunyS86}
\bibfield{author}{\bibinfo{person}{Michel Mauny} {and}
  \bibinfo{person}{Asc{\'{a}}nder Su{\'{a}}rez}.}
  \bibinfo{year}{1986}\natexlab{}.
\newblock \showarticletitle{Implementing Functional Languages in the
  Categorical Abstract Machine}. In \bibinfo{booktitle}{\emph{Proceedings of
  the 1986 {ACM} Conference on {LISP} and Functional Programming, {LFP} 1986,
  Cambridge, Massachusetts, USA, August 4-6, 1986}},
  \bibfield{editor}{\bibinfo{person}{William~L. Scherlis},
  \bibinfo{person}{John~H. Williams}, {and} \bibinfo{person}{Richard~P.
  Gabriel}} (Eds.). \bibinfo{publisher}{{ACM}}, \bibinfo{pages}{266--278}.
\newblock
\urldef\tempurl%
\url{https://doi.org/10.1145/319838.319869}
\showDOI{\tempurl}


\bibitem[Minamide et~al\mbox{.}(1996)]%
        {DBLP:conf/popl/MinamideMH96}
\bibfield{author}{\bibinfo{person}{Yasuhiko Minamide},
  \bibinfo{person}{J.~Gregory Morrisett}, {and} \bibinfo{person}{Robert
  Harper}.} \bibinfo{year}{1996}\natexlab{}.
\newblock \showarticletitle{Typed Closure Conversion}. In
  \bibinfo{booktitle}{\emph{Conference Record of POPL'96: The 23rd {ACM}
  {SIGPLAN-SIGACT} Symposium on Principles of Programming Languages, Papers
  Presented at the Symposium, St. Petersburg Beach, Florida, USA, January
  21-24, 1996}}, \bibfield{editor}{\bibinfo{person}{Hans{-}Juergen Boehm} {and}
  \bibinfo{person}{Guy L.~Steele Jr.}} (Eds.). \bibinfo{publisher}{{ACM}
  Press}, \bibinfo{pages}{271--283}.
\newblock
\urldef\tempurl%
\url{https://doi.org/10.1145/237721.237791}
\showDOI{\tempurl}


\bibitem[Morrisett et~al\mbox{.}(1999)]%
        {DBLP:journals/toplas/MorrisettWCG99}
\bibfield{author}{\bibinfo{person}{J.~Gregory Morrisett},
  \bibinfo{person}{David Walker}, \bibinfo{person}{Karl Crary}, {and}
  \bibinfo{person}{Neal Glew}.} \bibinfo{year}{1999}\natexlab{}.
\newblock \showarticletitle{From system {F} to typed assembly language}.
\newblock \bibinfo{journal}{\emph{{ACM} Trans. Program. Lang. Syst.}}
  \bibinfo{volume}{21}, \bibinfo{number}{3} (\bibinfo{year}{1999}),
  \bibinfo{pages}{527--568}.
\newblock
\urldef\tempurl%
\url{https://doi.org/10.1145/319301.319345}
\showDOI{\tempurl}


\bibitem[Necula(1997)]%
        {DBLP:conf/popl/Necula97}
\bibfield{author}{\bibinfo{person}{George~C. Necula}.}
  \bibinfo{year}{1997}\natexlab{}.
\newblock \showarticletitle{Proof-Carrying Code}. In
  \bibinfo{booktitle}{\emph{Conference Record of POPL'97: The 24th {ACM}
  {SIGPLAN-SIGACT} Symposium on Principles of Programming Languages, Papers
  Presented at the Symposium, Paris, France, 15-17 January 1997}},
  \bibfield{editor}{\bibinfo{person}{Peter Lee}, \bibinfo{person}{Fritz
  Henglein}, {and} \bibinfo{person}{Neil~D. Jones}} (Eds.).
  \bibinfo{publisher}{{ACM} Press}, \bibinfo{pages}{106--119}.
\newblock
\urldef\tempurl%
\url{https://doi.org/10.1145/263699.263712}
\showDOI{\tempurl}


\bibitem[Nielsen(2000)]%
        {nielsen2000denotational}
\bibfield{author}{\bibinfo{person}{Lasse~R Nielsen}.}
  \bibinfo{year}{2000}\natexlab{}.
\newblock \showarticletitle{A denotational investigation of
  defunctionalization}.
\newblock \bibinfo{journal}{\emph{BRICS Report Series}} \bibinfo{volume}{7},
  \bibinfo{number}{47} (\bibinfo{year}{2000}).
\newblock


\bibitem[Norell(2008)]%
        {DBLP:conf/afp/Norell08}
\bibfield{author}{\bibinfo{person}{Ulf Norell}.}
  \bibinfo{year}{2008}\natexlab{}.
\newblock \showarticletitle{Dependently Typed Programming in Agda}. In
  \bibinfo{booktitle}{\emph{Advanced Functional Programming, 6th International
  School, {AFP} 2008, Heijen, The Netherlands, May 2008, Revised Lectures}}
  \emph{(\bibinfo{series}{Lecture Notes in Computer Science},
  Vol.~\bibinfo{volume}{5832})}, \bibfield{editor}{\bibinfo{person}{Pieter
  W.~M. Koopman}, \bibinfo{person}{Rinus Plasmeijer}, {and}
  \bibinfo{person}{S.~Doaitse Swierstra}} (Eds.).
  \bibinfo{publisher}{Springer}, \bibinfo{pages}{230--266}.
\newblock
\urldef\tempurl%
\url{https://doi.org/10.1007/978-3-642-04652-0\_5}
\showDOI{\tempurl}


\bibitem[Paraskevopoulou and Appel(2019)]%
        {DBLP:journals/pacmpl/Paraskevopoulou19}
\bibfield{author}{\bibinfo{person}{Zoe Paraskevopoulou} {and}
  \bibinfo{person}{Andrew~W. Appel}.} \bibinfo{year}{2019}\natexlab{}.
\newblock \showarticletitle{Closure conversion is safe for space}.
\newblock \bibinfo{journal}{\emph{Proc. {ACM} Program. Lang.}}
  \bibinfo{volume}{3}, \bibinfo{number}{{ICFP}} (\bibinfo{year}{2019}),
  \bibinfo{pages}{83:1--83:29}.
\newblock
\urldef\tempurl%
\url{https://doi.org/10.1145/3341687}
\showDOI{\tempurl}


\bibitem[Patrignani et~al\mbox{.}(2019)]%
        {DBLP:journals/csur/PatrignaniAC19}
\bibfield{author}{\bibinfo{person}{Marco Patrignani}, \bibinfo{person}{Amal
  Ahmed}, {and} \bibinfo{person}{Dave Clarke}.}
  \bibinfo{year}{2019}\natexlab{}.
\newblock \showarticletitle{Formal Approaches to Secure Compilation: {A} Survey
  of Fully Abstract Compilation and Related Work}.
\newblock \bibinfo{journal}{\emph{{ACM} Comput. Surv.}} \bibinfo{volume}{51},
  \bibinfo{number}{6} (\bibinfo{year}{2019}), \bibinfo{pages}{125:1--125:36}.
\newblock
\urldef\tempurl%
\url{https://doi.org/10.1145/3280984}
\showDOI{\tempurl}


\bibitem[Pettyjohn et~al\mbox{.}(2005)]%
        {generalized-stack-inspection}
\bibfield{author}{\bibinfo{person}{Greg Pettyjohn}, \bibinfo{person}{John
  Clements}, \bibinfo{person}{Joe Marshall}, \bibinfo{person}{Shriram
  Krishnamurthi}, {and} \bibinfo{person}{Matthias Felleisen}.}
  \bibinfo{year}{2005}\natexlab{}.
\newblock \showarticletitle{Continuations from generalized stack inspection}.
  In \bibinfo{booktitle}{\emph{Proceedings of the 10th {ACM} {SIGPLAN}
  International Conference on Functional Programming, {ICFP} 2005, Tallinn,
  Estonia, September 26-28, 2005}}, \bibfield{editor}{\bibinfo{person}{Olivier
  Danvy} {and} \bibinfo{person}{Benjamin~C. Pierce}} (Eds.).
  \bibinfo{publisher}{{ACM}}, \bibinfo{pages}{216--227}.
\newblock
\urldef\tempurl%
\url{https://doi.org/10.1145/1086365.1086393}
\showDOI{\tempurl}


\bibitem[Podlovics et~al\mbox{.}(2021)]%
        {grin}
\bibfield{author}{\bibinfo{person}{Peter Podlovics}, \bibinfo{person}{Csaba
  Hruska}, {and} \bibinfo{person}{Andor Pénzes}.}
  \bibinfo{year}{2021}\natexlab{}.
\newblock \showarticletitle{A Modern Look at {GRIN}, an Optimizing Functional
  Language Back End}.
\newblock \bibinfo{journal}{\emph{Acta Cybernetica}} (\bibinfo{date}{Feb.}
  \bibinfo{year}{2021}).
\newblock
\urldef\tempurl%
\url{https://doi.org/10.14232/actacyb.282969}
\showDOI{\tempurl}


\bibitem[Pottier and Gauthier(2004)]%
        {DBLP:conf/popl/PottierG04}
\bibfield{author}{\bibinfo{person}{Fran{\c{c}}ois Pottier} {and}
  \bibinfo{person}{Nadji Gauthier}.} \bibinfo{year}{2004}\natexlab{}.
\newblock \showarticletitle{Polymorphic typed defunctionalization}. In
  \bibinfo{booktitle}{\emph{Proceedings of the 31st {ACM} {SIGPLAN-SIGACT}
  Symposium on Principles of Programming Languages, {POPL} 2004, Venice, Italy,
  January 14-16, 2004}}, \bibfield{editor}{\bibinfo{person}{Neil~D. Jones}
  {and} \bibinfo{person}{Xavier Leroy}} (Eds.). \bibinfo{publisher}{{ACM}},
  \bibinfo{pages}{89--98}.
\newblock
\urldef\tempurl%
\url{https://doi.org/10.1145/964001.964009}
\showDOI{\tempurl}


\bibitem[Pottier and Gauthier(2006)]%
        {DBLP:journals/lisp/PottierG06}
\bibfield{author}{\bibinfo{person}{Fran{\c{c}}ois Pottier} {and}
  \bibinfo{person}{Nadji Gauthier}.} \bibinfo{year}{2006}\natexlab{}.
\newblock \showarticletitle{Polymorphic typed defunctionalization and
  concretization}.
\newblock \bibinfo{journal}{\emph{High. Order Symb. Comput.}}
  \bibinfo{volume}{19}, \bibinfo{number}{1} (\bibinfo{year}{2006}),
  \bibinfo{pages}{125--162}.
\newblock
\urldef\tempurl%
\url{https://doi.org/10.1007/s10990-006-8611-7}
\showDOI{\tempurl}


\bibitem[Reynolds(1972)]%
        {DBLP:conf/acm/Reynolds72}
\bibfield{author}{\bibinfo{person}{John~C. Reynolds}.}
  \bibinfo{year}{1972}\natexlab{}.
\newblock \showarticletitle{Definitional interpreters for higher-order
  programming languages}. In \bibinfo{booktitle}{\emph{Proceedings of the {ACM}
  annual conference, {ACM} 1972, 1972, Volume 2}},
  \bibfield{editor}{\bibinfo{person}{John~J. Donovan} {and}
  \bibinfo{person}{Rosemary Shields}} (Eds.). \bibinfo{publisher}{{ACM}},
  \bibinfo{pages}{717--740}.
\newblock
\urldef\tempurl%
\url{https://doi.org/10.1145/800194.805852}
\showDOI{\tempurl}


\bibitem[Siek(2012)]%
        {essence-closure-conversion}
\bibfield{author}{\bibinfo{person}{Jeremy Siek}.}
  \bibinfo{year}{2012}\natexlab{}.
\newblock \bibinfo{title}{The Essence of Closure Conversion}.
\newblock
  \bibinfo{howpublished}{\url{http://siek.blogspot.com/2012/07/essence-of-closure-conversion.html}}.
\newblock


\bibitem[Steele~Jr(1978)]%
        {rabbit}
\bibfield{author}{\bibinfo{person}{Guy~Lewis Steele~Jr}.}
  \bibinfo{year}{1978}\natexlab{}.
\newblock \emph{\bibinfo{title}{RABBIT: A Compiler for SCHEME}}.
\newblock \bibinfo{thesistype}{Master's\ thesis}. \bibinfo{school}{MIT}.
\newblock


\bibitem[Sulzmann et~al\mbox{.}(2007)]%
        {DBLP:conf/tldi/SulzmannCJD07}
\bibfield{author}{\bibinfo{person}{Martin Sulzmann}, \bibinfo{person}{Manuel
  M.~T. Chakravarty}, \bibinfo{person}{Simon L.~Peyton Jones}, {and}
  \bibinfo{person}{Kevin Donnelly}.} \bibinfo{year}{2007}\natexlab{}.
\newblock \showarticletitle{System {F} with type equality coercions}. In
  \bibinfo{booktitle}{\emph{Proceedings of TLDI'07: 2007 {ACM} {SIGPLAN}
  International Workshop on Types in Languages Design and Implementation, Nice,
  France, January 16, 2007}}, \bibfield{editor}{\bibinfo{person}{Fran{\c{c}}ois
  Pottier} {and} \bibinfo{person}{George~C. Necula}} (Eds.).
  \bibinfo{publisher}{{ACM}}, \bibinfo{pages}{53--66}.
\newblock
\urldef\tempurl%
\url{https://doi.org/10.1145/1190315.1190324}
\showDOI{\tempurl}


\bibitem[Tarditi et~al\mbox{.}(1996)]%
        {DBLP:conf/pldi/TarditiMCSHL96}
\bibfield{author}{\bibinfo{person}{David Tarditi}, \bibinfo{person}{J.~Gregory
  Morrisett}, \bibinfo{person}{Perry Cheng}, \bibinfo{person}{Christopher~A.
  Stone}, \bibinfo{person}{Robert Harper}, {and} \bibinfo{person}{Peter Lee}.}
  \bibinfo{year}{1996}\natexlab{}.
\newblock \showarticletitle{{TIL:} {A} Type-Directed Optimizing Compiler for
  {ML}}. In \bibinfo{booktitle}{\emph{Proceedings of the {ACM} SIGPLAN'96
  Conference on Programming Language Design and Implementation (PLDI),
  Philadephia, Pennsylvania, USA, May 21-24, 1996}},
  \bibfield{editor}{\bibinfo{person}{Charles~N. Fischer}} (Ed.).
  \bibinfo{publisher}{{ACM}}, \bibinfo{pages}{181--192}.
\newblock
\urldef\tempurl%
\url{https://doi.org/10.1145/231379.231414}
\showDOI{\tempurl}


\bibitem[Timany and Sozeau(2017)]%
        {DBLP:journals/corr/abs-1710-03912}
\bibfield{author}{\bibinfo{person}{Amin Timany} {and} \bibinfo{person}{Matthieu
  Sozeau}.} \bibinfo{year}{2017}\natexlab{}.
\newblock \showarticletitle{Consistency of the Predicative Calculus of
  Cumulative Inductive Constructions {(pCuIC)}}.
\newblock \bibinfo{journal}{\emph{CoRR}}  \bibinfo{volume}{abs/1710.03912}
  (\bibinfo{year}{2017}).
\newblock
\showeprint[arXiv]{1710.03912}
\urldef\tempurl%
\url{http://arxiv.org/abs/1710.03912}
\showURL{%
\tempurl}


\bibitem[Tolmach and Oliva(1998)]%
        {DBLP:journals/jfp/TolmachO98}
\bibfield{author}{\bibinfo{person}{Andrew~P. Tolmach} {and}
  \bibinfo{person}{Dino Oliva}.} \bibinfo{year}{1998}\natexlab{}.
\newblock \showarticletitle{From {ML} to {Ada}: Strongly-Typed Language
  Interoperability via Source Translation}.
\newblock \bibinfo{journal}{\emph{J. Funct. Program.}} \bibinfo{volume}{8},
  \bibinfo{number}{4} (\bibinfo{year}{1998}), \bibinfo{pages}{367--412}.
\newblock
\urldef\tempurl%
\url{http://journals.cambridge.org/action/displayAbstract?aid=44181}
\showURL{%
\tempurl}


\bibitem[Vouillon and Balat(2014)]%
        {DBLP:journals/spe/VouillonB14}
\bibfield{author}{\bibinfo{person}{J{\'{e}}r{\^{o}}me Vouillon} {and}
  \bibinfo{person}{Vincent Balat}.} \bibinfo{year}{2014}\natexlab{}.
\newblock \showarticletitle{From bytecode to JavaScript: the Js{\_}of{\_}ocaml
  compiler}.
\newblock \bibinfo{journal}{\emph{Softw. Pract. Exp.}} \bibinfo{volume}{44},
  \bibinfo{number}{8} (\bibinfo{year}{2014}), \bibinfo{pages}{951--972}.
\newblock
\urldef\tempurl%
\url{https://doi.org/10.1002/spe.2187}
\showDOI{\tempurl}


\bibitem[Wadler and Blott(1989)]%
        {DBLP:conf/popl/WadlerB89}
\bibfield{author}{\bibinfo{person}{Philip Wadler} {and}
  \bibinfo{person}{Stephen Blott}.} \bibinfo{year}{1989}\natexlab{}.
\newblock \showarticletitle{How to Make ad-hoc Polymorphism Less ad-hoc}. In
  \bibinfo{booktitle}{\emph{Conference Record of the Sixteenth Annual {ACM}
  Symposium on Principles of Programming Languages, Austin, Texas, USA, January
  11-13, 1989}}. \bibinfo{publisher}{{ACM} Press}, \bibinfo{pages}{60--76}.
\newblock
\urldef\tempurl%
\url{https://doi.org/10.1145/75277.75283}
\showDOI{\tempurl}


\bibitem[Wang and Appel(2001)]%
        {DBLP:conf/popl/WangA01}
\bibfield{author}{\bibinfo{person}{Daniel~C. Wang} {and}
  \bibinfo{person}{Andrew~W. Appel}.} \bibinfo{year}{2001}\natexlab{}.
\newblock \showarticletitle{Type-preserving garbage collectors}. In
  \bibinfo{booktitle}{\emph{Conference Record of {POPL} 2001: The 28th {ACM}
  {SIGPLAN-SIGACT} Symposium on Principles of Programming Languages, London,
  UK, January 17-19, 2001}}, \bibfield{editor}{\bibinfo{person}{Chris Hankin}
  {and} \bibinfo{person}{Dave Schmidt}} (Eds.). \bibinfo{publisher}{{ACM}},
  \bibinfo{pages}{166--178}.
\newblock
\urldef\tempurl%
\url{https://doi.org/10.1145/360204.360218}
\showDOI{\tempurl}


\bibitem[Weeks(2006)]%
        {mlton}
\bibfield{author}{\bibinfo{person}{Stephen Weeks}.}
  \bibinfo{year}{2006}\natexlab{}.
\newblock \showarticletitle{Whole-Program Compilation in MLton}. In
  \bibinfo{booktitle}{\emph{Proceedings of the 2006 Workshop on ML}} (Portland,
  Oregon, USA) \emph{(\bibinfo{series}{ML '06})}.
  \bibinfo{publisher}{Association for Computing Machinery},
  \bibinfo{address}{New York, NY, USA}, \bibinfo{pages}{1}.
\newblock
\showISBNx{1595934839}
\urldef\tempurl%
\url{https://doi.org/10.1145/1159876.1159877}
\showDOI{\tempurl}


\bibitem[Weirich and Casinghino(2010)]%
        {DBLP:conf/ssgip/WeirichC10}
\bibfield{author}{\bibinfo{person}{Stephanie Weirich} {and}
  \bibinfo{person}{Chris Casinghino}.} \bibinfo{year}{2010}\natexlab{}.
\newblock \showarticletitle{Generic Programming with Dependent Types}. In
  \bibinfo{booktitle}{\emph{Generic and Indexed Programming - International
  Spring School, {SSGIP} 2010, Oxford, UK, March 22-26, 2010, Revised
  Lectures}} \emph{(\bibinfo{series}{Lecture Notes in Computer Science},
  Vol.~\bibinfo{volume}{7470})}, \bibfield{editor}{\bibinfo{person}{Jeremy
  Gibbons}} (Ed.). \bibinfo{publisher}{Springer}, \bibinfo{pages}{217--258}.
\newblock
\urldef\tempurl%
\url{https://doi.org/10.1007/978-3-642-32202-0\_5}
\showDOI{\tempurl}


\bibitem[Xi and Harper(2001)]%
        {DBLP:conf/icfp/XiH01}
\bibfield{author}{\bibinfo{person}{Hongwei Xi} {and} \bibinfo{person}{Robert
  Harper}.} \bibinfo{year}{2001}\natexlab{}.
\newblock \showarticletitle{A Dependently Typed Assembly Language}. In
  \bibinfo{booktitle}{\emph{Proceedings of the Sixth {ACM} {SIGPLAN}
  International Conference on Functional Programming {(ICFP} '01), Firenze
  (Florence), Italy, September 3-5, 2001}},
  \bibfield{editor}{\bibinfo{person}{Benjamin~C. Pierce}} (Ed.).
  \bibinfo{publisher}{{ACM}}, \bibinfo{pages}{169--180}.
\newblock
\urldef\tempurl%
\url{https://doi.org/10.1145/507635.507657}
\showDOI{\tempurl}


\bibitem[Yallop and White(2014)]%
        {DBLP:conf/flops/YallopW14}
\bibfield{author}{\bibinfo{person}{Jeremy Yallop} {and} \bibinfo{person}{Leo
  White}.} \bibinfo{year}{2014}\natexlab{}.
\newblock \showarticletitle{Lightweight Higher-Kinded Polymorphism}. In
  \bibinfo{booktitle}{\emph{Functional and Logic Programming - 12th
  International Symposium, {FLOPS} 2014, Kanazawa, Japan, June 4-6, 2014.
  Proceedings}} \emph{(\bibinfo{series}{Lecture Notes in Computer Science},
  Vol.~\bibinfo{volume}{8475})}, \bibfield{editor}{\bibinfo{person}{Michael
  Codish} {and} \bibinfo{person}{Eijiro Sumii}} (Eds.).
  \bibinfo{publisher}{Springer}, \bibinfo{pages}{119--135}.
\newblock
\urldef\tempurl%
\url{https://doi.org/10.1007/978-3-319-07151-0\_8}
\showDOI{\tempurl}


\end{thebibliography}

\appendix

\section{Dependent defunctionalization in Agda}
\label{appendix:agda-defun}

\begin{code}%
\>[0]\AgdaSymbol{\{-\#}\AgdaSpace{}%
\AgdaKeyword{OPTIONS}\AgdaSpace{}%
\AgdaPragma{--type-in-type}\AgdaSpace{}%
\AgdaSymbol{\#-\}}\<%
\end{code}

\begin{code}%
\>[0]\AgdaKeyword{data}\AgdaSpace{}%
\AgdaDatatype{Π}\AgdaSpace{}%
\AgdaSymbol{:}\AgdaSpace{}%
\AgdaSymbol{(}\AgdaBound{A}\AgdaSpace{}%
\AgdaSymbol{:}\AgdaSpace{}%
\AgdaPrimitive{Set}\AgdaSymbol{)}\AgdaSpace{}%
\AgdaSymbol{→}\AgdaSpace{}%
\AgdaSymbol{(}\AgdaBound{A}\AgdaSpace{}%
\AgdaSymbol{→}\AgdaSpace{}%
\AgdaPrimitive{Set}\AgdaSymbol{)}\AgdaSpace{}%
\AgdaSymbol{→}\AgdaSpace{}%
\AgdaPrimitive{Set}\<%
\\
\>[0]\AgdaKeyword{infixl}\AgdaSpace{}%
\AgdaNumber{9}\AgdaSpace{}%
\AgdaOperator{\AgdaFunction{\AgdaUnderscore{}\$\AgdaUnderscore{}}}\<%
\\
\>[0]\AgdaOperator{\AgdaFunction{\AgdaUnderscore{}\$\AgdaUnderscore{}}}\AgdaSpace{}%
\AgdaSymbol{:}\AgdaSpace{}%
\AgdaSymbol{∀}\AgdaSpace{}%
\AgdaSymbol{\{}\AgdaBound{A}\AgdaSpace{}%
\AgdaSymbol{:}\AgdaSpace{}%
\AgdaPrimitive{Set}\AgdaSymbol{\}}\AgdaSpace{}%
\AgdaSymbol{\{}\AgdaBound{p}\AgdaSpace{}%
\AgdaSymbol{:}\AgdaSpace{}%
\AgdaBound{A}\AgdaSpace{}%
\AgdaSymbol{→}\AgdaSpace{}%
\AgdaPrimitive{Set}\AgdaSymbol{\}}\AgdaSpace{}%
\AgdaSymbol{→}\AgdaSpace{}%
\AgdaDatatype{Π}\AgdaSpace{}%
\AgdaBound{A}\AgdaSpace{}%
\AgdaBound{p}\AgdaSpace{}%
\AgdaSymbol{→}\AgdaSpace{}%
\AgdaSymbol{(}\AgdaBound{x}\AgdaSpace{}%
\AgdaSymbol{:}\AgdaSpace{}%
\AgdaBound{A}\AgdaSymbol{)}\AgdaSpace{}%
\AgdaSymbol{→}\AgdaSpace{}%
\AgdaBound{p}\AgdaSpace{}%
\AgdaBound{x}\<%
\\
\\[\AgdaEmptyExtraSkip]%
\>[0]\AgdaSymbol{\{-\#}\AgdaSpace{}%
\AgdaKeyword{NO\AgdaUnderscore{}POSITIVITY\AgdaUnderscore{}CHECK}\AgdaSpace{}%
\AgdaSymbol{\#-\}}\<%
\\
\>[0]\AgdaKeyword{data}\AgdaSpace{}%
\AgdaDatatype{Π}%
\>[8]\AgdaKeyword{where}\<%
\\
\>[0][@{}l@{\AgdaIndent{0}}]%
\>[3]\AgdaInductiveConstructor{F1}\AgdaSpace{}%
\AgdaSymbol{:}%
\>[41I]\AgdaDatatype{Π}\AgdaSpace{}%
\AgdaPrimitive{Set}\AgdaSpace{}%
\AgdaSymbol{(λ}\AgdaSpace{}%
\AgdaBound{A}\AgdaSpace{}%
\AgdaSymbol{→}\<%
\\
\>[.][@{}l@{}]\<[41I]%
\>[8]\AgdaDatatype{Π}\AgdaSpace{}%
\AgdaSymbol{(}\AgdaDatatype{Π}\AgdaSpace{}%
\AgdaBound{A}\AgdaSpace{}%
\AgdaSymbol{(λ}\AgdaSpace{}%
\AgdaBound{\AgdaUnderscore{}}\AgdaSpace{}%
\AgdaSymbol{→}\AgdaSpace{}%
\AgdaPrimitive{Set}\AgdaSymbol{))}\AgdaSpace{}%
\AgdaSymbol{(λ}\AgdaSpace{}%
\AgdaBound{B}\AgdaSpace{}%
\AgdaSymbol{→}\<%
\\
\>[8]\AgdaDatatype{Π}\AgdaSpace{}%
\AgdaSymbol{(}\AgdaDatatype{Π}\AgdaSpace{}%
\AgdaBound{A}\AgdaSpace{}%
\AgdaSymbol{(λ}\AgdaSpace{}%
\AgdaBound{x}\AgdaSpace{}%
\AgdaSymbol{→}\AgdaSpace{}%
\AgdaDatatype{Π}\AgdaSpace{}%
\AgdaSymbol{(}\AgdaBound{B}\AgdaSpace{}%
\AgdaOperator{\AgdaFunction{\$}}\AgdaSpace{}%
\AgdaBound{x}\AgdaSymbol{)}\AgdaSpace{}%
\AgdaSymbol{(λ}\AgdaSpace{}%
\AgdaBound{\AgdaUnderscore{}}\AgdaSpace{}%
\AgdaSymbol{→}\AgdaSpace{}%
\AgdaPrimitive{Set}\AgdaSymbol{)))}\AgdaSpace{}%
\AgdaSymbol{(λ}\AgdaSpace{}%
\AgdaBound{C}\AgdaSpace{}%
\AgdaSymbol{→}\<%
\\
\>[8]\AgdaDatatype{Π}\AgdaSpace{}%
\AgdaSymbol{(}\AgdaDatatype{Π}\AgdaSpace{}%
\AgdaBound{A}\AgdaSpace{}%
\AgdaSymbol{(λ}\AgdaSpace{}%
\AgdaBound{y}\AgdaSpace{}%
\AgdaSymbol{→}\AgdaSpace{}%
\AgdaDatatype{Π}\AgdaSpace{}%
\AgdaSymbol{(}\AgdaBound{B}\AgdaSpace{}%
\AgdaOperator{\AgdaFunction{\$}}\AgdaSpace{}%
\AgdaBound{y}\AgdaSymbol{)}\AgdaSpace{}%
\AgdaSymbol{(λ}\AgdaSpace{}%
\AgdaBound{z}\AgdaSpace{}%
\AgdaSymbol{→}\AgdaSpace{}%
\AgdaBound{C}\AgdaSpace{}%
\AgdaOperator{\AgdaFunction{\$}}\AgdaSpace{}%
\AgdaBound{y}\AgdaSpace{}%
\AgdaOperator{\AgdaFunction{\$}}\AgdaSpace{}%
\AgdaBound{z}\AgdaSymbol{)))}\AgdaSpace{}%
\AgdaSymbol{(λ}\AgdaSpace{}%
\AgdaBound{f}\AgdaSpace{}%
\AgdaSymbol{→}\<%
\\
\>[8]\AgdaDatatype{Π}\AgdaSpace{}%
\AgdaSymbol{(}\AgdaDatatype{Π}\AgdaSpace{}%
\AgdaBound{A}\AgdaSpace{}%
\AgdaSymbol{(λ}\AgdaSpace{}%
\AgdaBound{x}\AgdaSpace{}%
\AgdaSymbol{→}\AgdaSpace{}%
\AgdaBound{B}\AgdaSpace{}%
\AgdaOperator{\AgdaFunction{\$}}\AgdaSpace{}%
\AgdaBound{x}\AgdaSymbol{))}\AgdaSpace{}%
\AgdaSymbol{(λ}\AgdaSpace{}%
\AgdaBound{g}\AgdaSpace{}%
\AgdaSymbol{→}\<%
\\
\>[8]\AgdaDatatype{Π}\AgdaSpace{}%
\AgdaBound{A}\AgdaSpace{}%
\AgdaSymbol{(λ}\AgdaSpace{}%
\AgdaBound{x}\AgdaSpace{}%
\AgdaSymbol{→}\<%
\\
\>[8]\AgdaBound{C}\AgdaSpace{}%
\AgdaOperator{\AgdaFunction{\$}}\AgdaSpace{}%
\AgdaBound{x}\AgdaSpace{}%
\AgdaOperator{\AgdaFunction{\$}}\AgdaSpace{}%
\AgdaSymbol{(}\AgdaBound{g}\AgdaSpace{}%
\AgdaOperator{\AgdaFunction{\$}}\AgdaSpace{}%
\AgdaBound{x}\AgdaSymbol{)))))))}\<%
\\
\>[3]\AgdaInductiveConstructor{F2}\AgdaSpace{}%
\AgdaSymbol{:}%
\>[113I]\AgdaSymbol{(}\AgdaBound{A}\AgdaSpace{}%
\AgdaSymbol{:}\AgdaSpace{}%
\AgdaPrimitive{Set}\AgdaSymbol{)}\AgdaSpace{}%
\AgdaSymbol{→}\<%
\\
\>[.][@{}l@{}]\<[113I]%
\>[8]\AgdaDatatype{Π}\AgdaSpace{}%
\AgdaSymbol{(}\AgdaDatatype{Π}\AgdaSpace{}%
\AgdaBound{A}\AgdaSpace{}%
\AgdaSymbol{(λ}\AgdaSpace{}%
\AgdaBound{\AgdaUnderscore{}}\AgdaSpace{}%
\AgdaSymbol{→}\AgdaSpace{}%
\AgdaPrimitive{Set}\AgdaSymbol{))}\AgdaSpace{}%
\AgdaSymbol{(λ}\AgdaSpace{}%
\AgdaBound{B}\AgdaSpace{}%
\AgdaSymbol{→}\<%
\\
\>[8]\AgdaDatatype{Π}\AgdaSpace{}%
\AgdaSymbol{(}\AgdaDatatype{Π}\AgdaSpace{}%
\AgdaBound{A}\AgdaSpace{}%
\AgdaSymbol{(λ}\AgdaSpace{}%
\AgdaBound{x}\AgdaSpace{}%
\AgdaSymbol{→}\AgdaSpace{}%
\AgdaDatatype{Π}\AgdaSpace{}%
\AgdaSymbol{(}\AgdaBound{B}\AgdaSpace{}%
\AgdaOperator{\AgdaFunction{\$}}\AgdaSpace{}%
\AgdaBound{x}\AgdaSymbol{)}\AgdaSpace{}%
\AgdaSymbol{(λ}\AgdaSpace{}%
\AgdaBound{\AgdaUnderscore{}}\AgdaSpace{}%
\AgdaSymbol{→}\AgdaSpace{}%
\AgdaPrimitive{Set}\AgdaSymbol{)))}\AgdaSpace{}%
\AgdaSymbol{(λ}\AgdaSpace{}%
\AgdaBound{C}\AgdaSpace{}%
\AgdaSymbol{→}\<%
\\
\>[8]\AgdaDatatype{Π}\AgdaSpace{}%
\AgdaSymbol{(}\AgdaDatatype{Π}\AgdaSpace{}%
\AgdaBound{A}\AgdaSpace{}%
\AgdaSymbol{(λ}\AgdaSpace{}%
\AgdaBound{y}\AgdaSpace{}%
\AgdaSymbol{→}\AgdaSpace{}%
\AgdaDatatype{Π}\AgdaSpace{}%
\AgdaSymbol{(}\AgdaBound{B}\AgdaSpace{}%
\AgdaOperator{\AgdaFunction{\$}}\AgdaSpace{}%
\AgdaBound{y}\AgdaSymbol{)}\AgdaSpace{}%
\AgdaSymbol{(λ}\AgdaSpace{}%
\AgdaBound{z}\AgdaSpace{}%
\AgdaSymbol{→}\AgdaSpace{}%
\AgdaBound{C}\AgdaSpace{}%
\AgdaOperator{\AgdaFunction{\$}}\AgdaSpace{}%
\AgdaBound{y}\AgdaSpace{}%
\AgdaOperator{\AgdaFunction{\$}}\AgdaSpace{}%
\AgdaBound{z}\AgdaSymbol{)))}\AgdaSpace{}%
\AgdaSymbol{(λ}\AgdaSpace{}%
\AgdaBound{f}\AgdaSpace{}%
\AgdaSymbol{→}\<%
\\
\>[8]\AgdaDatatype{Π}\AgdaSpace{}%
\AgdaSymbol{(}\AgdaDatatype{Π}\AgdaSpace{}%
\AgdaBound{A}\AgdaSpace{}%
\AgdaSymbol{(λ}\AgdaSpace{}%
\AgdaBound{x}\AgdaSpace{}%
\AgdaSymbol{→}\AgdaSpace{}%
\AgdaBound{B}\AgdaSpace{}%
\AgdaOperator{\AgdaFunction{\$}}\AgdaSpace{}%
\AgdaBound{x}\AgdaSymbol{))}\AgdaSpace{}%
\AgdaSymbol{(λ}\AgdaSpace{}%
\AgdaBound{g}\AgdaSpace{}%
\AgdaSymbol{→}\<%
\\
\>[8]\AgdaDatatype{Π}\AgdaSpace{}%
\AgdaBound{A}\AgdaSpace{}%
\AgdaSymbol{(λ}\AgdaSpace{}%
\AgdaBound{x}\AgdaSpace{}%
\AgdaSymbol{→}\<%
\\
\>[8]\AgdaBound{C}\AgdaSpace{}%
\AgdaOperator{\AgdaFunction{\$}}\AgdaSpace{}%
\AgdaBound{x}\AgdaSpace{}%
\AgdaOperator{\AgdaFunction{\$}}\AgdaSpace{}%
\AgdaSymbol{(}\AgdaBound{g}\AgdaSpace{}%
\AgdaOperator{\AgdaFunction{\$}}\AgdaSpace{}%
\AgdaBound{x}\AgdaSymbol{))))))}\<%
\\
\>[3]\AgdaInductiveConstructor{F3}\AgdaSpace{}%
\AgdaSymbol{:}%
\>[184I]\AgdaSymbol{(}\AgdaBound{A}\AgdaSpace{}%
\AgdaSymbol{:}\AgdaSpace{}%
\AgdaPrimitive{Set}\AgdaSymbol{)}\AgdaSpace{}%
\AgdaSymbol{→}\<%
\\
\>[.][@{}l@{}]\<[184I]%
\>[8]\AgdaSymbol{(}\AgdaBound{B}\AgdaSpace{}%
\AgdaSymbol{:}\AgdaSpace{}%
\AgdaDatatype{Π}\AgdaSpace{}%
\AgdaBound{A}\AgdaSpace{}%
\AgdaSymbol{(λ}\AgdaSpace{}%
\AgdaBound{\AgdaUnderscore{}}\AgdaSpace{}%
\AgdaSymbol{→}\AgdaSpace{}%
\AgdaPrimitive{Set}\AgdaSymbol{))}\AgdaSpace{}%
\AgdaSymbol{→}\<%
\\
\>[8]\AgdaDatatype{Π}\AgdaSpace{}%
\AgdaSymbol{(}\AgdaDatatype{Π}\AgdaSpace{}%
\AgdaBound{A}\AgdaSpace{}%
\AgdaSymbol{(λ}\AgdaSpace{}%
\AgdaBound{x}\AgdaSpace{}%
\AgdaSymbol{→}\AgdaSpace{}%
\AgdaDatatype{Π}\AgdaSpace{}%
\AgdaSymbol{(}\AgdaBound{B}\AgdaSpace{}%
\AgdaOperator{\AgdaFunction{\$}}\AgdaSpace{}%
\AgdaBound{x}\AgdaSymbol{)}\AgdaSpace{}%
\AgdaSymbol{(λ}\AgdaSpace{}%
\AgdaBound{\AgdaUnderscore{}}\AgdaSpace{}%
\AgdaSymbol{→}\AgdaSpace{}%
\AgdaPrimitive{Set}\AgdaSymbol{)))}\AgdaSpace{}%
\AgdaSymbol{(λ}\AgdaSpace{}%
\AgdaBound{C}\AgdaSpace{}%
\AgdaSymbol{→}\<%
\\
\>[8]\AgdaDatatype{Π}\AgdaSpace{}%
\AgdaSymbol{(}\AgdaDatatype{Π}\AgdaSpace{}%
\AgdaBound{A}\AgdaSpace{}%
\AgdaSymbol{(λ}\AgdaSpace{}%
\AgdaBound{y}\AgdaSpace{}%
\AgdaSymbol{→}\AgdaSpace{}%
\AgdaDatatype{Π}\AgdaSpace{}%
\AgdaSymbol{(}\AgdaBound{B}\AgdaSpace{}%
\AgdaOperator{\AgdaFunction{\$}}\AgdaSpace{}%
\AgdaBound{y}\AgdaSymbol{)}\AgdaSpace{}%
\AgdaSymbol{(λ}\AgdaSpace{}%
\AgdaBound{z}\AgdaSpace{}%
\AgdaSymbol{→}\AgdaSpace{}%
\AgdaBound{C}\AgdaSpace{}%
\AgdaOperator{\AgdaFunction{\$}}\AgdaSpace{}%
\AgdaBound{y}\AgdaSpace{}%
\AgdaOperator{\AgdaFunction{\$}}\AgdaSpace{}%
\AgdaBound{z}\AgdaSymbol{)))}\AgdaSpace{}%
\AgdaSymbol{(λ}\AgdaSpace{}%
\AgdaBound{f}\AgdaSpace{}%
\AgdaSymbol{→}\<%
\\
\>[8]\AgdaDatatype{Π}\AgdaSpace{}%
\AgdaSymbol{(}\AgdaDatatype{Π}\AgdaSpace{}%
\AgdaBound{A}\AgdaSpace{}%
\AgdaSymbol{(λ}\AgdaSpace{}%
\AgdaBound{x}\AgdaSpace{}%
\AgdaSymbol{→}\AgdaSpace{}%
\AgdaBound{B}\AgdaSpace{}%
\AgdaOperator{\AgdaFunction{\$}}\AgdaSpace{}%
\AgdaBound{x}\AgdaSymbol{))}\AgdaSpace{}%
\AgdaSymbol{(λ}\AgdaSpace{}%
\AgdaBound{g}\AgdaSpace{}%
\AgdaSymbol{→}\<%
\\
\>[8]\AgdaDatatype{Π}\AgdaSpace{}%
\AgdaBound{A}\AgdaSpace{}%
\AgdaSymbol{(λ}\AgdaSpace{}%
\AgdaBound{x}\AgdaSpace{}%
\AgdaSymbol{→}\<%
\\
\>[8]\AgdaBound{C}\AgdaSpace{}%
\AgdaOperator{\AgdaFunction{\$}}\AgdaSpace{}%
\AgdaBound{x}\AgdaSpace{}%
\AgdaOperator{\AgdaFunction{\$}}\AgdaSpace{}%
\AgdaSymbol{(}\AgdaBound{g}\AgdaSpace{}%
\AgdaOperator{\AgdaFunction{\$}}\AgdaSpace{}%
\AgdaBound{x}\AgdaSymbol{)))))}\<%
\\
\>[3]\AgdaInductiveConstructor{F4}\AgdaSpace{}%
\AgdaSymbol{:}%
\>[254I]\AgdaSymbol{(}\AgdaBound{A}\AgdaSpace{}%
\AgdaSymbol{:}\AgdaSpace{}%
\AgdaPrimitive{Set}\AgdaSymbol{)}\AgdaSpace{}%
\AgdaSymbol{→}\<%
\\
\>[.][@{}l@{}]\<[254I]%
\>[8]\AgdaSymbol{(}\AgdaBound{B}\AgdaSpace{}%
\AgdaSymbol{:}\AgdaSpace{}%
\AgdaDatatype{Π}\AgdaSpace{}%
\AgdaBound{A}\AgdaSpace{}%
\AgdaSymbol{(λ}\AgdaSpace{}%
\AgdaBound{\AgdaUnderscore{}}\AgdaSpace{}%
\AgdaSymbol{→}\AgdaSpace{}%
\AgdaPrimitive{Set}\AgdaSymbol{))}\AgdaSpace{}%
\AgdaSymbol{→}\<%
\\
\>[8]\AgdaSymbol{(}\AgdaBound{C}\AgdaSpace{}%
\AgdaSymbol{:}\AgdaSpace{}%
\AgdaSymbol{(}\AgdaDatatype{Π}\AgdaSpace{}%
\AgdaBound{A}\AgdaSpace{}%
\AgdaSymbol{(λ}\AgdaSpace{}%
\AgdaBound{x}\AgdaSpace{}%
\AgdaSymbol{→}\AgdaSpace{}%
\AgdaDatatype{Π}\AgdaSpace{}%
\AgdaSymbol{(}\AgdaBound{B}\AgdaSpace{}%
\AgdaOperator{\AgdaFunction{\$}}\AgdaSpace{}%
\AgdaBound{x}\AgdaSymbol{)}\AgdaSpace{}%
\AgdaSymbol{(λ}\AgdaSpace{}%
\AgdaBound{\AgdaUnderscore{}}\AgdaSpace{}%
\AgdaSymbol{→}\AgdaSpace{}%
\AgdaPrimitive{Set}\AgdaSymbol{))))}\AgdaSpace{}%
\AgdaSymbol{→}\<%
\\
\>[8]\AgdaDatatype{Π}\AgdaSpace{}%
\AgdaSymbol{(}\AgdaDatatype{Π}\AgdaSpace{}%
\AgdaBound{A}\AgdaSpace{}%
\AgdaSymbol{(λ}\AgdaSpace{}%
\AgdaBound{y}\AgdaSpace{}%
\AgdaSymbol{→}\AgdaSpace{}%
\AgdaDatatype{Π}\AgdaSpace{}%
\AgdaSymbol{(}\AgdaBound{B}\AgdaSpace{}%
\AgdaOperator{\AgdaFunction{\$}}\AgdaSpace{}%
\AgdaBound{y}\AgdaSymbol{)}\AgdaSpace{}%
\AgdaSymbol{(λ}\AgdaSpace{}%
\AgdaBound{z}\AgdaSpace{}%
\AgdaSymbol{→}\AgdaSpace{}%
\AgdaBound{C}\AgdaSpace{}%
\AgdaOperator{\AgdaFunction{\$}}\AgdaSpace{}%
\AgdaBound{y}\AgdaSpace{}%
\AgdaOperator{\AgdaFunction{\$}}\AgdaSpace{}%
\AgdaBound{z}\AgdaSymbol{)))}\AgdaSpace{}%
\AgdaSymbol{(λ}\AgdaSpace{}%
\AgdaBound{f}\AgdaSpace{}%
\AgdaSymbol{→}\<%
\\
\>[8]\AgdaDatatype{Π}\AgdaSpace{}%
\AgdaSymbol{(}\AgdaDatatype{Π}\AgdaSpace{}%
\AgdaBound{A}\AgdaSpace{}%
\AgdaSymbol{(λ}\AgdaSpace{}%
\AgdaBound{x}\AgdaSpace{}%
\AgdaSymbol{→}\AgdaSpace{}%
\AgdaBound{B}\AgdaSpace{}%
\AgdaOperator{\AgdaFunction{\$}}\AgdaSpace{}%
\AgdaBound{x}\AgdaSymbol{))}\AgdaSpace{}%
\AgdaSymbol{(λ}\AgdaSpace{}%
\AgdaBound{g}\AgdaSpace{}%
\AgdaSymbol{→}\<%
\\
\>[8]\AgdaDatatype{Π}\AgdaSpace{}%
\AgdaBound{A}\AgdaSpace{}%
\AgdaSymbol{(λ}\AgdaSpace{}%
\AgdaBound{x}\AgdaSpace{}%
\AgdaSymbol{→}\<%
\\
\>[8]\AgdaBound{C}\AgdaSpace{}%
\AgdaOperator{\AgdaFunction{\$}}\AgdaSpace{}%
\AgdaBound{x}\AgdaSpace{}%
\AgdaOperator{\AgdaFunction{\$}}\AgdaSpace{}%
\AgdaSymbol{(}\AgdaBound{g}\AgdaSpace{}%
\AgdaOperator{\AgdaFunction{\$}}\AgdaSpace{}%
\AgdaBound{x}\AgdaSymbol{))))}\<%
\\
\>[3]\AgdaInductiveConstructor{F5}\AgdaSpace{}%
\AgdaSymbol{:}%
\>[323I]\AgdaSymbol{(}\AgdaBound{A}\AgdaSpace{}%
\AgdaSymbol{:}\AgdaSpace{}%
\AgdaPrimitive{Set}\AgdaSymbol{)}\AgdaSpace{}%
\AgdaSymbol{→}\<%
\\
\>[.][@{}l@{}]\<[323I]%
\>[8]\AgdaSymbol{(}\AgdaBound{B}\AgdaSpace{}%
\AgdaSymbol{:}\AgdaSpace{}%
\AgdaDatatype{Π}\AgdaSpace{}%
\AgdaBound{A}\AgdaSpace{}%
\AgdaSymbol{(λ}\AgdaSpace{}%
\AgdaBound{\AgdaUnderscore{}}\AgdaSpace{}%
\AgdaSymbol{→}\AgdaSpace{}%
\AgdaPrimitive{Set}\AgdaSymbol{))}\AgdaSpace{}%
\AgdaSymbol{→}\<%
\\
\>[8]\AgdaSymbol{(}\AgdaBound{C}\AgdaSpace{}%
\AgdaSymbol{:}\AgdaSpace{}%
\AgdaSymbol{(}\AgdaDatatype{Π}\AgdaSpace{}%
\AgdaBound{A}\AgdaSpace{}%
\AgdaSymbol{(λ}\AgdaSpace{}%
\AgdaBound{x}\AgdaSpace{}%
\AgdaSymbol{→}\AgdaSpace{}%
\AgdaDatatype{Π}\AgdaSpace{}%
\AgdaSymbol{(}\AgdaBound{B}\AgdaSpace{}%
\AgdaOperator{\AgdaFunction{\$}}\AgdaSpace{}%
\AgdaBound{x}\AgdaSymbol{)}\AgdaSpace{}%
\AgdaSymbol{(λ}\AgdaSpace{}%
\AgdaBound{\AgdaUnderscore{}}\AgdaSpace{}%
\AgdaSymbol{→}\AgdaSpace{}%
\AgdaPrimitive{Set}\AgdaSymbol{))))}\AgdaSpace{}%
\AgdaSymbol{→}\<%
\\
\>[8]\AgdaSymbol{(}\AgdaBound{f}\AgdaSpace{}%
\AgdaSymbol{:}\AgdaSpace{}%
\AgdaSymbol{(}\AgdaDatatype{Π}\AgdaSpace{}%
\AgdaBound{A}\AgdaSpace{}%
\AgdaSymbol{(λ}\AgdaSpace{}%
\AgdaBound{y}\AgdaSpace{}%
\AgdaSymbol{→}\AgdaSpace{}%
\AgdaDatatype{Π}\AgdaSpace{}%
\AgdaSymbol{(}\AgdaBound{B}\AgdaSpace{}%
\AgdaOperator{\AgdaFunction{\$}}\AgdaSpace{}%
\AgdaBound{y}\AgdaSymbol{)}\AgdaSpace{}%
\AgdaSymbol{(λ}\AgdaSpace{}%
\AgdaBound{z}\AgdaSpace{}%
\AgdaSymbol{→}\AgdaSpace{}%
\AgdaBound{C}\AgdaSpace{}%
\AgdaOperator{\AgdaFunction{\$}}\AgdaSpace{}%
\AgdaBound{y}\AgdaSpace{}%
\AgdaOperator{\AgdaFunction{\$}}\AgdaSpace{}%
\AgdaBound{z}\AgdaSymbol{))))}\AgdaSpace{}%
\AgdaSymbol{→}\<%
\\
\>[8]\AgdaDatatype{Π}\AgdaSpace{}%
\AgdaSymbol{(}\AgdaDatatype{Π}\AgdaSpace{}%
\AgdaBound{A}\AgdaSpace{}%
\AgdaSymbol{(λ}\AgdaSpace{}%
\AgdaBound{x}\AgdaSpace{}%
\AgdaSymbol{→}\AgdaSpace{}%
\AgdaBound{B}\AgdaSpace{}%
\AgdaOperator{\AgdaFunction{\$}}\AgdaSpace{}%
\AgdaBound{x}\AgdaSymbol{))}\AgdaSpace{}%
\AgdaSymbol{(λ}\AgdaSpace{}%
\AgdaBound{g}\AgdaSpace{}%
\AgdaSymbol{→}\<%
\\
\>[8]\AgdaDatatype{Π}\AgdaSpace{}%
\AgdaBound{A}\AgdaSpace{}%
\AgdaSymbol{(λ}\AgdaSpace{}%
\AgdaBound{x}\AgdaSpace{}%
\AgdaSymbol{→}\<%
\\
\>[8]\AgdaBound{C}\AgdaSpace{}%
\AgdaOperator{\AgdaFunction{\$}}\AgdaSpace{}%
\AgdaBound{x}\AgdaSpace{}%
\AgdaOperator{\AgdaFunction{\$}}\AgdaSpace{}%
\AgdaSymbol{(}\AgdaBound{g}\AgdaSpace{}%
\AgdaOperator{\AgdaFunction{\$}}\AgdaSpace{}%
\AgdaBound{x}\AgdaSymbol{)))}\<%
\\
\>[3]\AgdaInductiveConstructor{F6}\AgdaSpace{}%
\AgdaSymbol{:}%
\>[391I]\AgdaSymbol{(}\AgdaBound{A}\AgdaSpace{}%
\AgdaSymbol{:}\AgdaSpace{}%
\AgdaPrimitive{Set}\AgdaSymbol{)}\AgdaSpace{}%
\AgdaSymbol{→}\<%
\\
\>[.][@{}l@{}]\<[391I]%
\>[8]\AgdaSymbol{(}\AgdaBound{B}\AgdaSpace{}%
\AgdaSymbol{:}\AgdaSpace{}%
\AgdaDatatype{Π}\AgdaSpace{}%
\AgdaBound{A}\AgdaSpace{}%
\AgdaSymbol{(λ}\AgdaSpace{}%
\AgdaBound{\AgdaUnderscore{}}\AgdaSpace{}%
\AgdaSymbol{→}\AgdaSpace{}%
\AgdaPrimitive{Set}\AgdaSymbol{))}\AgdaSpace{}%
\AgdaSymbol{→}\<%
\\
\>[8]\AgdaSymbol{(}\AgdaBound{C}\AgdaSpace{}%
\AgdaSymbol{:}\AgdaSpace{}%
\AgdaSymbol{(}\AgdaDatatype{Π}\AgdaSpace{}%
\AgdaBound{A}\AgdaSpace{}%
\AgdaSymbol{(λ}\AgdaSpace{}%
\AgdaBound{x}\AgdaSpace{}%
\AgdaSymbol{→}\AgdaSpace{}%
\AgdaDatatype{Π}\AgdaSpace{}%
\AgdaSymbol{(}\AgdaBound{B}\AgdaSpace{}%
\AgdaOperator{\AgdaFunction{\$}}\AgdaSpace{}%
\AgdaBound{x}\AgdaSymbol{)}\AgdaSpace{}%
\AgdaSymbol{(λ}\AgdaSpace{}%
\AgdaBound{\AgdaUnderscore{}}\AgdaSpace{}%
\AgdaSymbol{→}\AgdaSpace{}%
\AgdaPrimitive{Set}\AgdaSymbol{))))}\AgdaSpace{}%
\AgdaSymbol{→}\<%
\\
\>[8]\AgdaSymbol{(}\AgdaBound{f}\AgdaSpace{}%
\AgdaSymbol{:}\AgdaSpace{}%
\AgdaSymbol{(}\AgdaDatatype{Π}\AgdaSpace{}%
\AgdaBound{A}\AgdaSpace{}%
\AgdaSymbol{(λ}\AgdaSpace{}%
\AgdaBound{y}\AgdaSpace{}%
\AgdaSymbol{→}\AgdaSpace{}%
\AgdaDatatype{Π}\AgdaSpace{}%
\AgdaSymbol{(}\AgdaBound{B}\AgdaSpace{}%
\AgdaOperator{\AgdaFunction{\$}}\AgdaSpace{}%
\AgdaBound{y}\AgdaSymbol{)}\AgdaSpace{}%
\AgdaSymbol{(λ}\AgdaSpace{}%
\AgdaBound{z}\AgdaSpace{}%
\AgdaSymbol{→}\AgdaSpace{}%
\AgdaBound{C}\AgdaSpace{}%
\AgdaOperator{\AgdaFunction{\$}}\AgdaSpace{}%
\AgdaBound{y}\AgdaSpace{}%
\AgdaOperator{\AgdaFunction{\$}}\AgdaSpace{}%
\AgdaBound{z}\AgdaSymbol{))))}\AgdaSpace{}%
\AgdaSymbol{→}\<%
\\
\>[8]\AgdaSymbol{(}\AgdaBound{g}\AgdaSpace{}%
\AgdaSymbol{:}\AgdaSpace{}%
\AgdaSymbol{(}\AgdaDatatype{Π}\AgdaSpace{}%
\AgdaBound{A}\AgdaSpace{}%
\AgdaSymbol{(λ}\AgdaSpace{}%
\AgdaBound{x}\AgdaSpace{}%
\AgdaSymbol{→}\AgdaSpace{}%
\AgdaBound{B}\AgdaSpace{}%
\AgdaOperator{\AgdaFunction{\$}}\AgdaSpace{}%
\AgdaBound{x}\AgdaSymbol{)))}\AgdaSpace{}%
\AgdaSymbol{→}\<%
\\
\>[8]\AgdaDatatype{Π}\AgdaSpace{}%
\AgdaBound{A}\AgdaSpace{}%
\AgdaSymbol{(λ}\AgdaSpace{}%
\AgdaBound{x}\AgdaSpace{}%
\AgdaSymbol{→}\<%
\\
\>[8]\AgdaBound{C}\AgdaSpace{}%
\AgdaOperator{\AgdaFunction{\$}}\AgdaSpace{}%
\AgdaBound{x}\AgdaSpace{}%
\AgdaOperator{\AgdaFunction{\$}}\AgdaSpace{}%
\AgdaSymbol{(}\AgdaBound{g}\AgdaSpace{}%
\AgdaOperator{\AgdaFunction{\$}}\AgdaSpace{}%
\AgdaBound{x}\AgdaSymbol{))}\<%
\\
\\[\AgdaEmptyExtraSkip]%
\>[0]\AgdaSymbol{\{-\#}\AgdaSpace{}%
\AgdaKeyword{TERMINATING}\AgdaSpace{}%
\AgdaSymbol{\#-\}}\<%
\\
\>[0]\AgdaInductiveConstructor{F1}\AgdaSpace{}%
\AgdaOperator{\AgdaFunction{\$}}\AgdaSpace{}%
\AgdaBound{A}\AgdaSpace{}%
\AgdaSymbol{=}\AgdaSpace{}%
\AgdaInductiveConstructor{F2}\AgdaSpace{}%
\AgdaBound{A}\<%
\\
\>[0]\AgdaInductiveConstructor{F2}\AgdaSpace{}%
\AgdaBound{A}\AgdaSpace{}%
\AgdaOperator{\AgdaFunction{\$}}\AgdaSpace{}%
\AgdaBound{B}\AgdaSpace{}%
\AgdaSymbol{=}\AgdaSpace{}%
\AgdaInductiveConstructor{F3}\AgdaSpace{}%
\AgdaBound{A}\AgdaSpace{}%
\AgdaBound{B}\<%
\\
\>[0]\AgdaInductiveConstructor{F3}\AgdaSpace{}%
\AgdaBound{A}\AgdaSpace{}%
\AgdaBound{B}\AgdaSpace{}%
\AgdaOperator{\AgdaFunction{\$}}\AgdaSpace{}%
\AgdaBound{C}\AgdaSpace{}%
\AgdaSymbol{=}\AgdaSpace{}%
\AgdaInductiveConstructor{F4}\AgdaSpace{}%
\AgdaBound{A}\AgdaSpace{}%
\AgdaBound{B}\AgdaSpace{}%
\AgdaBound{C}\<%
\\
\>[0]\AgdaInductiveConstructor{F4}\AgdaSpace{}%
\AgdaBound{A}\AgdaSpace{}%
\AgdaBound{B}\AgdaSpace{}%
\AgdaBound{C}\AgdaSpace{}%
\AgdaOperator{\AgdaFunction{\$}}\AgdaSpace{}%
\AgdaBound{f}\AgdaSpace{}%
\AgdaSymbol{=}\AgdaSpace{}%
\AgdaInductiveConstructor{F5}\AgdaSpace{}%
\AgdaBound{A}\AgdaSpace{}%
\AgdaBound{B}\AgdaSpace{}%
\AgdaBound{C}\AgdaSpace{}%
\AgdaBound{f}\<%
\\
\>[0]\AgdaInductiveConstructor{F5}\AgdaSpace{}%
\AgdaBound{A}\AgdaSpace{}%
\AgdaBound{B}\AgdaSpace{}%
\AgdaBound{C}\AgdaSpace{}%
\AgdaBound{f}\AgdaSpace{}%
\AgdaOperator{\AgdaFunction{\$}}\AgdaSpace{}%
\AgdaBound{g}\AgdaSpace{}%
\AgdaSymbol{=}\AgdaSpace{}%
\AgdaInductiveConstructor{F6}\AgdaSpace{}%
\AgdaBound{A}\AgdaSpace{}%
\AgdaBound{B}\AgdaSpace{}%
\AgdaBound{C}\AgdaSpace{}%
\AgdaBound{f}\AgdaSpace{}%
\AgdaBound{g}\<%
\\
\>[0]\AgdaInductiveConstructor{F6}\AgdaSpace{}%
\AgdaBound{A}\AgdaSpace{}%
\AgdaBound{B}\AgdaSpace{}%
\AgdaBound{C}\AgdaSpace{}%
\AgdaBound{f}\AgdaSpace{}%
\AgdaBound{g}\AgdaSpace{}%
\AgdaOperator{\AgdaFunction{\$}}\AgdaSpace{}%
\AgdaBound{x}\AgdaSpace{}%
\AgdaSymbol{=}\AgdaSpace{}%
\AgdaBound{f}\AgdaSpace{}%
\AgdaOperator{\AgdaFunction{\$}}\AgdaSpace{}%
\AgdaBound{x}\AgdaSpace{}%
\AgdaOperator{\AgdaFunction{\$}}\AgdaSpace{}%
\AgdaSymbol{(}\AgdaBound{g}\AgdaSpace{}%
\AgdaOperator{\AgdaFunction{\$}}\AgdaSpace{}%
\AgdaBound{x}\AgdaSymbol{)}\<%
\end{code}

\section{Defunctionalization of fully-dependent \texttt{compose}}
\label{appendix:compose-defunctionalization}

\noindent
The fully dependent compose.
\[
\begin{array}{rcl}
  \text{compose} &::=& \lam{A}{U_0}
  {\enspace \lam{B}{(\pitype{x}{A}{U_0})}
  {\enspace \lam{C}{(\pitype{x}{A}{\pitype{y}{B\ x}{U_0}})}{} }}\\
    && \quad \lam{f}{(\pitype{y}{A}{(\pitype{z}{B\ y}{C\ y\ z})})}
    {\enspace \lam{g}{(\pitype{x}{A}{B\ x})}{} }\\
      && \qquad \lam{x}{A}{\enspace f\ x\ (g\ x)}
\end{array}
\]

\noindent
Defunctionalized non-dependent compose.

\[
\begin{array}{rcl}
  [[D]] &::=& 
\target{\flabel_2}(\{\target{B}:\target{U_0}, \target{C}:\target{U_0}, \target{A}:\target{U_0}, \target{f}:(\target{\Pi}\target{\_}:\target{B}.\target{C}), \target{g}:(\target{\Pi}\target{\_}:\target{A}.\target{B})\}, \target{x}:\target{A} \mapsto \target{f} \target{@} (\target{g} \target{@} \target{x}):\target{C}),\\
&&\target{\flabel_1}(\{\target{B}:\target{U_0}, \target{C}:\target{U_0}, \target{A}:\target{U_0}, \target{f}:(\target{\Pi}\target{\_}:\target{B}.\target{C})\}, \\
&& \quad \target{g}:(\target{\Pi}\target{\_}:\target{A}.\target{B}) \mapsto \target{\flabel_2}\{\target{B}, \target{C}, \target{A}, \target{f}, \target{g}\}:\target{\Pi}\target{x}:\target{A}.\target{C}),\\
&&\target{\flabel_0}(\{\target{B}:\target{U_0}, \target{C}:\target{U_0}, \target{A}:\target{U_0}\}, \\
&& \quad \target{f}:(\target{\Pi}\target{\_}:\target{B}.\target{C}) \mapsto \target{\flabel_1}\{\target{B}, \target{C}, \target{A}, \target{f}\}:\target{\Pi}\target{g}:(\target{\Pi}\target{\_}:\target{A}.\target{B}).\target{\Pi}\target{x}:\target{A}.\target{C})\\
  [[denv]] &::=& \target{A}:\target{U_0}, \target{B}:\target{U_0}, \target{C}:\target{U_0}\\
  \targettext{compose} &::=& [[LL_[0] {dnone}]]\\
  \targettext{composeT} &::=& \target{\Pi}\target{f}:(\target{\Pi}\target{\_}:\target{B}.\target{C}).\target{\Pi}\target{g}:(\target{\Pi}\target{\_}:\target{A}.\target{B}).\target{\Pi}\target{x}:\target{A}.\target{C}
\end{array}
\]

\noindent
Defunctionalized fully dependent compose.

\[
\begin{array}{rcl}
  [[D]]                 &::=& 
\target{\flabel_5}(\{\target{A}:\target{U_0}, \target{B}:(\target{\Pi}\target{\_}:\target{A}.\target{U_0}), \target{C}:(\target{\Pi}\target{x}:\target{A}.\target{\Pi}\target{\_}:(\target{B} \target{@} \target{x}).\target{U_0}), \\
&& \quad \target{f}:(\target{\Pi}\target{x}:\target{A}.\target{\Pi}\target{y}:(\target{B} \target{@} \target{x}).(\target{C} \target{@} \target{x}) \target{@} \target{y}), \target{g}:(\target{\Pi}\target{x}:\target{A}.\target{B} \target{@} \target{x})\},\\
&& \quad \target{x}:\target{A} \mapsto (\target{f} \target{@} \target{x}) \target{@} (\target{g} \target{@} \target{x}):(\target{C} \target{@} \target{x}) \target{@} (\target{g} \target{@} \target{x})),\\

&&\target{\flabel_4}(\{\target{A}:\target{U_0}, \target{B}:(\target{\Pi}\target{\_}:\target{A}.\target{U_0}), \target{C}:(\target{\Pi}\target{x}:\target{A}.\target{\Pi}\target{\_}:(\target{B} \target{@} \target{x}).\target{U_0}),\\
&& \quad \target{f}:(\target{\Pi}\target{x}:\target{A}.\target{\Pi}\target{y}:(\target{B} \target{@} \target{x}).(\target{C} \target{@} \target{x}) \target{@} \target{y})\},\\
&& \quad \target{g}:(\target{\Pi}\target{x}:\target{A}.\target{B} \target{@} \target{x}) \mapsto \target{\flabel_5}\{\target{A}, \target{B}, \target{C}, \target{f}, \target{g}\}:\target{\Pi}\target{x}:\target{A}.(\target{C} \target{@} \target{x}) \target{@} (\target{g} \target{@} \target{x})),\\

&&\target{\flabel_3}(\{\target{A}:\target{U_0}, \target{B}:(\target{\Pi}\target{\_}:\target{A}.\target{U_0}), \target{C}:(\target{\Pi}\target{x}:\target{A}.\target{\Pi}\target{\_}:(\target{B} \target{@} \target{x}).\target{U_0})\},\\
&& \quad  \target{f}:(\target{\Pi}\target{x}:\target{A}.\target{\Pi}\target{y}:(\target{B} \target{@} \target{x}).(\target{C} \target{@} \target{x}) \target{@} \target{y}) \mapsto \target{\flabel_4}\{\target{A}, \target{B}, \target{C}, \target{f}\}:\\
&& \quad \target{\Pi}\target{g}:(\target{\Pi}\target{x}:\target{A}.\target{B} \target{@} \target{x}).\target{\Pi}\target{x}:\target{A}.(\target{C} \target{@} \target{x}) \target{@} (\target{g} \target{@} \target{x})),\\

&&\target{\flabel_2}(\{\target{A}:\target{U_0}, \target{B}:(\target{\Pi}\target{\_}:\target{A}.\target{U_0})\}, \target{C}:(\target{\Pi}\target{x}:\target{A}.\target{\Pi}\target{\_}:(\target{B} \target{@} \target{x}).\target{U_0}) \mapsto \target{\flabel_3}\{\target{A}, \target{B}, \target{C}\}:\\
&& \quad \target{\Pi}\target{f}:(\target{\Pi}\target{x}:\target{A}.\target{\Pi}\target{y}:(\target{B} \target{@} \target{x}).(\target{C} \target{@} \target{x}) \target{@} \target{y}).\target{\Pi}\target{g}:(\target{\Pi}\target{x}:\target{A}.\target{B} \target{@} \target{x}).\target{\Pi}\target{x}:\target{A}.(\target{C} \target{@} \target{x}) \target{@} (\target{g} \target{@} \target{x})),\\

&&\target{\flabel_1}(\{\target{A}:\target{U_0}\}, \target{B}:(\target{\Pi}\target{\_}:\target{A}.\target{U_0}) \mapsto \target{\flabel_2}\{\target{A}, \target{B}\}: \quad \target{\Pi}\target{C}:(\target{\Pi}\target{x}:\target{A}.\target{\Pi}\target{\_}:(\target{B} \target{@} \target{x}).\target{U_0}).\\
&& \quad \target{\Pi}\target{f}:(\target{\Pi}\target{x}:\target{A}.\target{\Pi}\target{y}:(\target{B} \target{@} \target{x}).(\target{C} \target{@} \target{x}) \target{@} \target{y}).\target{\Pi}\target{g}:(\target{\Pi}\target{x}:\target{A}.\target{B} \target{@} \target{x}).\target{\Pi}\target{x}:\target{A}.(\target{C} \target{@} \target{x}) \target{@} (\target{g} \target{@} \target{x})),\\

&&\target{\flabel_0}(\{\}, \target{A}:\target{U_0} \mapsto \target{\flabel_1}\{\target{A}\}:\target{\Pi}\target{B}:(\target{\Pi}\target{\_}:\target{A}.\target{U_0}).\target{\Pi}\target{C}:(\target{\Pi}\target{x}:\target{A}.\target{\Pi}\target{\_}:(\target{B} \target{@} \target{x}).\target{U_0}).\\
&& \quad \target{\Pi}\target{f}:(\target{\Pi}\target{x}:\target{A}.\target{\Pi}\target{y}:(\target{B} \target{@} \target{x}).(\target{C} \target{@} \target{x}) \target{@} \target{y}).\target{\Pi}\target{g}:(\target{\Pi}\target{x}:\target{A}.\target{B} \target{@} \target{x}).\target{\Pi}\target{x}:\target{A}.(\target{C} \target{@} \target{x}) \target{@} (\target{g} \target{@} \target{x}))\\

  \targettext{compose}  &::=& [[LL_[0] {dnone} ]]\\
  \targettext{composeT} &::=& 
  [[Pi dxA : dU[0] . dnone]] \enspace
  [[Pi dxB : (!Pi dx : dxA . dU[0]!) . dnone]] \enspace
  [[Pi dxC : (!Pi dx : dxA . (Pi dy : dxB @ dx . dU[0])!) . dnone ]]\\
  && [[Pi dxf : (!Pi dx : dxA . (Pi dy : dxB @ dx . (! dxC @ dx @ dy !)) !) . dnone ]]\\
  && [[Pi dxg : (!Pi dx : dxA . (dxB @ dx)!) . dnone ]]\\
  && [[Pi dx : dxA . ((!dxC @ dx!) @ dxg @ dx)]]
\end{array}
\]

A sketched derivation of the defunctionalization transformation.

\begin{prooftree}
\AxiomC{$A, B, C, f, g \vdash \lambda^5x.f\;x\;(g\;x)
\leadsto_d \target{\flabel_5}(\{\target{A},\target{B},\target{C},\target{f},\target{g}\}, \target{x}:\_ \mapsto [[(!dxf @ dx!) @ dxg @ dx]] : \_ )
$}
\RightLabel{\rref*{d-Lambda}}
\UnaryInfC{$A, B, C, f \vdash \lambda^4g.\lambda^5x.f\;x\;(g\;x)
\leadsto_d \cdots, \target{\flabel_4}(\{\target{A},\target{B},\target{C},\target{f}\}, \target{g}:\_ \mapsto [[LL_[5] {dxA,dxB,dxC,dxf,dxg}]] : \_ )
$}
\RightLabel{\rref*{d-Lambda}}
\UnaryInfC{$A, B, C \vdash \lambda^3f.\lambda^4g.\lambda^5x.f\;x\;(g\;x)
\leadsto_d \cdots, \target{\flabel_3}(\{\target{A},\target{B},\target{C}\}, \target{f}:\_ \mapsto [[LL_[4] {dxA,dxB,dxC,dxf}]] : \_ )
$}
\RightLabel{\rref*{d-Lambda}}
\UnaryInfC{$A, B\vdash \lambda^2C.\lambda^3f.\lambda^4g.\lambda^5x.f\;x\;(g\;x)
\leadsto_d \cdots, \target{\flabel_2}(\{\target{A},\target{B}\}, \target{C}:\_ \mapsto [[LL_[3] {dxA,dxB,dxC}]] : \_ )
$}
\RightLabel{\rref*{d-Lambda}}
\UnaryInfC{$A \vdash \lambda^1B.\lambda^2C.\lambda^3f.\lambda^4g.\lambda^5x.f\;x\;(g\;x)
\leadsto_d \cdots, \target{\flabel_1}(\{\target{A}\}, \target{B}:\_ \mapsto [[LL_[2] {dxA,dxB}]] : \_ )
$}
\RightLabel{\rref*{d-Lambda}}
\UnaryInfC{$\cdot \vdash \lambda^0A.\lambda^1B.\lambda^2C.\lambda^3f.\lambda^4g.\lambda^5x.f\;x\;(g\;x)
\leadsto_d \cdots, \target{\flabel_0}(\{\}, \target{A}:\_ \mapsto [[LL_[1] {dxA}]] : \_ )
$}
\end{prooftree}

\renewcommand{\appendix}{}
\end{document}